# Geoneutrinos and geoscience:
# an intriguing joint-venture

G. Bellini[1], K. Inoue[2,3], F. Mantovani[4], A. Serafini[4], V. Strati[4], H. Watanabe[2,3]

[1]*Department of Physics, University of Milan and INFN, 20133 Milan, Italy*
[2]*Research Center for Neutrino Science, Tohoku University, Sendai 980-8578, Japan*
[3]*Institute for the Physics and Mathematics of the Universe, Tokyo University, Kashiwa 277-8568, Japan*
[4]*Department of Physics and Earth Sciences, University of Ferrara and INFN, 44122 Ferrara, Italy*

**Abstract**

The review is conceived to help the reader to interpret present geoneutrino results in the framework of Earth's energetics and composition. Starting from the comprehension of antineutrino production, propagation, and detection, the status of the geoneutrino field is presented through the description of the experimental and technological features of the Borexino and KamLAND ongoing experiments. The current understanding of the energetical, geophysical and geochemical traits of our planet is examined in a critical analysis of the currently available models. By combining theoretical models and experimental results, the mantle geoneutrino signal extracted from the results of the two experiments demonstrates the effectiveness in investigating deep earth radioactivity through geoneutrinos from different sites. The obtained results are discussed and framed in the puzzle of the diverse classes of formulated Bulk Silicate Earth models, analyzing their implications on planetary heat budget and composition. As a final remark, we present the engaging technological challenges and the future experiments envisaged for the next decade in the geoneutrino field.

# Index

# 1 Introduction

The ephemeral properties of neutrinos have always been the blessing and the curse for their employment in the comprehension of the Universe. Their weak interaction with matter makes them particularly elusive particles to detect, but also precious probes for exploring the most remote parts of the Earth, Sun and stars. The use of neutrinos for the real-time monitoring of the thermonuclear fusion processes occurring in the Sun's core has represented a major astrophysics milestone of the past fifty years. Concurrently, starting from the mid-twentieth century, the electron antineutrinos originating from $\beta^-$ emitters inside our planet, geoneutrinos, were proposed as a precious tool for exploring the inner Earth. Since from the first formulation of this hypothesis, it was clear to everyone that the low expected flux and the small cross section would represent an arduous technological challenge for future experiments. The lack of stringent constraints on the radiogenic contribution to terrestrial heat power and the impossibility of having direct geochemical insights of the deep Earth, make neutrino geoscience a particularly multifaceted and convoluted discipline.

Why a review about geoneutrinos at the beginning of the third decade of the 21$^{st}$ century? After the claims of the first geoneutrino observations by the KamLAND collaboration in 2005 [1] followed by the Borexino collaboration in 2010 [2], the results published with greater statistical significance in the second decade of the 21$^{st}$ century highlighted an unavoidable necessity of geophysical and geological models for understanding geoneutrino signals (Figure 1).

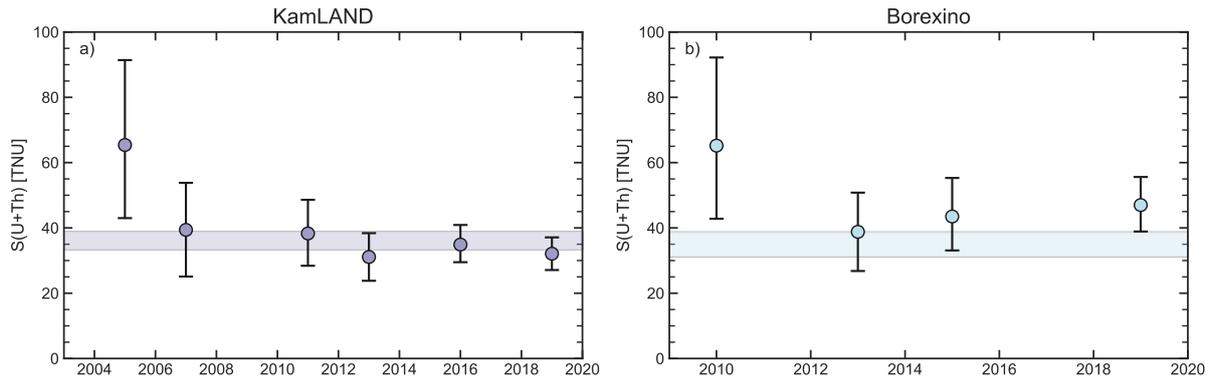

**Figure 1** - Comparison of the geoneutrino signal S(U+Th) (in TNU) measured by **(a)** KamLAND and **(b)** Borexino during the last decades. KamLAND and Borexino results are taken from [1,3-7] and from [2,8-10] respectively. Horizontal bars represent the 1σ C.L. of the signal expected following the arguments presented in Section 9.

The next decade will open the era of "multi-site detection" of geoneutrinos. By 2030, humanity would collect the results from four experiments spread over three different continents: KamLAND and JUNO in Asia, Borexino in Europe and SNO + in America.

All these direct measurements of antineutrinos produced by U and Th decay chains in the bowels of the Earth will pose stringent constraints to fundamental questions about the formation, thermal history, dynamics and composition of our planet. This monumental experimental effort will be in vain unless followed by an equally enormous joint effort between particle physicists and Earth scientists in understanding each other's paradigms and methods. The study of uncertainties and their correlations is the new frontiers of geoneutrino science: no experimental data will be fully understood without the development of a common ground among different disciplines.

This review is a small step in the direction of an ever-deeper integration between experimental observations and theoretical modeling. Each section can be savored independently without compulsive reading of the previous or subsequent parts. The key ingredients for understanding the production, propagation and detection of geoneutrinos are presented in Section 2. The results of the only two experiments that so far measured geoneutrinos, KamLAND and Borexino, are presented in Sections 3 and 4 and will allow the reader to appreciate their technological and scientific value. Since mantle geoneutrinos potentially bring to the surface geochemical and energetic information about the unexplored deep Earth, a combined analysis of the two experimental signals is performed in Section 9 and discussed in terms of impact on radiogenic heat power and mantle composition in Section 10. The reader who is not familiar with the geophysical structure of our planet can find an overview of its main features in Section 5. Section 6 presents the current understanding of Earth's heat budget, including the modeling of the heat flow at surface, secular cooling and radiogenic heat production, which can be directly investigated measuring geoneutrinos which escape freely and instantaneously from Earth's interior. The compositional models envisaged for the outermost shell of our planet (i.e., the lithosphere) and the deep Earth are the essential key for interpreting experimental results: they are critically reviewed in Section 7 and 8 with the aim of guiding the reader among the numerous models published in the last decades. Section 11 offers a glance at the future presenting new-generation conventional detectors for $^{238}$U and $^{232}$Th geoneutrinos measurements and innovative techniques for investigating $^{40}$K geoneutrinos and the direction of incoming neutrinos.

## 2   What are geoneutrinos?

Geoneutrinos are antineutrinos produced by the decay of radioactive isotopes present inside the Earth [11]. They belong to the families with a half-life comparable to or longer than the Earth's age (4.543 Gyr) with $^{238}$U, $^{232}$Th, $^{40}$K, which are naturally present in the Earth as progenitors[1] (Table 1). While decaying, these isotopes produce not only antineutrinos, but also energy, the so-called *radiogenic heat* of the present Earth. For this reason, these elements are usually referred to as *Heat-Producing Elements* (HPEs). The production of antineutrinos and radiogenic heat occurs in a well-fixed ratio (Section 6.2.1) and for this reason a measurement of the geoneutrino flux can be used as a probe for estimating the radiogenic heat production of the inaccessible Earth.

**Table 1** – Main properties of $^{238}$U, $^{232}$Th decay chains and $^{40}$K β⁻ decay. For each parent nucleus the table presents the natural isotopic abundance $X_{iso}$ and the half-life $T_{1/2}$ from the most recent Nuclear Data Sheets [12-14], where the numbers in brackets represent the uncertainty on the last digits of the value. The number N of antineutrinos emitted per decay of the parent nucleus represents the number of β⁻ decays needed to reach a stable nucleus: for $^{40}$K this is represented by its β⁻ branching ratio [14]. The maximal energy of the emitted antineutrino $E_{max}$ are taken from [15], while the antineutrino production rates for unit mass of the isotope ($\epsilon_{\bar{\nu}}$) and for unit mass at natural isotopic abundance ($\epsilon'_{\bar{\nu}}$) are derived from $X_{iso}$, $T_{1/2}$ and N. Production rates $\epsilon_{\bar{\nu}}$ and $\epsilon'_{\bar{\nu}}$ for $^{40}$K are higher than previous reports because of the recently updated $^{40}$K's half-life $T_{1/2}$ estimation [14], which is respectively 2.6%, 2.3% and 1.1% lower than what reported in [15], [16] and [17].

| Decay family | $X_{iso}$ [mole fraction] | $T_{1/2}$ [Gyr] | N [decay⁻¹] | $E_{max}$ [MeV] | $\epsilon_{\bar{\nu}}$ [kg⁻¹ s⁻¹] | $\epsilon'_{\bar{\nu}}$ [kg⁻¹ s⁻¹] |
|---|---|---|---|---|---|---|
| $^{238}U \rightarrow {}^{206}Pb + 8\alpha + 6e^- + 6\bar{\nu}_e$ | 0.992742 (10) | 4.468 (6) | 6 | 3.26* | 7.46 × 10⁷ | 7.41 × 10⁷ |
| $^{232}Th \rightarrow {}^{208}Pb + 6\alpha + 4e^- + 4\bar{\nu}_e$ | 1 | 14.0 (1) | 4 | 2.25 | 1.63 × 10⁷ | 1.63 × 10⁷ |
| $^{40}K \rightarrow {}^{40}Ca + e^- + \bar{\nu}_e$ (89.28%) | 1.17 × 10⁻⁴ (1) | 1.248 (3) | 0.8928 (11) | 1.311 | 2.37 × 10⁸ | 2.83 × 10⁴ |

*The $^{238}$U decay chain contains β⁻ decays (e.g. $^{210}$Tl) producing geoneutrinos with energies >3.26 MeV. Historically, these are not considered for geoneutrino analyses because of their low intensities, impossible to observe through current experiments.

The unoscillated geoneutrino flux at position $\bar{r}$ on Earth's surface ($\Phi(i, \bar{r})$) depends only on the abundance and distribution of the HPEs inside the planet and can be modelled as:

$$\Phi(i, \bar{r}) = \epsilon'_{i,\bar{\nu}} \cdot \int_0^{E_{max}} dE_{\bar{\nu}} \cdot Sp(i, E_{\bar{\nu}}) \int d^3r' \cdot \frac{a(i, \bar{r}') \cdot \rho(\bar{r}')}{4\pi |\bar{r} - \bar{r}'|^2} \qquad (1)$$

where $a(i, \bar{r}')$ is the abundance of the *i*-th HPE as a function of its position $\bar{r}'$ inside the Earth, $\rho(\bar{r}')$ is Earth's mass density function, $|\bar{r} - \bar{r}'|$ is the distance between the antineutrino production point and the detector, $\epsilon'_{i,\bar{\nu}}$ is the antineutrino production rate for unit mass of the element *i* at natural isotopic composition and $Sp(i, E_{\bar{\nu}})$ is the energy spectrum of the produced geoneutrino (Figure 2) which results in N (Table 1) once integrated over the entire antineutrino energy.

The detection technique permitting to current and future experiments to measure this flux is based on Inverse Beta Decay (IBD) on free protons, a charged current interaction which makes detectable only electron-flavoured antineutrinos ($\bar{\nu}_e$). Since geoneutrinos, as all neutrinos and antineutrinos, undergo a phenomenon called neutrino oscillation, the effective geoneutrino flux $\Phi'(i, \bar{r})$ observed at detector site appears reduced by this flavor oscillation. The transformation matrix ruling these oscillations is called Pontecorvo–Maki–Nakagawa–Sakata matrix (PMNS matrix) [18], a presumably unitary matrix [19] depending on the mixing angles between neutrinos eigenstates $\theta_{12}$, $\theta_{13}$ and $\theta_{23}$, the square mass differences $\delta m^2$ and $\Delta m^2$ between these states and the δ phase accounting for the possible CP violation (Table 2).

---
[1] Geoneutrinos emitters include $^{40}$K, $^{87}$Rb, $^{113}$Cd, $^{115}$In, $^{138}$La, $^{176}$Lu, $^{187}$Re and the elements belonging to the decay chains of $^{232}$Th, $^{235}$U and $^{238}$U. Because of their longer half-lives or higher abundances, the most important emitters in terms of luminosity are $^{40}$K and the ones belonging to $^{232}$Th and $^{238}$U decay chains, with only the latter two observable with present detection techniques. Differently from the other mentioned isotopes (which only undergo β⁻ decays), $^{40}$K can produce both neutrinos and antineutrinos (Section 11.6). However, the detection of neutrinos is prevented by their low energy and the overwhelming solar neutrino flux which is nearly three orders of magnitude higher.

**Table 2** - Updated oscillation parameters for neutrino oscillation together with their 1σ range. Values are taken from [18], considering normal ordering in the mass hierarchy.

|  | Best fit | 1σ range |
|---|---|---|
| $\delta m^2$ | 7.34 × 10⁻⁵ eV² | [7.20 – 7.51] × 10⁻⁵ eV² |
| $sin\ ^2\theta_{12}$ | 3.04 × 10⁻¹ | [2.91 – 3.18] × 10⁻¹ |
| $sin\ ^2\theta_{13}$ | 2.14 × 10⁻² | [2.07 – 2.23] × 10⁻² |
| $|\Delta m^2|$ | 2.455 × 10⁻³ eV² | [2.423 – 2.490] × 10⁻³ eV² |
| $sin\ ^2\theta_{23}$ | 5.51 × 10⁻¹ | [4.81 – 5.70] × 10⁻¹ |
| δ | 1.32 π | [1.14 π – 1.55 π] |

By taking into account this flavor oscillation, the *survival probability*, namely the probability of observing an antineutrino of energy $E_{\bar{\nu}}$ which travelled a distance $L$ from its emission point as still electron-flavored, can be approximated to [20]:

$$P_{ee}(L, E_{\bar{\nu}}) \sim cos\ ^4\theta_{13}\left(1 - sin\ ^22\theta_{12} sin\ ^2\left(\frac{\delta m^2 L}{4 E_{\bar{\nu}}}\right)\right) + sin\ ^4\theta_{13} \qquad (2)$$

Hence, by making use of this equation, the effective *oscillated flux* ($\Phi'(i, \bar{r})$) detectable by IBD-based experiments can be modelled as:

$$\Phi'(i, \bar{r}) = \epsilon'_{i,\bar{\nu}} \cdot \int dE_{\bar{\nu}} \cdot Sp(i, E_{\bar{\nu}}) \int d^3 r' \cdot P_{ee}(|\bar{r} - \bar{r}'|, E_{\bar{\nu}}) \cdot \frac{a(i, \bar{r}') \cdot \rho(\bar{r}')}{4\pi |\bar{r} - \bar{r}'|^2} \qquad (3)$$

By inputting in this equation model-dependent assumptions based on geochemistry and geophysics evidences ($a(i, \bar{r}') \cdot \rho(\bar{r}')$) and integrating on Earth's volume, it is possible, starting from the geoneutrino energy spectra $Sp(i, E_{\bar{\nu}})$, to estimate the flux energy spectra $\Phi'(\bar{r}, E_{\bar{\nu}})$ at a given position (Figure 2). Typically, the expected geoneutrino flux at surface is ~10⁶ cm⁻² s⁻¹, which is dominated by the crustal contributions (Section 8). In order to recover any kind of information on Earth's radiogenic heat production from a flux measurement, the final goal is the extraction of $a(i, \bar{r}')$. However, recovering this information is not straightforward, since the experimentally measured geoneutrino flux represents a volume integral weighted by the inverse square distance, and modulated by the $P_{ee}$ oscillation probability. While the latter two ingredients are known with good accuracy, the volume distribution of Th and U is subjected to relatively large uncertainties, especially in the mantle (Section 5.3). Hence, to disentangle interesting pieces of information from geoneutrino measurements, scientists need an interdisciplinary approach capable of including supplementary constraints and assumptions from Earth science.

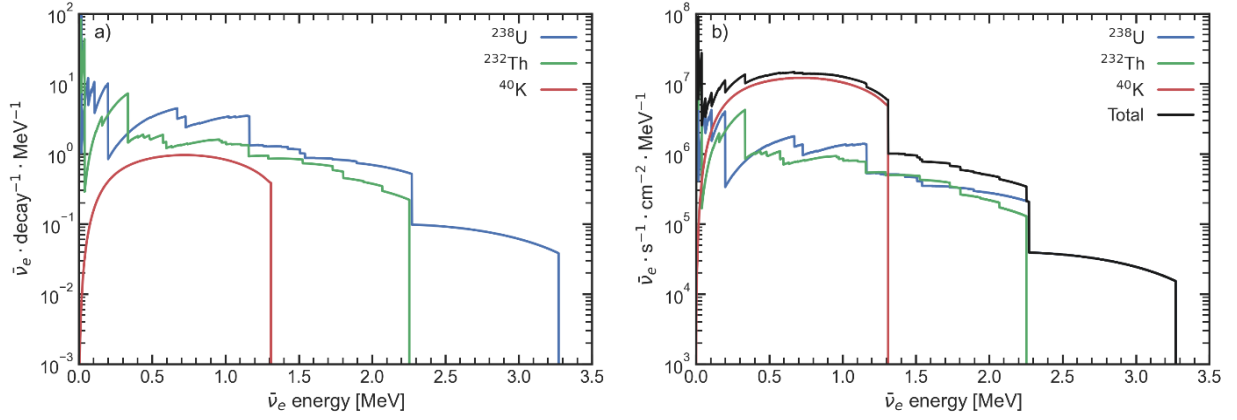

**Figure 2 – (a)** Geoneutrino energy spectra of $^{238}$U (in blue), $^{232}$Th (in green) and $^{40}$K (in red) from **[21]**. All spectra are normalized to one decay of the head element of the chain, leading to a total of 6, 4 and 0.89 geoneutrinos for the energy spectra of $^{238}$U, $^{232}$Th and $^{40}$K, respectively. **(b)** Estimated oscillated fluxes of $^{238}$U (in blue), $^{232}$Th (in green), $^{40}$K (in red) geoneutrinos and their sum (in black) at Laboratori Nazionali del Gran Sasso (LNGS) as a function of the geoneutrino energy.

Because of the homogeneity of geoneutrino production inside the Earth and the wide energy range of their spectra, most authors make the reasonable approximation of oscillation-averaged $P_{ee}$. Under this approximation, the oscillating term $\sin^2\left(\frac{\delta m^2 L}{4E}\right)$ is averaged over energy and distance at its average integrated value of 1/2. Hence, the survival probability for electron flavored antineutrinos can be simplified to [20]:

$$\langle P_{ee} \rangle \cong \cos^4\theta_{13}\left(1 - \frac{1}{2}\sin^2 2\theta_{12}\right) + \sin^4\theta_{13} \tag{4}$$

The survival probability rapidly converges to its approximated value $\langle P_{ee} \rangle$ for distances >100 km. Instead, for distances <100 km, the survival probability has a stronger impact in the flux estimation and for this reason the Earth's region in the vicinity of the detector needs a refined modeling (Section 8.2). A functional approach to flux estimations consists in utilizing the precise survival probability for local regions, and the average survival probability for the rest of the Earth. The *a posteriori* assessment of the effect of this approximation [20] justifies its use, as the impact on the estimated signal is <0.2%, well below the experimental sensitivity reached so far (Figure 3).

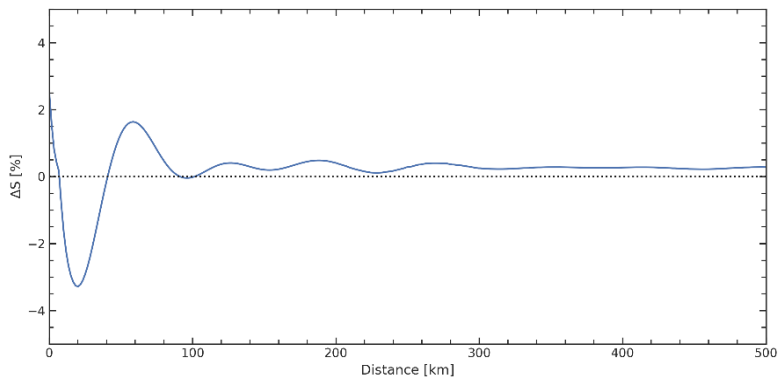

**Figure 3 –** Percentage difference between the signal calculated by employing non-averaged (Eq. (2)) and averaged (Eq. (4)) survival probability normalized to the former signal as a function of the distance at which the survival probability is approximated at its average value of 0.55. Oscillation probability is calculated using best fit coefficients from Table 2.

The increasingly important achievements in the field of neutrino physics led to a continued refinement of the oscillation parameters. Because of the different values associated to $\theta_{12}$ and $\theta_{13}$, the average survival probability $\langle P_{ee} \rangle$ used by the different authors changed along the years (Table 3), leading to slightly different flux estimates.

**Table 3** – Oscillation parameters ($\sin^2 \theta_{12}$, $\sin^2 \theta_{13}$) and derived average survival probability ($\langle P_{ee} \rangle$) used by different authors in the calculation of geoneutrino signals.

| Reference | $\sin^2 \theta_{12}$ | $\sin^2 \theta_{13}$ | $\langle P_{ee} \rangle$ |
|---|---|---|---|
| Mantovani et al., 2004 [22] | 0.315 ± 0.035 | - | 0.57 |
| Fogli et al., 2006 [23] | 0.31$^{+0.06}_{-0.05}$ | 0.009$^{+0.023}_{-0.009}$ | 0.57 |
| Enomoto et al., 2007 [3] | 0.29$^{+0.05}_{-0.04}$ | - | 0.595 |
| Dye, 2010 [24] | 0.32$^{+0.03}_{-0.02}$ | | 0.56 ± 0.02 |
| Fiorentini et al., 2012 [20] | 0.306 ± 0.017 | 0.021 ± 0.007 | 0.551 ± 0.015 |
| Huang et al., 2013 [25] | - | - | 0.55 |
| Šrámek et al., 2016 [26] | - | - | 0.553 |

After 2012, the adopted value for $\langle P_{ee} \rangle$ has converged to 0.55. When comparing recent estimates with signals obtained before 2012, careful attention must be given to rescaling the results to the up-to-date average survival probability.

The only two running experiments in the world which until now measured geoneutrinos are Borexino and KamLAND. Both experiments make use of the IBD reaction on free protons to detect antineutrinos:

$$\bar{\nu} + p \rightarrow e^+ + n - 1.806 \; MeV \tag{5}$$

In this reaction, the incoming antineutrino collides with a proton, producing a positron and a neutron (Figure 4). The outgoing positron promptly annihilates, producing two 511 keV gammas, usually referred as *prompt signal*. The outgoing neutron takes a mean time of ~≥200 μs (254.5 ± 1.8 μs in Borexino and 207.5 ± 2.8 μs in KamLAND) to thermalize and then to be captured by a proton, producing a deuteron with the emission of a 2.2 MeV gamma (*delayed signal*), with an 1.1% of neutrons captured by a $^{12}$C nucleus, with the emission of a 4.95 MeV gamma. Whereas these prompt and delayed signals are both time and space correlated, background-induced signals are not. Hence, this delayed coincidence method provides an extremely powerful background suppression, working as a very effective tagging technique for antineutrino interactions. KamLAND and Borexino employ the hydrogen atoms (protons) attached to hydrocarbon molecules as targets for IBD. Both use pseudocumene (1,2,4-trimethylbenzene) as scintillator solvent and PPO (2,5-diphenyloxazole) as fluor, but with slightly different percentages: in Borexino the PPO has a concentration of 1.5 g/l, while in KamLAND the concentration is 1.36 g/l. In addition, in KamLAND an 80% of an oil, dodecane, is added.

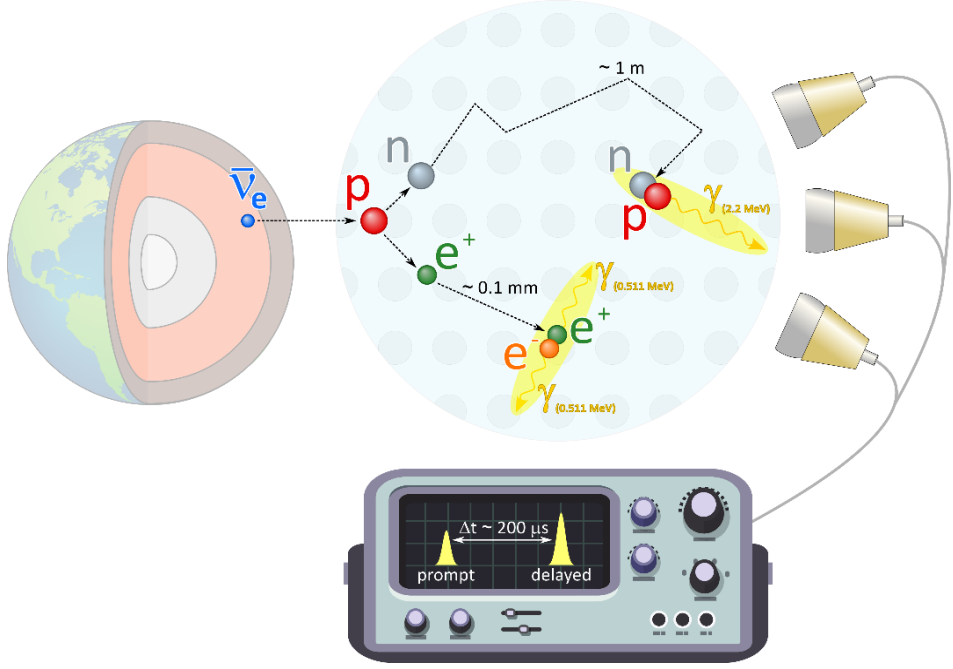

**Figure 4** – A scheme of the IBD reaction. An incoming antineutrino collides with a proton (p) inside the liquid scintillator, producing a neutron (n) and a positron (e⁺). The positron readily annihilates producing two 0.511 MeV photons (γ) which excites the scintillator molecules, producing scintillating light seen by the photomultiplier tubes (PMTs) as a prompt signal. After thermalizing (typically in ~≥200 μs), the neutron is later captured by a proton producing a 2.2 MeV photon, seen by the PMTs as a delayed signal.

The IBD, whose cross-section $\sigma$ is known with 0.4% uncertainty, allowed KamLAND and Borexino to collect data even in periods when it is impossible to study neutrino interactions, as during operations in the detector or with the detector equipped for other scientific purposes. However, this reaction has a kinematic threshold of 1.806 MeV due the mass difference between the produced neutron and the target proton. Consequently, only 38% of the geoneutrinos emitted by the $^{238}$U decay chain and 15% from $^{232}$Th one remain above this threshold, while those from $^{40}$K remain undetectable.

The U and Th geoneutrino signals rates $S(U,\bar{r})$ and $S(Th,\bar{r})$ observed by a detector at position $\bar{r}$ can be calculated convolving the differential oscillated geoneutrino fluxes with the IBD cross section $\sigma$:

$$S(U,\bar{r}) = N_p \cdot T \cdot \epsilon'_{U,\bar{\nu}} \cdot \eta \cdot \int dE_{\bar{\nu}} \cdot \sigma(E_{\bar{\nu}}) \cdot Sp(U, E_{\bar{\nu}}) \cdot \int d^3r' \cdot P_{ee}(|\bar{r}-\bar{r}'|, E_{\bar{\nu}}) \cdot \frac{a(U,\bar{r}') \cdot \rho(\bar{r}')}{4\pi|\bar{r}-\bar{r}'|^2}$$

$$S(Th,\bar{r}) = N_p \cdot T \cdot \epsilon'_{Th,\bar{\nu}} \cdot \eta \cdot \int dE_{\bar{\nu}} \cdot \sigma(E_{\bar{\nu}}) \cdot Sp(Th, E_{\bar{\nu}}) \cdot \int d^3r' \cdot P_{ee}(|\bar{r}-\bar{r}'|, E_{\bar{\nu}}) \cdot \frac{a(Th,\bar{r}') \cdot \rho(\bar{r}')}{4\pi|\bar{r}-\bar{r}'|^2}$$

(6)

where $E_{\bar{\nu}}$ is the antineutrino energy integrated from 0 up to the endpoint of the antineutrino spectra, $N_p$ is the number of proton targets available in the detector, $T$ is the exposure time and $\eta$ is the detector efficiency. Historically, geoneutrinos signals have been measured in *Terrestrial Neutrino Units* (TNU), where 1 TNU corresponds to 1 antineutrino event measured over 1 year by a detector containing $10^{32}$ free protons target, assuming 100% detection efficiency. By convolving the oscillated fluxes shown in Figure 2 with the IBD cross section $\sigma$, it is possible to estimate the signal energy spectra expected at detector position (Figure 5).

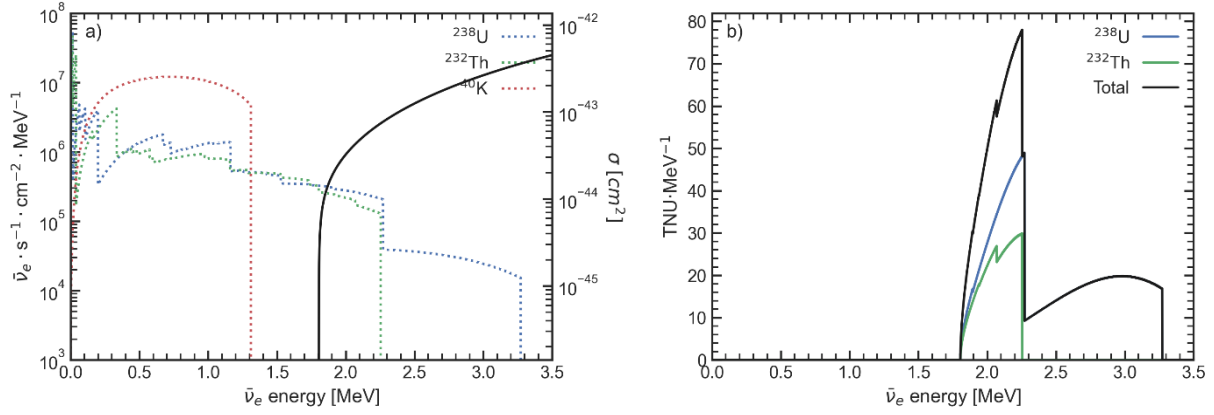

**Figure 5 - (a)** IBD cross section σ (in black, expressed in cm²) as a function of geoneutrino energy (in MeV). The dotted lines in the background show the oscillated fluxes of $^{238}$U (in blue), $^{232}$Th (in green) and $^{40}$K (in red) geoneutrinos at LNGS (Figure 2). Because of the IBD kinematic threshold at 1.806 MeV, only 38% and 15% of the $^{238}$U and $^{232}$Th fluxes are detectable. The entire $^{40}$K flux is instead below threshold. **(b)** $^{238}$U (in blue), $^{232}$Th (in green) and total (in black) geoneutrino signals (in TNU) expected at LNGS. These spectra are the results of the convolution between geoneutrino fluxes and IBD cross-section depicted in (a).

As we did for the fluxes, the formulas obtained for the geoneutrino signals can be simplified with some assumptions. Indeed, by substituting in Eq. (6) the average survival probability $\langle P_{ee} \rangle$ it is possible to separately integrate on volume (hence recovering the cumulative unoscillated antineutrino flux $\Phi$) and on energy. The energy integral $\int \sigma(E_{\bar{\nu}}) \cdot Sp_i(E_{\bar{\nu}}) \cdot dE_{\bar{\nu}}$ is usually referred to as the integrated IBD cross section $\langle \sigma \rangle$, which can be easily calculated from the IBD cross section. For U and Th spectra, $\langle \sigma \rangle$ assumes the values $\langle \sigma(U) \rangle = 2.8\ TNU \cdot 10^{-6} s \cdot cm^2$ and $\langle \sigma(Th) \rangle = 4.04\ TNU \cdot 10^{-6} s \cdot cm^2$ [10], hence permitting to express the geoneutrino signals as:

$$S(U, \bar{r}) \cong \langle \sigma(U) \rangle \cdot \langle P_{ee} \rangle \cdot \Phi(U, \bar{r}) = 12.8 \cdot \langle P_{ee} \rangle \cdot \Phi(U, \bar{r})$$
$$S(Th, \bar{r}) \cong \langle \sigma(Th) \rangle \cdot \langle P_{ee} \rangle \cdot \Phi(Th, \bar{r}) = 4.04 \cdot \langle P_{ee} \rangle \cdot \Phi(Th, \bar{r})$$
(7)

where signals $S$ and the unoscillated fluxes $\Phi$ are expressed in TNU and $10^6 s^{-1} \cdot cm^{-2}$.

## 3　KamLAND

### 3.1　KamLAND experiment

KamLAND (Kamioka Liquid scintillator Anti-Neutrino Detector) is a multi-purpose experiment with very low radio-impurity and large target volume liquid scintillator. Because of these detector features, KamLAND has a low energy threshold and can distinguish electron antineutrinos from the other types via IBD reaction. This method is beneficial in strongly suppressing background interference and enables the detector to address a variety of scientific topics.

The primary goal of KamLAND is the detection of $\bar{\nu}_e$ from nuclear power reactors and to demonstrate the oscillation nature of neutrino flavor transformation. In 2002, KamLAND observed the significant deficit of $\bar{\nu}_e$ from reactors and the oscillatory function of the flavor changing probability which is the most characteristic feature of neutrino oscillation [27,28]. The neutrino oscillation frequency is related to the mass-squared differences and its amplitude with mixing angles. These parameters were measured using a mixture of results from experiments involving solar, reactor, atmospheric, and accelerator neutrinos. KamLAND determined the precise value for the oscillation parameters combined with other experiments, and the latest result presented a combined three-flavor analysis of solar and KamLAND, which exhibited best fit values for the oscillation parameters of $\tan^2 \theta_{12} = 0.436^{+0.029}_{-0.025}$, $\Delta m^2 = (7.53 \pm 0.18) \times 10^{-5} eV^2$, and $\sin^2 \theta_{12} = 0.023 \pm 0.002$, including constraints on $\theta_{13}$ from accelerator and short-baseline reactor neutrino experiments [5].

While we continue to explore the neutrino properties, we began to utilize neutrinos as a tool to look into astronomical objects, such as the Earth emitting geoneutrinos. In 2005, KamLAND performed the first experimental study of geoneutrinos, and this dataset based on a total detector live-time of 749.1 ± 0.5 days (March 7, 2002-October 30, 2004), gave $25^{+19}_{-18}$ geoneutrinos candidates from the $^{238}$U and $^{232}$Th decay chains [1]. This result was consistent with our reference model expectation (19 events). At that time, the energy range of geoneutrinos was dominated by $\bar{\nu}_e$ from commercial nuclear reactors and the $^{13}$C(α, n)$^{16}$O reaction initiated by the decay of radioactive contaminations in the detector. Two-time purification in the period March - August 2007 and June - February 2008 improved the purity of the liquid scintillator. As a result, background rate from the $^{13}$C(α, n)$^{16}$O reaction went down by a factor of ~20. The measurement presented in 2011 was based on 2135 days live-time data (March 7, 2002-November 4, 2009), and $106^{+29}_{-28}$ events excess (expected signal is 106 events) was observed in the Geoneutrino Energy Region (GER, from 1.8 MeV to 3.3 MeV) [4].

In this section, we will present the recent results of geoneutrino measurements based on the 2019 preliminary results. The data reported here are based on a total live-time of 4397 days (March 7, 2002-April 15, 2018), including the recent reactor-off period. The reactor $\bar{\nu}_e$ flux was significantly reduced because of the shutdown of all Japanese reactors following the March 2011 earthquake in Japan. In September 2011, the KamLAND-Zen neutrinoless double beta-decay search was launched [29], and a 3.08 m diameter Inner Balloon (IB) was installed at the center of the detector. We can continue to measure $\bar{\nu}_e$ outside the IB as normal.  Dataset is divided into three periods depending on detector situation:
- Period 1 (1486 days live-time, -May 2007): before liquid scintillator purification campaign.
- Period 2 (1151 days live-time, -September 2011): during and after liquid scintillator purification campaign.
- Period 3 (1760 days live-time, -April 2018) : after IB installation, corresponds to the reactor-off period.

Because 40% of dataset was collected in Period 3, our data increased power to measure geoneutrinos with low-reactor background comparing with past publications. Figure 6 shows time variation of reactor $\bar{\nu}_e$ flux at KamLAND detector site.

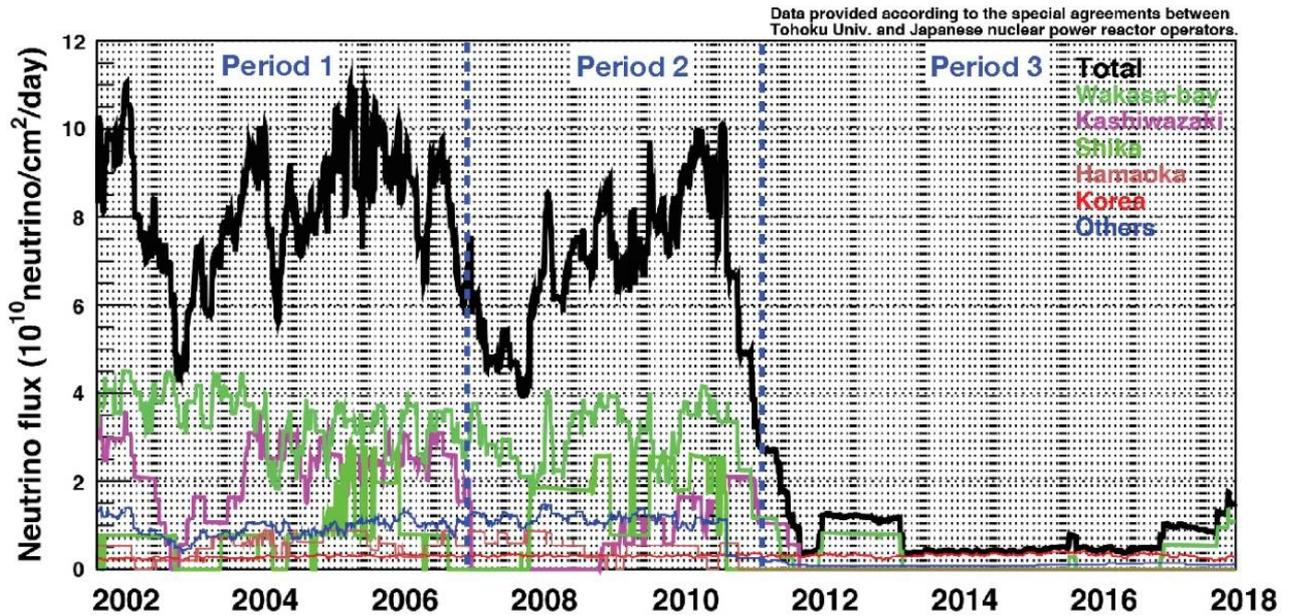

**Figure 6 -** Time variation of reactor antineutrino flux at Kamioka site. Black line shows the total antineutrino flux. Colored lines show contributions from the different areas hosting several reactor plants including Korea (red) and other world reactors (blue). Following the Fukushima nuclear reactor accident in March 2011, many of Japanese nuclear industries were shutdown. The total neutrino flux at Kamioka site was suddenly decreased to ~5 % of under typical operation. Dataset categories (Period 1-3) are shown in the figure.

KamLAND is located in a rock cavern in the Kamioka mine, ~1000 m below the summit of Mt. Ikenoyama in Gifu, Japan. The 2700-meter water equivalent overburden reduces the cosmic ray flux by a factor of roughly $10^{-5}$ compared to the surface flux. The cosmic muon rate is about 0.34 Hz in the inner detector. The KamLAND detector is marked by the ability to detect low energy $\bar{\nu}_e$ signals at liquid scintillator compared with water Cherenkov detector.

The KamLAND detector occupies the former site of the Kamiokande experiment [30] in the Kamioka mine. The construction commenced in 1997, and data were collected from 2002. Figure 7a shows an overview of the detector and experimental site. The distance between the Kamioka mine entrance and the KamLAND area is approximately 3 km.

Figure 7b shows a schematic view of the KamLAND detector. The detector consists of an 18-m-diameter stainless steel spherical tank which defines the boundary between the inner and outer detectors (ID and OD, respectively). A 13.0-m-diameter spherical balloon is suspended inside the stainless-steel tank and filled with 1000 tons (1171 ± 25 m$^3$) of ultra-pure liquid scintillator. The liquid scintillator consists of 80.2% dodecane ($C_{12}H_{26}$) and 19.8 % pseudocumene (1,2,4-trimethylbenzene, $C_9H_{12}$) by volume, with 1.36 ± 0.03 g/L of PPO (2,5-diphenylox-azole, $C_{15}H_{11}NO$) as a primary fluor. The ID section is designed for $\bar{\nu}_e$ detection. The OD section acts as a cosmic ray active veto and attenuates γ radiation from the surrounding rocks. The light produced in the ID and OD is detected by the PMTs (1325 fast 17-inch aperture Hamamatsu PMTs custom-designed for KamLAND, and 554 20-inch PMTs inherited from Kamiokande). Waveforms from PMTs which are read out as voltage are recorded and later used to reconstruct the energies and position of events. To test the performance of energy and position reconstructions, radioactive calibration sources with known energies are deployed at exactly known positions.

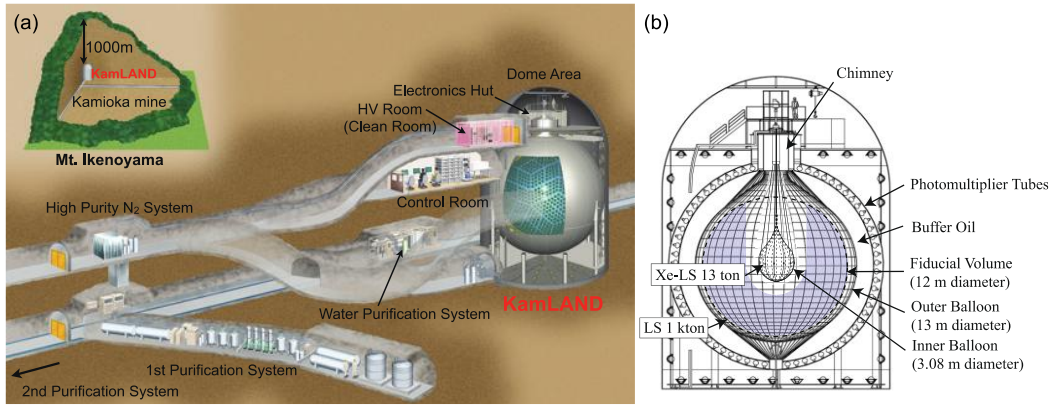

**Figure 7 -** **(a)** KamLAND experimental site and **(b)** the schematic design of the detector. **(a)** There are 2 types of liquid scintillator purification system (liquid-liquid extraction and distillation). The water purification is operated to produce pure water filled in the OD. **(b)** The shaded region in the LS shows the fiducial volume for $\bar{\nu}_e$ in Period 3.

### 3.2 Data analysis and backgrounds

#### 3.2.1 Event reconstruction

The event reconstruction is a process of interpreting the timing and charge information produced by each PMT to infer physics events. The fast time response of scintillation light emission (~10 ns) makes it possible to detect the physics events in real time and to reconstruct the position and energy of the events. The reconstructed events can be classified into two types: 'point-like' and 'track-like'. The former events deposit their entire energy in a single point of the scintillator. As a result, scintillation light appears to be generated in a specific point of the detector. Most of the events in the low energy region (<10 MeV) belong to this category of events. On the other hand, track-like events deposit their energy along their path and are therefore characterized by a light emission appearing as a track. These events are dominated by the cosmic-ray muons (~0.3 Hz at KamLAND). The reconstructed events also include non-physics events caused by electrical noise and broken PMTs etc. Therefore, reconstruction quality is finally evaluated from the time and charge information of PMTs. The procedure of event reconstruction is as follows:

a. waveform analysis: the digital waveforms are converted to time and charge information.
b. time and charge calibration of PMTs: each PMT has a particular time and charge information. The transit time of each PMT is calibrated with a pulse dye-laser. The charge for 17-inch PMTs makes one photoelectron (1 p.e.) distribution in low energy condition. To improve the energy and vertex resolution, the dead or unstable PMT channels should be eliminated.
c. identification of events: the reconstructed events are classified into two types, point-like or track-like events. Most muons pass through the liquid scintillator: these are selected using high energy event selection criteria. Some of the muons only cross the buffer oil emitting Cherenkov light, but do not interact with the scintillator. The number of these muons are ~1/20 of the scintillation light events and these are tagged by their OD hits.
d. event reconstruction: once identified, events are analyzed to reconstruct their main features. For track-like events such as muons, the track of the event is extrapolated by the spatial and temporal information collected. For point-like events both the vertex position and the energy information are reconstructed.
e. reconstruction quality check: finally, the reconstruction quality is evaluated from the time and charge information recorded by the PMTs. If the event reconstruction is succeeded, real physics events have a good correlation between the reconstructed vertex, PMT position, time, and charge. The bad quality events such as the noise and flasher events are eliminated in advance.

### 3.2.2 Detector calibration

The location of interactions inside the detector (vertex) is determined from PMT hits time distribution, which is called 'pulse shape'. The pulse shape generally depends on the PMT types, distance from the source to PMT, the intensity of the signal, the origin of the signal (gamma, beta, neutrino, positron) and the distance traveled through the liquid scintillator. In the vertex reconstruction process, the actual experimental pulse shapes obtained via the source calibration are used for producing the likelihood function. The maximum likelihood method is applied to determine the vertex.

The energy is determined as a measurement of hit and charge information of PMTs in the detector as visible energy, which corresponds to the scintillation light emission. The relation between the visible energy and deposited energy is expressed as the non-linear function due to the dark hit charge, the detection efficiency of single photoelectron, the quenching effect of the liquid scintillator, and Cherenkov light effect. These non-linear sources generate the uncertainties of energy scale.

The reconstructed vertex and energy are checked using calibration sources which have known energy of radioactive γ-ray ($^{60}$Co, $^{68}$Ge, $^{203}$Hg, $^{65}$Zn, $^{137}$Cs, $^{241}$Am, $^{210}$Po) and spallation events induced by cosmic ray muons which are uniformly distributed in the detector. The vertex and energy reconstruction qualities are regularly monitored by the source calibrations from the detector construction. The vertex biases in the detector are evaluated as < 3 cm (Period 1) and < 5 cm (Period 2 and 3). The achieved vertex resolution is ~12 cm/ √ E(MeV) and the energy resolution is 6.4 %/ √ $E$(MeV).

In 2007, we successfully built and operated an 'off-axis' source deployment system for the KamLAND detector [31]. The system was used for positioning radioactive sources throughout the detector. The calibration data obtained with this system were used to fully characterize detector position and energy reconstruction biases.

Using the 'off-axis' calibration data, we achieved a factor of two reduction in the size of the fiducial volume uncertainty and 1.8% is assigned to 6.0 m radius $\bar{\nu}_e$ analysis for the before purification data. Furthermore, the fiducial volume uncertainty for the after-purification data is estimated by incorporating a study of muon-induced $^{12}$B/$^{12}$N decays which are distributed uniformly in the detector. Figure 8 shows the time variation of $^{12}$B/$^{12}$N events ratio between all volume and 6.0 m radius region with a vertical central cylinder cut at the upper hemisphere. The data points are stable within expected ratio. Since there is <1.8% difference between before and after purification data, the fiducial volume uncertainty for after purification data is evaluated to 2.5%.

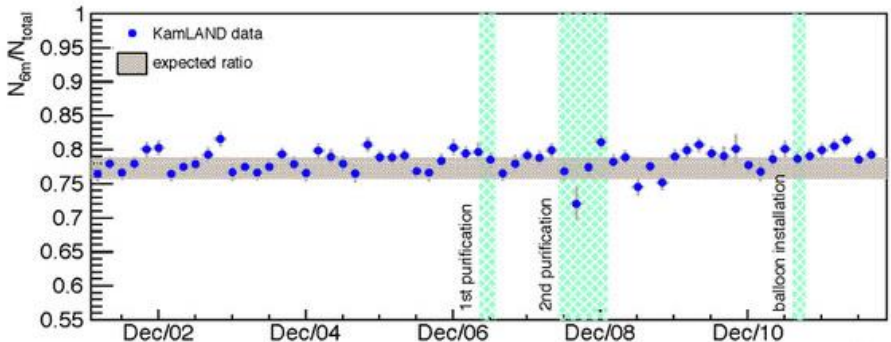

**Figure 8** - $^{12}$B/$^{12}$N event ratio time variation for checking the fiducial volume uncertainty. Vertical central cylinder cut at the upper hemisphere is applied to eliminate backgrounds from the KamLAND-Zen material. Gray shaded area shows the expected ratio, which is calculated by measured liquid scintillator volume, 0.773 ± 0.016. The ratio was stable from the start of data taking, and kept around the expected level.

### 3.2.3 Data selection

The selection criteria for $\bar{\nu}_e$ events are summarized in Table 4. The $\bar{\nu}_e$ delayed coincidence (DC) events are characterized by two spatially and timely correlated signals. To reject the muon and its related background, spallation cut and some kinds of veto are applied. Given the IB contains Xe-loaded liquid scintillator whose composition and emission are different from those of KamLAND, the quality of the event reconstruction is not uniform within the detector. The IB and the corrugated tube connecting to the chimney area are made by clean enough materials for KamLAND-Zen experiment, but gamma rays from very low level $^{238}$U and $^{232}$Th contaminations become accidental background for the $\bar{\nu}_e$ events. In the Period 3 dataset, the vertical centered cylindrical cut of the upper hemisphere is applied to eliminate backgrounds and effects due to KamLAND-Zen material. In order to increase the ratio of signal to accidental background, a second-level cut is applied using a likelihood selection method. This analysis procedure is as follows:

1. select $\bar{\nu}_e$ DC pairs (1st-level cut, summarized in Table 4);
2. construct Probability Density Function (PDF) for $\bar{\nu}_e$ DC pairs ($f_{\bar{\nu}_e}$) and accidental DC pairs ($f_{acci}$) The $\bar{\nu}_e$ events $f_{\bar{\nu}_e}$ is constructed from GEANT4 simulation. The accidental events $f_{acci}$ is evaluated directly from the data of accidental backgrounds. The PDFs are based on the six cut parameters ($E_p$, $E_d$, $\Delta R$, $\Delta T$, $R_p$, $R_d$);
3. construct prompt energy dependent likelihood ratio ($L_{ratio}(E_p) = f_{\bar{\nu}_e}/( f_{\bar{\nu}_e} + f_{acci})$) to maximize separation quality of the $\bar{\nu}_e$ events and accidental backgrounds. $L_{ratio}(E_p)$ is used for setting selection parameter by maximizing the figure-of-merit (FOM) (FOM($E_p$) = $S(L_{ratio})/ \sqrt{(S(L_{ratio})/ + B_{acci}(L_{ratio}))}$ ) for each 0.1 MeV interval in prompt energy. $S$ is the number of $\bar{\nu}_e$ events expected from the oscillation-free reactor antineutrinos and geoneutrinos. $B_{acci}$ is the number of observed accidental background which selected with almost same cut conditions for $\bar{\nu}_e$ candidates, except for the time correlation criterion (10 ms < $\Delta T_{acci}$ < 20 sec). $B_{acci}$ is scaled by the time window ratio, $\Delta T_{\bar{\nu}_e}/\Delta T_{acci} = 5.0 \times 10^{-2}$;
4. apply second level cut for $\bar{\nu}_e$ DC pairs using selected $L_{ratio}$, namely $L_{cut}$.

Table 4 - Selection criteria for antineutrino candidate events. To select spatially and timely correlated events, the first level selections are applied. The delayed coincidence method has strong power to suppress background event. The cylinder cut around Inner Balloon is applied during Period 3 to avoid accidental background from radioactive contaminations (e.g. $^{238}$U and $^{232}$Th) and the effect coming from the use of different types of liquid scintillator and materials of the Inner Balloon. Second level cuts are applied to efficiently discern antineutrino candidate events from accidental coincidences.

|  | Parameters | Criteria |
|---|---|---|
| First level | Prompt Energy | 0.9 < $E_p$ < 8.5 MeV |
|  | Delayed Energy | 1.8 < $E_d$ < 2.6 MeV (capture on proton) |
|  |  | 4.4 < $E_d$ < 5.6 MeV (capture on $^{12}$C) |
|  | Space Correlation | $\Delta R$ < 2.0 m |
|  | Time Correlation | 0.5 < $\Delta T$ < 1000 μs |
|  | Fiducial Volume | $R_p$, $R_d$ < 6.0 m |
|  | (Period 3) | $R_d$ > 2.5 m and $\rho_d$ > 2.5 m, $Z_d$ > 0 m |
| Second level | $L_{ratio}$ ($E_p$) | $L_{ratio}$ ($E_p$) > $L_{cut}$ ($E_p$) |

### 3.2.4 Backgrounds

There are various background sources in the detector, such as radioactive impurities and spallation neutrons caused by cosmic muons (Table 5). The background events can be classified into two types. The first is correlated events related to neutrons produced by muons, spallation products with neutron emitters, short-lived nuclei, spontaneous fission of nuclei, and other $\bar{\nu}_e$ sources. The second is uncorrelated events, such as accidental coincidences. In case of geoneutrino measurement, the $^{238}$U and $^{232}$Th decay chains emit $\bar{\nu}_e$ with energies below 3.4 MeV, so that the reactor $\bar{\nu}_e$ events with similar energies pose a background for this signal.

**Reactor antineutrinos**

Nuclear reactors of Japanese power plants had been the primary source of $\bar{\nu}_e$ for KamLAND since the detector was constructed, and we demonstrated neutrino oscillation phenomena by measuring such well-controlled $\bar{\nu}_e$. Commercial nuclear power plants emit $\bar{\nu}_e$ of ~2 × 10$^{20}$ GW$_{th}^{-1}$ s$^{-1}$. For information of reactor operation, records are required to predict the reactor $\bar{\nu}_e$ flux. A consortium of Japanese electric power companies provides information, such as the thermal power variation, fuel burn-up, when and which reactor's fuel rods are exchanged and reshuffled, to the KamLAND collaboration with sufficient accuracy. That makes the commercial reactor a convenient and powerful source of $\bar{\nu}_e$ with well-known relevant characteristics. The absolute thermal power used to normalize the fission rates is measured within 2%.

The four main isotopes contributing to > 99.9% of the reactor $\bar{\nu}_e$ spectra are $^{235}$U, $^{238}$U, $^{239}$Pu and $^{241}$Pu. We use the $\bar{\nu}_e$ energy spectrum from each sources provided in [32] ($^{238}$U) and [33] ($^{235}$U, $^{239}$Pu and $^{241}$Pu). These studies show that the $\bar{\nu}_e$ spectra per fission of these isotopes become ~3% higher than the previous calculation [34,35]. Recent short-baseline reactor experiments measured the event excess around 4-6 MeV by about ~10%. To avoid uncertainties from the reactor spectrum modeling, we use the $\bar{\nu}_e$ spectrum measured by DayaBay experiment in a model-independent way [36]. The normalization of the cross section per fission for each reactor is adjusted to reproduce that Bugey4 result [37] which measured reactor $\bar{\nu}_e$ spectra with 15 m baseline length.

We consider other small effects: the contribution from Korean reactors and other reactors around the world are estimated to be ~5% and ~1%, respectively. There are some long-lived beta decay nuclei produced by the fission of the main 4 isotopes and by $^{106}$Ru, $^{144}$Nd, and $^{90}$Zr. It has ~0.7% level of additional contribution on reactor $\bar{\nu}_e$ spectra according to their lifetime.

There are 56 nuclear power reactors in Japan. In July 2007, Kashiwazaki-Kariwa nuclear station, which is located ~160 km far from KamLAND with the largest thermal power in the world, was stopped due to an earthquake. This running condition reduced $\bar{\nu}_e$ flux by about half of normal operation. Following March 2011 earthquake, the entire Japanese nuclear power plants were subjected to protracted shutdown. This unexpected situation caused a substantial reduction in $\bar{\nu}_e$ flux to ~5%. This reactor-off data yields improved sensitivity for geoneutrino flux and provided a unique opportunity to confirm and constrain backgrounds for $\bar{\nu}_e$ analysis.

**Accidental coincidences**

The cause of the accidental coincidence is mainly radioactive impurities. The events around the balloon surface come from outside of the detector, and that would become candidates for the accidental background. Fiducial volume cut is effective in rejecting this background. Period 2 and 3 datasets are dominated by this background.

### ⁹Li/⁸He background

⁸He and ⁹Li are neutron emitters, and their lifetimes are 171.7 ms and 257.2 ms, respectively. There is time correlation between delayed coincidence events and muons.

### (α,n) background

The main (α,n) reaction is $^{13}C(\alpha,n)^{16}O$, and the α source is $^{210}Po$. The purity of KamLAND liquid scintillator was improved eliminating most of the $^{210}Pb$ that used to feed the decay chain responsible for the production of α-particles from $^{210}Po$ decay. This in turn dramatically reduced the (α, n) backgrounds. In Period 1, (α,n) reaction is the dominant background.

### Fast neutrons

Fast neutron events are detected as delayed coincidence events tagged with a muon which passed only in the outer detector. The prompt signal is the proton which is recoiled by neutron. Since the OD has small inefficiency, the not tagged events become background for $\bar{\nu}_e$. The contribution from a muon which passed only in the surrounding rocks are considered separately.

### Atmospheric neutrinos

The conservative background rates from the atmospheric neutrinos are estimated by the typical flux calculation model.

**Table 5 -** Estimated backgrounds for geoneutrinos in the energy range between 0.9 MeV and 2.6 MeV after event selection cuts. Fast neutron and atmospheric neutrino backgrounds are assumed to have flat energy spectra.

| Backgrounds | [events] |
|---|---|
| ⁹Li | 4.4 ± 0.1 |
| Accidental | 121.9 ± 0.1 |
| Fast neutron and atmospheric neutrino | < 4.1 |
| $^{13}C(\alpha,n)^{16}O$ | 211.6 ± 23.3 |
| Reactor $\bar{\nu}_e$ | 629.0 ± 34.4 |
| Total | 966.9 ± 41.8 |

### 3.3    Results

Reactor antineutrinos present the largest background in the KamLAND measurement because their energy spectrum partially overlaps that of geoneutrinos. There is difference between the reactor and the geoneutrino spectra. The reactor $\bar{\nu}_e$ event rate varies with the output of the nuclear power plants and the geoneutrinos have constant contribution to the event rate. For the above reasons, a simultaneous fit to both sources can be used to extract the neutrino oscillation parameters and geoneutrino fluxes.

The most sensitive analysis is performed by using the unbinned maximum likelihood method which takes into account the event rate, the prompt energy spectrum shape, including their time variations, in the range $0.9 < E_p < 8.5$ MeV. The principle of the method is to find out the set of parameters that gives the maximum probability density of observing data. The $\chi^2$ is defined as:

$$\chi^2 = \chi^2(\theta_{12}, \theta_{13}, \Delta m_{21}^2, N_{BG\ 1\to 5}, N_{U,Th}^{geo}, \alpha_{1\to 4})$$
$$- 2\ln L_{shape}(\theta_{12}, \theta_{13}, \Delta m_{21}^2, N_{BG\ 1\to 5}, N_{U,Th}^{geo}, \alpha_{1\to 4}) \quad (8)$$
$$+ \chi_{BG}^2(N_{BG\ 1\to 5}) + \chi_{syst}^2(\alpha_{1\to 4}) + \chi_{osci}^2(\theta_{12}, \theta_{13}, \Delta m_{21}^2)$$

The χ² terms are, in order of contribution:
  a. the time-variating event rate
  b. the time-variating prompt energy spectrum shape
  c. a penalty term for backgrounds
  d. a penalty term for systematic uncertainties
  e. a penalty term for the oscillation parameters

$N_{BG\ 1\to 5}$ are the expected number of backgrounds, and they are allowed to vary in the fit, but are constrained with the penalty term (c). $N_{U,Th}^{geo}$ are the flux normalization parameters for U and Th geoneutrinos, and allow for the Earth-model-independent analysis. $\chi_{syst}^2(\alpha_{1\to 4})$ parametrizes the uncertainties: the reactor $\bar{\nu}_e$ spectrum, the energy scale, the event rate, and the energy-dependent detection efficiency, which are allowed to vary in the fit but are constrained with the penalty term (d). The total systematic uncertainties are 3.5 % and 4.0 % for the pre-purification (Period 1) and the post- purification (Period 2 and 3) datasets. $\chi_{osci}^2(\theta_{12}, \theta_{13}, \Delta m_{21}^2)$ is the constraint from solar, accelerator, and short-baseline reactor neutrino experiments.

Based on the fit discussed above, the best-fit yields $123.3_{-39.1}^{+41.2}$ and $41.6_{-24.7}^{+24.6}$ geoneutrinos from $^{238}$U and $^{232}$Th decays when the contributions of the two isotopes are left to vary independently. As shown in Figure 9a, the confidence intervals in the parameter space of $^{238}$U and $^{232}$Th events have anti-correlation. Figure 9b and Figure 9c show the Δχ² profiles of $^{238}$U, $^{232}$Th events and total geoneutrino events. These Δχ² profiles have two-type analysis results which have constraints on Th/U mass ratio or not. Th/U mass ratio is predicted by the geochemical model of [38] as 3.9. In the case of ratio free, the null hypothesis of no $^{238}$U and $^{232}$Th signals is disfavored at 3.51σ C.L. and 1.68σ C.L., respectively. Fixing the Th/U mass ratio, the total number of geoneutrinos is better constrained to $168.8_{-26.5}^{+26.3}$, which corresponds to a 15.6% uncertainty. This number of geoneutrinos corresponds to an oscillated electron antineutrino flux of $3.6_{-0.6}^{+0.6} \times 10^6$ cm²/s from $^{238}$U and $^{232}$Th at the Earth's surface. In TNU units, the geoneutrino signal at Kamioka is S(U+Th) = $32.1_{-5.0}^{+5.0}$ TNU. From the null hypothesis, the absence of geoneutrino events ($N_U+N_{Th}=0$) is rejected at 8.14σ C.L.

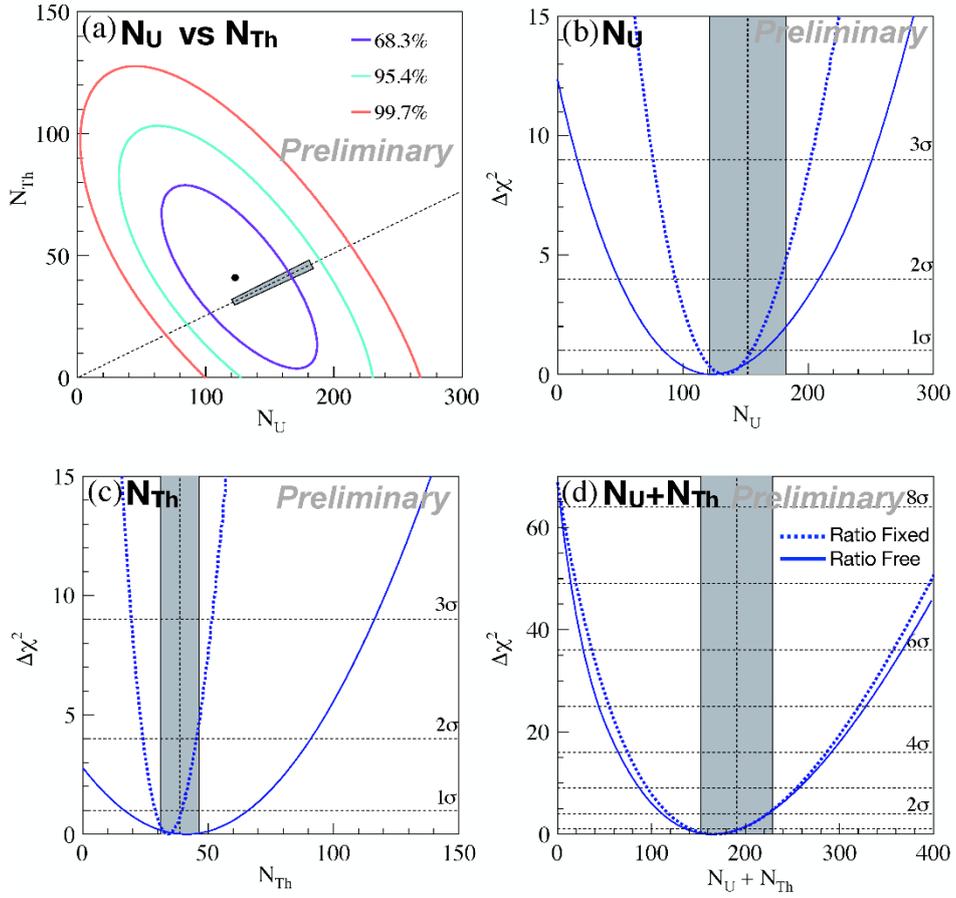

**Figure 9** - $\chi^2$ profiles for observed geoneutrinos events. The dashed black line represent the estimations predicted by the BSE reference model [38], which is characterized by a Th/U mass ratio of 3.9. The gray band represents the 20% uncertainty in U and Th abundances predicted by the reference model [3]. **(a)** $\chi^2$ displayed by the number N of observed geoneutrinos events from $^{238}$U and $^{232}$Th. **(b)** $\Delta\chi^2$ profile for the number of observed geoneutrino events from $^{238}$U. **(c)** $\Delta\chi^2$ profile for the number of observed geoneutrino events from $^{232}$Th **(d)** $\Delta\chi^2$ profile for the number of observed geoneutrino events from $^{238}$U and $^{232}$Th. Solid blue lines show the $\Delta\chi^2$ profiles obtained in the free Th/U case while dotted blue lines show the $\Delta\chi^2$ profiles obtained by constraining Th/U = 3.9.

Prompt energy spectrum of geo $\nu_e$ energy region, $0.9 < E_p < 2.6$ MeV, for all data-taking period is shown in Figure 10. The backgrounds and the geoneutrino contributions are based on the best-fit parameters incorporating all available constraints, such as oscillation parameter constraints from solar, accelerator and reactor neutrino experiments. Applying all selection cuts, 1167 antineutrino candidates are remained in the GER. Numbers of estimated backgrounds are summarized in Table 5. As shown in Figure 11, the energy spectrum for Period 3 only shows clear contributions from geoneutrinos which have peak around 1.3 MeV. The ratio between geoneutrinos and background is estimated as ~1.3. Measurement of clear geoneutrino energy spectrum enhances the chance to perform geoneutrino spectroscopy.

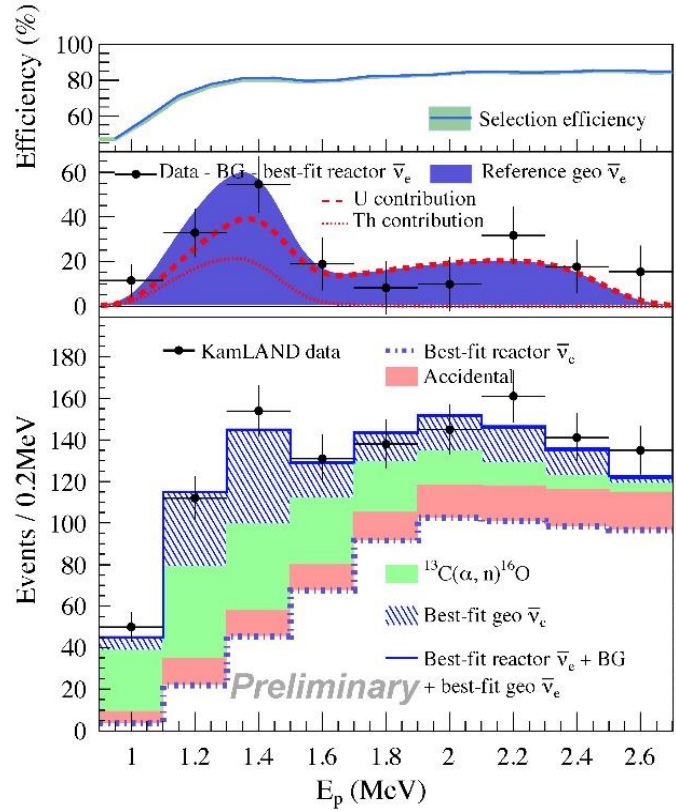

**Figure 10 -** Prompt energy spectrum of geoneutrino energy region for all data-taking period. Bottom panel shows the measured data points together with the best-fit backgrounds and geoneutrino contributions. Middle panel shows the background subtracted data. The blue shaded spectrum is the geoneutrino expectation from the reference Earth model [38], and the best-fit U (dashed) and Th (dotted) contributions. Top panel shows the energy-dependent selection efficiency.

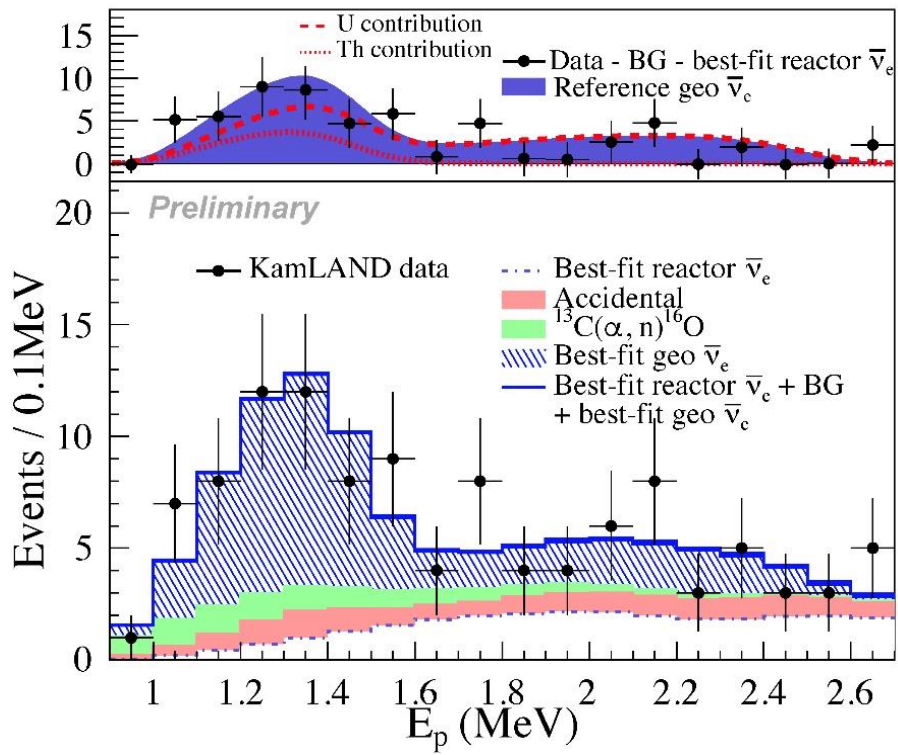

**Figure 11 -** Prompt energy spectrum of geoneutrino energy region for all Period 3. Backgrounds and geoneutrinos are the best-fit results of Period 3 only analysis.

**Table 6** – Preliminary results from KamLAND analysis adapted from [7]

| S(Th)/S(U) = chondritic ratio | | | |
|---|---|---|---|
| | Median ± (68% C.L. stat. & sys.) | | Median ± (68% C.L. stat. & sys.) |
| $N_{geo}$ [events] | $168.8^{+26.3}_{-26.5}$ | $S_{geo}$ [TNU] | $32.1^{+5.0}_{-5.0}$ |
| $N_U$ [events] | $138.0^{+22.3}_{-20.5}$ | $S_U$ [TNU] | $26.1^{+4.2}_{-3.9}$ |
| $N_{Th}$ [events] | $34.1^{+5.4}_{-5.1}$ | $S_{Th}$ [TNU] | $6.6^{+1.1}_{-1.0}$ |
| S(Th) and S(U) free and independent | | | |
| $N_{geo}$ [events] | $164.5^{+28.7}_{-27.6}$ | $S_{geo}$ [TNU] | $31.3^{+5.5}_{-5.2}$ |
| $N_U$ [events] | $123.3^{+41.2}_{-39.1}$ | $S_U$ [TNU] | $23.3^{+7.8}_{-7.4}$ |
| $N_{Th}$ [events] | $41.6^{+24.6}_{-24.7}$ | $S_{Th}$ [TNU] | $8.1^{+4.8}_{-4.8}$ |

## 4    Borexino

### 4.1    Borexino experiment

The Borexino detector was designed to measure solar neutrinos starting from their lowest energies. In particular, the first goal was the measurement of solar neutrinos produced in the fusion reactions involving $^7$Be (0.862 and 0.380 MeV). The first condition necessary for reaching this goal was to break down as much as possible the natural radioactivity present, first of all, in the liquid scintillator, chosen as detecting medium, and in the construction materials of the detector, while the radiations from the environment had to be efficiently shielded.

The first effort in Borexino was therefore the development of innovative methods for the reduction of the scintillator's radioactivity. After five years of R&D and tests in a detector, which represented a reduced and simplified version of Borexino (called Counting Test Facility-CTF), we succeeded in developing systems able to cut down the detector's radioactivity up to $5 \cdot 10^{-16}$ grams of contaminants on grams of pure matter (g/g), saturating the CTF sensitivity; then the construction of the detector was started [39]. After a first purification via distillation performed during the filling, between 2010 and 2011, a further purification in continuous via water extraction was performed.

With these processes an unprecedented radiopurity reaching three orders of magnitude ($10^{-18} - 10^{-19}$ g/g) below that required by the design was obtained [40-42]. Due to these achievements, Borexino is still today an experiment unique in the world and, up to now, is the only detector that succeeded in measuring the entire pp-chain and the carbon–nitrogen–oxygen (CNO) cycles [43,44].

Borexino also studied the geoneutrinos: such a low level of radiopurity like that reached by Borexino was not even necessary for the search of geoneutrinos because they have definitively higher energy than the lowest part of solar neutrino energy spectrum .

The Borexino detector [45] consists of various concentric layers, following the principle of graded shielding: the detecting material, i.e., the liquid scintillator, is deployed at the center of the detector and the closer the layer is to the center, the greater its radio purity. The 278 tons of scintillator are contained in a vessel (Inner Vessel-IV) with a radius of 4.25 m; the vessel is made of nylon 125 micron thick. During the analysis, a fiducial volume (FV) is defined whose virtual wall has a distance from that of the vessel, in order to absorb possible emissions from the IV nylon, even if the raw material had been carefully selected in order to have a radio-impurity of $10^{-12}$ g/g, and the IV has been prepared and assembled in radon free clean rooms of class 100. In addition, the thickness of the nylon was chosen in order to minimize the amount of material near the fiducial volume and thus reduce any residue backgrounds coming from the nylon itself. The size of the fiducial volume changes according to the different analyses.

The scintillator consists of an aromatic solvent, pseudocumene (PC, 1,2,4-trimethylbenzene) and a solute, PPO (2,5-diphenyloxazole, a fluorescent dye) at a concentration of 1.5 g/l. In the early '90s, when the detector was designed, pseudocumene was the preferred choice among various liquid scintillators, especially for large volume set ups [46].

The IV is surrounded by another nylon vessel (Outer Vessel-OV) with a radius of 5.5 m, whose function is to act as a barrier against the Radon ($^{222}$Rn), and in particular the one produced by the PMTs which are mounted on a 6.85-meter radius Stainless Steel Sphere (SSS), that surrounds the two vessels. The regions between IV and OV, and between OV and SSS are filled with a liquid (buffer liquid) consisting of pseudocumene with 5.0 g/l DMP (dimethylphthalate) - reduced later to 3.0 g/l- which acts as a quencher for the low luminescence produced by the solvent alone. The SSS encloses the central part of the detector and, as mentioned, acts as a support for 2212 PMTs [47,48].

The buffer liquid shields the residual external radiations (laboratory rocks and environment) that survive after crossing the 2100 tons of the highly purified water surrounding the SSS and contained in a tank, 16.9 m high and 9 m radius. Thus, the total liquid, shielding the IV from external radiations, is about 5.5-meter water equivalent; it consists cumulatively of 2.14 m of water, 1.25 m of buffer and the thickness of the liquid scintillator between the IV and the virtual wall defined by the FV.

The liquid scintillator contained in the IV is a little less dense than the buffer liquid, about 0.1% with the DMP at a concentration of 5 g/l; this small buoyancy was further reduced to 0.01% when the concentration of DMP was decreased to 3 g/l. Nevertheless, this small density generates the need for thin low background ropes made of ultra-high-density polyethylene in order to hold the nylon vessel in place. The reduction of the DMP from 5% to 3% g/l aims at preventing the outer scintillator flow into the buffer liquid due to a small leaking point that occurred on the surface of the IV, approximately one year after the data collection started. Even if this DMP reduction, carried out with the goal to minimize the density difference between the buffer and the scintillator, was successful in reducing the scintillator's leak, the IV shape is non-spherical and is changing in time. This forced the Borexino collaboration to a software reconstruction of the IV shape using the data themselves.

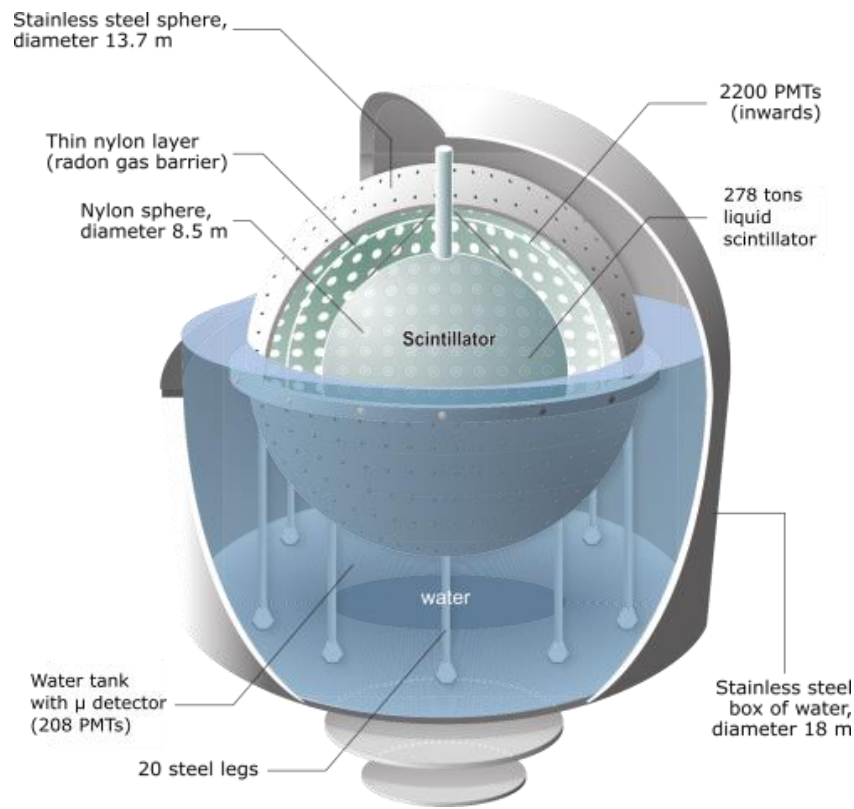

**Figure 12** – Scheme of the Borexino detector.

The PMTs, that read the light from the scintillator (ID) are, as mentioned, 2212 8"ETL 9251: the Borexino collaboration worked with the manufacturer in order to use a glass with low radioactivity. The PMTs, except 371, are equipped with light concentrators which increase the optical coverage for photons produced in the liquid scintillator; the purpose of 371 PMTs without concentrators is to study photons emitted by residual contaminants in the scintillator as well as to help in identifying muons that cross the buffer liquid and not the IV. However, the muons are primarily identified by 210 PMTs installed in the highly purified shielding water that fills the area between the SSS and the water tank in the OD [45].

The selection of the construction materials (stainless-steel, phototubes, cables, light concentrators, nylon, etc.) was made in order to have extremely low radioactivity; the components were manufactured and assembled with highly clean processes and the surfaces were pickled and passivated. The assemblies were performed in clean rooms as much as possible; the detector itself was equipped as a class 10000 clean room. One of the five clean rooms of the setup, class 100, was placed at the SSS entrance and everything mounted in SSS including the scaffolding received there the final cleaning. The purification system as well as the handling of the scintillator and of the buffer liquid were designed and installed to cope with the exceptional level of radiopurity required by Borexino. The welds were performed in nitrogen atmosphere.

Particular attention was paid to the production and transport of the PC, starting from the crude oil obtained from particularly old and very deep layers (to minimize the $^{12}$C content). A collaboration was therefore established with the manufacturer in Sardinia (Italy), which allowed to install an ad hoc loading station on their site in order to convey the product directly from the production column to the isotanks mounted on trucks, which were able to transport the PC to the underground laboratory in less than 18 hours in order to minimize the production of $^7$Be from the cosmic rays. Delivered at the underground laboratory, the PC was transferred by an unloading station to four tanks especially cleaned and treated like the other components, and installed in the so-called storage area. From there the PC was purified via distillation and then mixed with the PPO or DMP and directly inserted into the IV or into the buffer region.

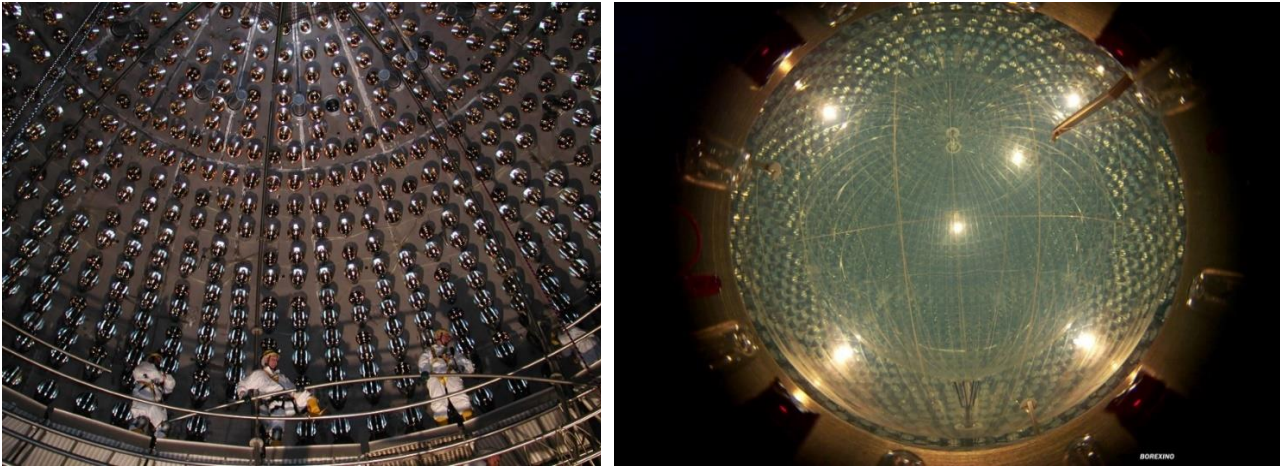

**Figure 13 – a)** The scaffolding, cleaned in the entrance clean room, in uninstallation phase. **b)** A picture of Borexino taken from cameras mounted inside the SSS.

The light yield of Borexino is approximately 500 detected photoelectrons (p.e.) per 1 MeV of deposited energy. The energy resolution is $\frac{5\%}{\sqrt{E(MeV)}}$, while the position is reconstructed using the time-of-flight technique with a resolution of about 10 cm at 1 MeV at the center of the detector; this resolution decreases a little for large radii positions. The uncertainty of the energy scale is less than 1.5% and that of the FV is $X^{+1.5}_{-0.5}$%. An extensive calibration was carried out in 2009 and 2010 using 11 different artificial sources, plus a vial filled with scintillator $^{222}$Rn loaded; they were deployed in various positions within the IV using a movable arm. This calibration allowed also to study the space uniformity of the detector energy response and to tune Monte Carlo code with 1.5% accuracy in the energy range 0.15-2.00 MeV. The source positions (about 200) were measured with laser and charge-coupled device, with a precision better than 2 cm [49].

One of these sources, $^{241}$Am-$^9$Be, is of particular interest, since the emitted neutrons closely represent the delayed signal of the IBD, the reaction that identifies the antineutrinos interactions (Section 2). Not invasive calibrations were also carried out inserting a $^{228}$Th ($\tau$ = 2.76 $y$) source, which emits a 2.615 MeV $\gamma$, in 9 detector inlets reaching the SSS external surface at different positions along a vertical plane. In addition, regular offline checks of the detector's stability and regular online PMTs' calibration are continuously carried out.

The overall muon detection efficiency is at least 99.992%, increased to 99.9969% when a Fast Analog Digital Converter (FADC) system for muon detection was switched on. The muon tagging system, and in particular the Water-Cherenkov OD, is of prime importance for the geoneutrino study because the undetected muons can be a relevant background source. The good reliability of the OD in the muon identification was confirmed by the successful measurement of the muon signal seasonal modulation over 10 years [50].

## 4.2 Data analysis and backgrounds

### 4.2.1 Data selection

The Borexino collaboration recently released [10] a comprehensive geoneutrino analysis which is the result of 3263 days of data collected from December 2007 until April 2019 with an exposure of $(1.29 \pm 0.05) \cdot 10^{32}$ protons per year. The first Borexino geoneutrino measurement was obtained in 2010 [2] with a geoneutrinos evidence at 4.2σ C.L., followed by a second paper [8] and a third [9] where the null observation was rejected at 5.9σ C.L.

The antineutrinos are detected via IBD (Section 2). The good tagging provided by the IBD allows to release somewhat the FV constraints with respect to the neutrino analysis. Taking into account the possible IV shape change, the FV is defined on the basis of the distance from the IV surface: the prompt signal must have a minimum distance from the IV wall of 10 cm. This distance takes into account the accuracy of the IV shape and is supported by the absence of IBD candidates excess close to the IV. The shape of the IV is reconstructed by means of the radioactive decays of $^{210}$Bi, $^{40}$K, $^{208}$Tl, which contaminate the nylon: only their products falling in the range 800 – 900 keV of energy were used.

The selection of the data consists of various steps: definition of the FV, of space and time correlations between prompt and delayed signals as well as of the cuts on their energy, the α/β discrimination, the muons vetoes and others concerning α and γ interactions.

The space correlation, useful as the time correlation to reject background, is based on the distance between the prompt and the delayed events, which is greater than the distance between their production points because of two effects: the γ interactions are not point like because each of them produces a shower of Compton scattering, the accuracy of the reconstruction (10 cm at 1 MeV) smears the reconstructed position. For these reasons, an optimized value was searched for by mean of thousands of Monte Carlo pseudo experiments, defining a distance of 1.3 m.

Similarly, the coincidence in time between the two events, prompt and delayed, was studied. A mean time of ~255 μs for the neutron, needed to thermalize and then be captured, was measured during the calibration campaign using the neutrons emitted by the source $^{241}$Am-$^{9}$Be. Two different event samples must be considered separately, depending on whether 1 or 2 events enter the data acquisition (DAQ) gate, 16 μs long plus 2-3 μs for the electronics dead time. The time window $dt$ between prompt and delayed signals is assumed to be different if the two signals are two separate triggers/events with single cluster each or fall in a single event with two clusters. In the first case the coincidence time window is assumed between $dt_{min}$ = 20 μs and $dt_{max}$ = 1280 μs (incorporating 91.8% of IBD interactions), in the second case the minimum and the maximum values of the coincidence are 2.5 μs and 12.5 μs, respectively.

Energy cuts are applied to prompt and delayed events, measured via an energy estimator based upon the number of photoelectrons (p.e.) which is obtained adding all the PMTs hits of an event. For the prompt event (two 511 keV γs) produced by the annihilation of the positron, a 0.8 MeV threshold was set without an upper limit definition. The choice of the thresholds is based on the $^{241}$Am-$^{9}$Be calibration data.

For the delayed event (2.22 MeV γ or 4.95 MeV with 1.1% of probability, Section 2) produced by the neutron capture, the situation is more complicated because, if the capture occurs at a great distance from the center, a part of the γ shower may exit the FV, or even the IV, and deposits in the buffer. In this case, of course, the measured energy is biased and also the average of the γ peak shower can be shifted towards lower values. To define a proper energy interval, it is necessary to take into account the possible interference of the $^{214}$Po decays to α plus γ (this background is correlated to the presence of Radon in the detector). Taking into account that a good α identification is possible via the α/β discrimination and that the α energy distribution ends approximately at 1.1 MeV, the energy interval was set between 1.41 (1.75 during the water extraction in continuous) and 5.5 MeV. 1.75 MeV is adopted when a purification or a refilling injects Radon in the detector.

The α/β discrimination is possible because the Borexino liquid scintillator shows a different decay time when crossed by particles α or β/γ; it is therefore possible to use this property to discriminate between these types of particles in two different ways. The first method relies on a parametric evaluation proposed by Emilio Gatti. The Gatti's G parameter is defined as:

$$G = \sum_i P_i S_i \qquad (9)$$

where $P_i = \frac{a_i - b_i}{a_i + b_i}$ where $a_i$ and $b_i$ are the reference shapes for α and β pulses obtained from a sequence, as for instance the $^{222}$Rn daughters $^{214}$Bi and $^{214}$Po, while $S_i$ are the number of p.e. for individual shape within a given Δt. $G$ is positive for α particles and negative for β/γ. This method is efficient if the number of p.e. is relatively high (350-400 p.e.), but becomes significantly less efficient with fewer p.e.. A much more efficient method can be based on the Multi-layers Perceptron (MLP) machine learning algorithm; this is a neural network approach with a certain number of α/β discriminating input variables, calculated event by event. The neural network can be trained for example by using again the $^{222}$Rn daughters as $^{214}$Bi and $^{214}$Po. This method is very efficient as shown, just as an example, in Figure 14 where this approach is applied to the Borexino events. Also, in this approach the efficiency is lower at lower energy because it depends on the pulse shape.

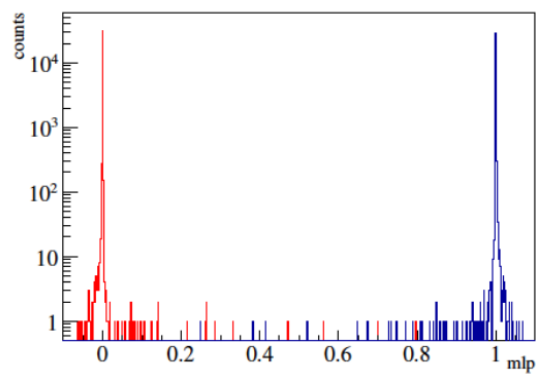

**Figure 14** - Distribution of MLP for α (red) and β (blue) [51].

The muon cut is a bit more complicated and various categories have to be considered. In general, muons can generate background because they produce spallation in connection with the emission of neutrons that can simulate IBD reactions.

When a muon is detected only by the OD, a 2 ms veto is imposed after the OD signal, a time eight times greater than the neutron capture one.

The muons crossing the IV can also create a background for the geoneutrinos due to the fast neutrons, as well as the isotopes $^9$Li, $^8$He, $^{12}$B; this hadronic background is dominated by $^9$Li, whose decay time is 0.25 s; the $^8$He production rate is estimated to be negligible in Borexino taking into account also a comparison with the far detector of Double Chooz. One decay channel of both $^9$Li and $^8$He is ($\beta^-$+n) [52]. $^{12}$B, which decays β can mimic the IBD interaction only if, after a first decay, a second one falls within the IBD time window (1.28 ms), but this window is much smaller than the $^{12}$B lifetime and its Q value largely extends over the geoneutrino and the reactor antineutrinos end point.

Also, cosmogenic fast neutrons are produced by cosmic muons. If the muons reach the scintillator can scatter off protons, which, together with the neutron eventually captured, can mimic the prompt and delayed IBD signals. Similarly, the untagged muons can present a background if they produce a neutron, or are followed by another untagged muon very close in time or through a possible spallation, because, having not been detected or recognized, no vetoes are placed.

A 2s cut is very effective against these backgrounds: it is eight times the lifetime of the isotope that lasts longest, but this means a 10 – 11% of exposure loss. So, with a more complicated selection it is possible to reduce the exposure loss to 2.2%, provided that various categories are distinguished:

a. the muons that give a signal in the ID but not in the OD - are 6.3% - and those which also give a signal in OD followed by at least one neutron detected - are 1.8%. For these two types of events, a 2 s veto is imposed on the entire detector;
b. the muons not passing through the scintillator, but only through the buffer. In this case the neutrons produced cannot be detected because they produce too weak light. A 2 ms veto is applied to these muons, which are 57.8%; in this way possible fast neutrons are suppressed;
c. the muons passing through the scintillator without producing neutrons have a good probability that a neutron from a potential $^9$Li would be detected. To them, which are 34.1%, a 1.6 s veto is applied. But the 27% has the trace well reconstructed identifying all four points, i.e., entry and exit in the OD and ID. In this case it is enough to apply a cylindrical veto around the muon track with a radius decided according to the lateral distance between muon and the IBD interaction for the prompt and delayed event. The remaining 7%, when not all four points are reconstructed, requests a 1.6 s veto for the whole detector.

A multiplicity cut is applied if further events with energy > 0.8 MeV fall within ± 2 ms around the prompt or the delayed candidate. This time window is about 8 times the one needed by the neutron to be thermalized and then captured. This cut reduces the exposure of only 0.01%.

**4.2.2 Backgrounds**

The background encountered in the analysis of geoneutrinos in Borexino can be classified as a background of antineutrinos and non-antineutrinos. By far the most important background is due to antineutrinos produced by nuclear reactors. Conversely, the background originated inside the detector after the cuts is almost negligible.

**Reactor antineutrinos**

Many nuclei produced by fission in nuclear reactors decay beta producing electron-antineutrinos. These antineutrinos have an energy spectrum that goes up to ~10 MeV and overlap the geoneutrinos energy spectrum (1.8 – 3.27 MeV); they come from the 440 nuclear power reactors operating in the world. Each fission process produces on average ~200 MeV of energy and six antineutrinos: it follows that each reactor, which typically produces 3 GW thermal power, emits $5.6 \cdot 10^{20}\, \bar{\nu}\, s^{-1}$.

An accurate determination of the expected signal was carried out and required to be aware of the nominal thermal power of each reactor, the distance of the single reactor from the detector, the power fraction of each fuel of the reactor ($^{235}$U, $^{238}$U, $^{239}$Pu, $^{241}$Pu), the energy produced as well as the antineutrinos spectra produced by the fission of every component, the cross section for IBD interactions at the different energies of the same spectrum, the $P_{ee}$. In this calculation, the weighted average of the *thermal load factor* per month for each reactor must also be considered, as far as the Borexino exposure for each of the 137 months of data taking concerning the present results (December 2007 – April 2019). The *thermal load factor* is defined as the ratio between the net thermal power produced during the considered period and what it would have been produced if the reactor would have operated continuously at the nominal power. Since this quantity is not available for each power plant, we have made the assumption that *thermal load factors* are identical to the electrical ones.

The needed information for this assessment is obtained from the Power Reactor Information System (PRIS) continuously updated by the International Atomic Energy Agency (IAEA) [53]. The measurements of reactor thermal power can give errors below 1%. In US and Japan the safety for operative reactors requires 2% as minimal accuracy [54]. Therefore we assume, conservatively, for the thermal power, an accuracy of 2% [10].

In order to know the power fraction, the different technologies adopted in the reactor design have to be also taken into account as far as the change of composition of the nuclear fuel, in particular the ratio between Pu and U. Since the change over time of the fuel composition is not available, representative values for the power fractions were used, which vary according to the reactor type. Pressurized Water Reactors, Boiled Water Reactors, Light Water Graphite Reactors, and Gas Cooled Reactors are assumed to use an enriched Uranium composition, with some differences for about 30 Pressurized Water Reactors using MOX and for the Pressurized Heavy Water Reactors (see [10]).

The definition of the energy spectrum of reactor antineutrinos is problematic. The results of Daya Bay, Double Chooz, RENO show a spectrum which deviates from the paradigmatic spectral prediction of [55-57] in the energy range 4 - 6 MeV: this deviation is called *5 MeV excess.* Borexino calculated the spectrum of antineutrinos with and without the *5 MeV excess*. The predicted signal varies by about 6% depending on whether or not the *5 MeV excess* is considered: $84.5^{+1.5}_{-1.4}\ TNU$ without and $79.6^{+1.4}_{-1.3}\ TNU$ with.

**Atmospheric neutrinos**

Atmospheric neutrinos can be a potential background for geoneutrinos. They consist of neutrinos and antineutrinos and their muonic flavor is about two times the electronic one. Interactions with $^{1}$H, $^{12}$C, $^{13}$C and many reactions with these last two can simulate IBD reactions. The Borexino collaboration developed a simulation code for these interactions using HKKM2014 model [58] and Fluka code. The number of background events due to the atmospheric neutrinos, once the optimized cuts are applied, is: $2.2 \pm 1.1$ in the geoneutrino energy window and 3.3 ± 1.6 in the reactor antineutrino window.

**Cosmogenic background**

In the previous section it is discussed the cuts imposed on muons after crossing the LNGS overburden, which in the hall C are 1.2 muons m$^{-2}$s$^{-1}$, and the cosmogenic background as the unstable nuclides $^{9}$Li, $^{8}$He, $^{12}$B that can be produced by the muons: only $^{9}$Li generates background, because, also considering the Double Chooz results, the rate of $^{8}$He is negligible and $^{12}$B does not emit neutrons though its decay. Once studied the time and spatial distributions of the $^{9}$Li candidates with respect to the parent muon, the different types of internal muons were considered and the respective cuts were applied as described in the previous section: in this way the expected $^{9}$Li background in the IBD candidates is evaluated to be 3.6 ± 1.0 events.

For the untagged muons, taking into account the (0.0013 ± 0.0005)% of the inefficiency of the muon flag, and the different configurations which could eventually generate background, the conclusion is that only (0.023 ± 0.007) could simulate IBD events.

Another possible background source can be produced by the fast neutrons due to ID undetected muons crossing the Water Tank or the surrounding rocks. The evaluation of the first source consequences were studied checking the contemporaneity of the scattered proton ID signal and the muon OD signal. It is looked for either a prompt signal not tagged by the ID and a delayed signal consisting of a neutron cluster, or an external muon with a cluster in the ID, followed, within 2 ms, by a point-like event. This evaluation gives an upper limit of 0.013 IBD like coincidences (90% C.L.). The background connected to a muon passing through the hall C rocks is evaluated via Monte Carlo, which brings to an upper limit <1.43 events (95% C.L.).

**Accidental coincidences**

To study accidental coincidences, it is necessary to increase the statistics and then consider a fairly large interval: the choice is 2s-20s. Then the number of coincidences is scaled taking into account the time window defined for the geoneutrinos, which in the case of a single event entering the gate is 20 μs-1280 μs, while in the case of 2 events is 2.5 μs-12.5 μs, therefore in total 1270 μs (see Section 4.2.1); in addition it has to be taken into account that the muon vetoes produce a suppression factor, which changes with the time interval between the delayed event and the prompt one, in particular it continues to increase up to 2s (> 0.99993), and then, as time increases, it remains constant (0.896 ± 0.0039). Taking into account these circumstances the accidental events in the IBD candidates are (3.846 ± 0.017) events.

**(α,n) background**

In the Borexino detector, the $^{210}$Po, which decaying produces α particles, has been always present and observed to be out of equilibrium with the other $^{238}$U chain components. Normally $^{210}$Po in the liquid scintillator increases during purification or refilling operations; its lifetime is 138.4 days. Therefore, during the years, the $^{210}$Po rate had significant variations and therefore for the evaluation of the background the mean value of (12.75 ± 0.08) events day$^{-1}$ ton$^{-1}$ were considered, averaged during the 3263 days of data taking considered in this analysis.

The probability that $^{210}$Po αs trigger the reaction (α, n) in the scintillator was calculated using the Talys software (NeuCBOT) [59] which simulates nuclear reactions. In this calculation, only the PC was considered, because the PPO makes a negligible contribution. The neutron yield was found to be (1.45±0.22) x 10$^{-7}$ neutron per every $^{210}$Po decay [10]. The calculation of the IBD-like coincidences triggered by the neutrons takes into account the total exposure, and the probability of an (α,n) producing an IBD-like coincidence is evaluated via a Monte Carlo study. The result is (0.81 ± 0.13) background events for the whole analysis period.

A possible background can be produced by two $^{210}$Po decays in the buffer and this probability is 0.23% for the inner buffer and fully negligible for the outer one. The $^{210}$Po contamination in the buffer is estimated, in a very conservative way, to be < 0.14 mBq/kg. This limit would agree with an upper limit of 2.6 background events.

**(γ,n) and possible induced background**

An energetic γ produced by neutron capture in detector materials or in rocks or even in radioactive decays can produce a reaction (γ, n) in the scintillator or in the buffer liquid. If the γ produces a Compton scattering or a shower and the neutron is captured, these two coinciding events perfectly mimic an IBD event. However, taking into account the energy threshold necessary for an interaction (γ, n), it can be concluded that only gammas with energies higher than 3 MeV can trigger two events similar to IBD interactions; ɣs cannot be produced at that energy in sizeable amount by natural radioactive chains decays. The 3 MeV ɣ can interact with the deuterons before to be absorbed and then it is possible to calculate this background, taking into account the number of ɣs, the deuteron density, the interaction cross section, the ɣ absorption length, the detection efficiency: an upper limit of 0.34 events at 95% C.L. was found. The ɣ capture on $^{13}$C and $^{12}$C has been also considered and their contribution to the background is absolutely negligible.

The possible contribution to the background that may be produced by spontaneous fissions of $^{238}$U present in the PMTs glass and dynodes, which are distant from the center of the detector of about 6.85 m, was evaluated taking into account the attenuation of the neutrons in the path from the PMTs to the IV, and the solid angle: this background was found to be < 0.057 events.

Other minor backgrounds, as the one produced by the Radon in the scintillator, have been studied and the number of possible events simulating IBD is fully negligible. In Table 7 the number of events after cuts for each source of background are summarized.

Table 7 – Number of events after cuts for each source of background [10].

| Background Type | Events |
|---|---|
| Reactor antineutrinos without "5 MeV excess" | $97.6^{+1.7}_{-1.6}$ |
| Reactor antineutrinos with "5 MeV excess" | $91.9^{+1.6}_{-1.5}$ |
| Atmospheric neutrinos | $3.3 \pm 1.6^+$ (~10*) |
| $^9$Li background | $3.6 \pm 1.0$ |
| Untagged muons | negligible |
| Fast $n$s (µ in WT) | negligible |
| Fast $n$s (µ in rocks) | < 1.43 (95% C.L.) |
| Accidental coincidences | $3.846 \pm 0.017$ |
| (α, n) in the scintillator | $0.81 \pm 0.13$ |
| (α, n) in the buffer liquid | < 2.6 |
| (ɣ, n) | < 0.34 |
| Fission in PMTs | negligible |

+over the reactor antineutrinos endpoint
*in the whole spectrum

### 4.2.3   Detection efficiency for antineutrinos

Extensive detector calibration was performed in 2009 and 2010 using 11 different artificial sources in addition to a vial containing $^{222}$Rn loaded scintillator. The most interesting source for the geoneutrinos study is the $^{241}$Am-$^9$Be which emits neutrons as the delayed signal of the IBD reactions. In this way it was possible to tune the Geant-4 based Monte Carlo, developing PDFs which include the detector response. The signal and background (total antineutrinos spectra and the single $^{232}$Th and $^{238}$U chains) were accurately simulated by including all the Borexino experimental conditions and applying the same cuts selecting the real data; the reactor antineutrinos spectra were entered with and without the "5 MeV excess".

In this way the detection efficiencies for geoneutrinos and reactor antineutrinos were obtained (Table 8) including also the FV reconstruction errors and the systematic uncertainties.

**Table 8 –** Detection efficiency after the optimized selection cut [10].

| Source | Efficiency [%] |
|---|---|
| Geoneutrinos | 87.0 ± 1.5 |
| $^{238}$U geoneutrinos | 87.6. ± 1.5 |
| $^{232}$Th geoneutrinos | 84.8 ± 1.5 |
| Reactor antineutrinos | 89.5 ± 1.5 |

### 4.2.4 Systematic uncertainties

A small contribution to systematics comes from the atmospheric neutrinos; in the entire spectrum about 10 events are expected in the energy range concerning geoneutrinos and reactor antineutrinos. The fit of Figure 15a was redone entering the background due to atmospheric neutrinos: the atmospheric neutrinos background was found to be 1.2 ± 4.1 events with very small systematic errors $^{+0.00}_{-0.38}$ % (Table 9). This last value was obtained through two fits: the first up to the end point of the spectrum of reactor antineutrinos, the second up to the end point of the atmospheric neutrinos that passed the IBD selection criteria. The results show that the number of geoneutrinos and that of atmospheric neutrinos are practically unchanged (see [10]).

The change of the reactor antineutrinos spectrum shape was also investigated by studying the differences with and without the "5 MeV excess" : the change of the number of reactor antineutrino events ($N_{rea}$) was found to be negligible.

A 5 cm error was conservatively assumed for the IV position, which changes time to time. In order to evaluate the possible systematic uncertainty on the IBD candidates due to the IV cut, the distance of each IBD candidate with respect to the IV wall was smeared with a gaussian function having a standard deviation of 5 cm, and the FV cut was applied to these smeared distances. The positive offset so found was assumed conservatively as systematic error (Table 9).

Also, the Monte Carlo efficiency arising from the event losses near the IV was calculated comparing the Monte Carlo simulations with the trigger efficiency for the 2.2 MeV ⍰ from the $^{241}$Am-$^9$Be calibration source. The efficiency uncertainty was included in the systematic error.

Finally, the systematic error due to the position reconstruction was evaluated using the photon arrival time. The uncertainty on the position reconstruction affects the FV error and the exposure. The study proceeded via the $^{222}$Rn and the $^{241}$Am-$^9$Be calibration sources displayed at many positions inside the scintillator and comparing them to the charge-coupled device cameras mounted inside the detector. The resulting uncertainties affecting the FV position are reported in Table 9.

**Table 9 -** Summary of the systematic uncertainties [10].

| Source | Geoneutrino error [%] | Reactor antineutrino error [%] |
|---|---|---|
| Atmospheric neutrinos | +0.00 <br> -0.38 | +0.00 <br> -3.90 |
| Shape of reactor spectrum | +0.00 <br> -0.57 | +0.04 <br> -0.00 |
| Vessel shape | +3.46 <br> -0.00 | +3.25 <br> -0.00 |
| Efficiency | 1.5 | 1.5 |
| Position reconstruction | 3.6 | 3.6 |
| **Total** | **+5.2** <br> **-4.0** | **+5.1** <br> **-5.5** |

## 4.3 Results

Over the 3262.74 days of data taking, 154 IBD candidates were selected; they were distributed evenly over time and in the FV. The distance of the prompt signals from the IV walls was studied because IV itself could produce some background: in this case an excess should be present near the walls, but no events excess was found in that region.

The charge of the prompt signal of the 154 candidates was used for the analysis and its distribution was fitted with an unbinned likelihood. In the fit, in addition to the geoneutrinos, also the reactor antineutrinos were left unconstrained, while the background due to cosmogenic $^9$Li, accidental coincidences and (α,n) reactions were constrained in the fit, according to the values listed in Table 7, with Gaussian pull terms.

Two approaches were adopted for the fit: (i) the Th/U ratio was fixed *a priori* equal to the chondritic ratio and (ii) the two contributions of Th and U were left free. The results of the first case, with the Th/U =3.9, are shown in Figure 15a and in Table 10, where the median value and the interval corresponding to ±1σ are quoted.

For the reactor antineutrinos, the best fit value is perfectly aligned with the expectation calculated both if we consider the "5 MeV excess" and if we exclude it. If the fit is redone constraining the reactor antineutrinos [10] to the expected numbers, practically the same values for geoneutrinos were obtained, with only 1.5% of difference; the reason for this stability, which proves the Borexino's reliability in measuring antineutrinos, is also due to the absence of non-antineutrinos background in the GER.

The results obtained with the fit with $^{238}$U and $^{232}$Th left free are compatible with the fit with the Th/U set at chondritic ratio, and the only difference relates to the errors which, in the first case, are greater: this compatibility is observed both for the geoneutrinos and for the reactor antineutrinos. The best fit values of the number of U geoneutrino events ($N_U$) vs Th ones ($N_{Th}$) are shown in Figure 16b, where they are compared with the chondritic ratio. Finally, in Figure 16a the total number of geoneutrino events ($N_{geo}$) vs $N_{rea}$ contour plot from the fit with the Th/U set at chondritic ratio is showed and the 1σ bands of the expected reactor antineutrino signal with and without the "5 MeV excess" is marked.

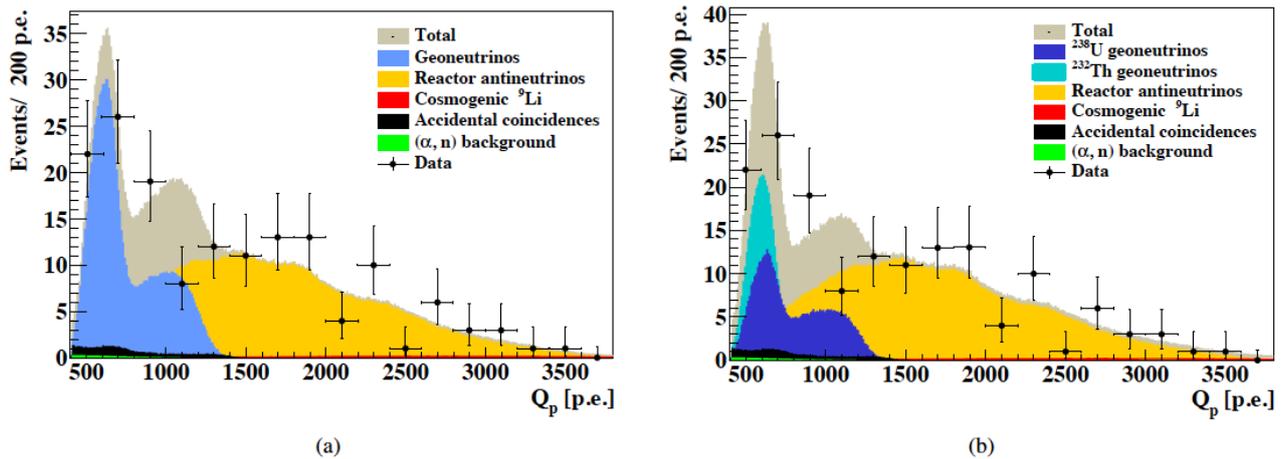

**Figure 15 – a)** Spectral fit assuming the chondritic Th/U ratio. **b)** Same fit but with $^{238}$U and $^{232}$Th as free and independent components [10].

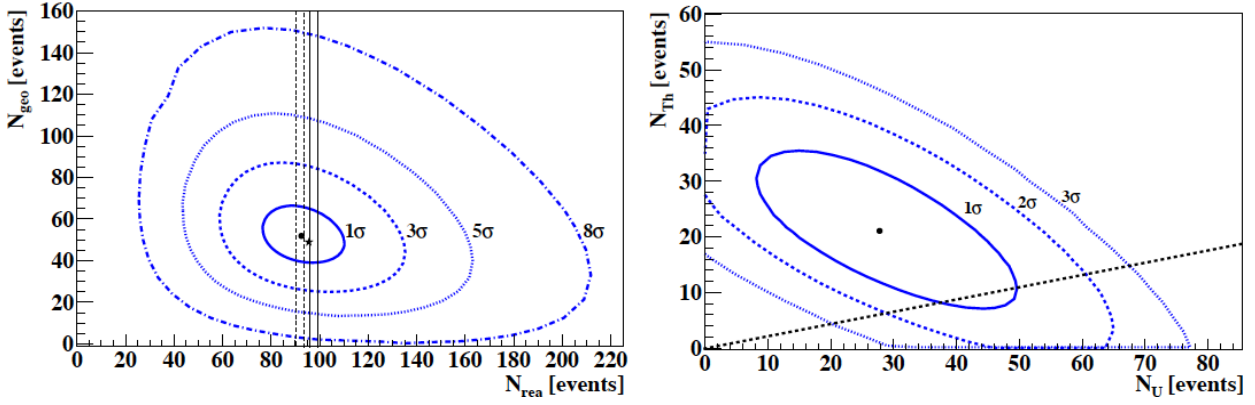

**Figure 16 - a)** $N_{geo}$ vs $N_{react}$. The black point is the best fit and the contours with 1, 3, 5, and 8σ are shown assuming Th/U at the chondritic ratio. The vertical lines with solid lines correspond to the 1σ band for the expected reactor antineutrinos signal without "5 MeV excess", while the dashed lines are related to the signal with "5 MeV excess". The star close to the center is the best fit with $^{232}$Th and $^{238}$U free and independent. **b)** $N_{Th}$ vs $N_U$. Black point: best fit with Th and U free and independent; the dashed line refers to chondritic Th/U ratio [10].

The results can be expressed in TNU following the equation:

$$S_{geo}\ [TNU] = \frac{N_i}{\frac{\epsilon_p\ \epsilon_{i-det.eff.}}{10^{32}}} \tag{10}$$

where in Borexino the exposure $\epsilon_p = (1.29 \pm 0.05) \times 10^{32}$ protons per year, $N_i$ refer to geoneutrinos/reactor antineutrinos and $^{232}$Th and $^{238}$U events number, while $\epsilon_{i-det.eff.}$ are the corresponding detector efficiencies quoted in Table 8. The results are quoted in Table 10.

Table 10 - Summary of the geoneutrino and reactor antineutrino events and the corresponding signal in TNU [10].

| | S(Th)/S(U) = chondritic ratio | | |
|---|---|---|---|
| | Median ± (68% C.L. stat. & sys.) | | Median ± (68% C.L. stat. & sys.) |
| $N_{geo}$ [events] | $52.6^{+9.6}_{-9.0}$ | $S_{geo}$ [TNU] | $47.0^{+8.6}_{-8.1}$ |
| $N_{rea}$ [events] | $93.4^{+12.4}_{-11.8}$ | $S_{rea}$ [TNU] | $80.5^{+10.7}_{-10.2}$ |
| $N_U$ [events] | $41.1^{+7.5}_{-7.1}$ | $S_U$ [TNU] | $36.3^{+6.7}_{-6.2}$ |
| $N_{Th}$ [events] | $11.5^{+2.2}_{-1.9}$ | $S_{Th}$ [TNU] | $10.5^{+2.1}_{-1.7}$ |
| | S(Th) and S(U) free and independent | | |
| $N_{geo}$ [events] | $50.4^{+23.6}_{-22.2}$ | $S_{geo}$ [TNU] | $45.0^{+21.1}_{-19.8}$ |
| $N_{rea}$ [events] | $96.7^{+13.4}_{-12.5}$ | $S_{rea}$ [TNU] | $83.4^{+11.5}_{-10.8}$ |
| $N_U$ [events] | $29.0^{+14.2}_{-13.0}$ | $S_U$ [TNU] | $25.7^{+12.5}_{-11.5}$ |
| $N_{Th}$ [events] | $21.4^{+9.4}_{-9.2}$ | $S_{Th}$ [TNU] | $19.5^{+8.5}_{-8.4}$ |

## 5  A picture of the Earth

### 5.1  The structure of the Earth

The major divisions of the Earth's interior in crust, mantle and core have been known from seismology for about 80 years. This knowledge is based on the reflection and refraction of primary (P) and secondary (S) body waves emitted by earthquakes and traveling through the interior of the Earth. Thanks to the linear relation between compressional P-wave velocity ($v_P$) and the density ($\rho$) of rocks and minerals [60], it was possible to accurately establish a density profile for our planet [61,62]. The obtained Earth's density profile can be further tested and constrained through the measurement of the terrestrial moment of inertia, which notoriously differs from that of a homogeneous sphere (Table 11). The picture that emerges is of a concentrically layered planet, characterized by a density increasing monotonically with depth, subdividable in a solid inner core (IC), a liquid outer core (OC) consistent with the absence of shear waves, a highly viscous mantle and an outer solid shell called lithosphere (LS).

Each Earth reservoir is separated from the others by sharp changes in density (Figure 17). These seismic discontinuities are the result of compositional boundaries or mineralogical phase changes and are often associated with the presence of transition zones. The most significant compositional boundary in the Earth is the core–mantle boundary (CMB), which is surmounted by a ~200 km thermal boundary layer called D'' zone, whose inhomogeneities determine some properties of hotspots and mantle convection (Section 5.3). Instead, the 410-660 km transition zone is a well-documented thermally controlled boundary layer which defines an Upper Mantle (UM) and a Lower Mantle (LM), possibly distinct in composition. Finally, the boundary between the crust and the mantle is called the Mohorovicic discontinuity (MOHO) and it may be a chemical change or a phase change or both. In its uppermost part, the UM is further subdivided by the Lithosphere–Asthenosphere Boundary (LAB), a seismic and electromagnetic transition whose depth (typically at ~175 km) is still a topic of debate. The solid and rigid UM underlying the continents contained between the LAB and the MOHO is usually referred to as Continental Lithospheric Mantle (CLM). The ductile mantle below the LAB is often called sublithospheric mantle, hereafter referred to simply as mantle (M). The silicate portion of our planet, corresponding to M and LS, is known as Bulk Silicate Earth (BSE).

The IC extends from Earth's center up to ~1220 km. It is solid and represents only 5% of the core's mass. The remaining portion of the core, the OC, is instead liquid and extends from 1220 km up to the CMB, situated at ~3480 km. Having a mass of $1.8 \cdot 10^{24}$ kg, the OC is the second main reservoir of our planet after the mantle. The mantle extends from the CMB up to the MOHO for a total mass of $4.01 \cdot 10^{24}$ kg. Most of this mass ($3.911 \cdot 10^{24}$ kg) is attributed to the sublithospheric mantle, which extends from the CMB to the LAB for a total of ~2800 km. The CLM, together with the rest of the oceanic crust (OCC) and continental crust (CC), forms the LS, the outer rocky shell of our planet.

**Table 11** – Average thickness, bulk masses [25], factor of inertia [63] of different Earth's reservoirs. Columns 5-7 list the average density ($\rho$) and the velocities of P ($v_p$) and S ($v_s$) seismic waves obtained from the seismic profile of the Earth [61]. Columns 8-10 report the order of magnitude of the expected abundances of U ($a$(U)) and Th ($a$(Th)) together with the main four elements composing the reservoir.

| Reservoir | Thickness [km] | Mass [kg] | Inertia factor | $\rho$ [kg/m³] | $v_p$ [km/s] | $v_s$ [km/s] | $a$(U) [µg/g] | $a$(Th) [µg/g] | Main elements |
|---|---|---|---|---|---|---|---|---|---|
| Oceans | 3.7 | $1.36 \cdot 10^{21}$ | - | 1025 | - | - | ~0.001 | ~0.000001 | O, H, Cl, Na |
| Athmosphere | 16 (480) | $5.15 \cdot 10^{18}$ | - | <1.2 | - | - | 0 | 0 | N, O, Ar, C |
| CC | 35-40 | $20.6 \cdot 10^{21}$ | - | 2861 | 5.7 | 3.1 | ~1 | ~1 | O, Si, Al, Fe |
| OCC | 7-10 | $6.6 \cdot 10^{21}$ | - | 2826 | | | ~0.1 | ~0.1 | |
| CLM | 140 | $97 \cdot 10^{21}$ | - | 3370 | - | - | ~0.01 | ~0.1 | O, Mg, Si, Fe |
| M | 2800 | $3.911 \cdot 10^{24}$ | 0.29215 | 4776 | 10.2 | 5.5 | ~0.01 | ~0.01 | O, Mg, Si, Fe |
| OC | 2260 | $1.835 \cdot 10^{24}$ | 0.03757 | 10832 | 9.5 | 0.14 | 0 | 0 | Fe, Si, Ni, S |
| IC | 1220 | $9.675 \cdot 10^{22}$ | 0.000235 | 12720 | 11.2 | 3.6 | 0 | 0 | |
| BSE | 2891 | $4.035 \cdot 10^{24}$ | 0.29217 | 4420 | 9.4 | 5.1 | ~0.01 | ~0.01 | O, Mg, Si, Fe |
| Earth | 6371 | $5.972 \cdot 10^{24}$ | 0.3299765 | 5510 | 9.5 | 4.4 | ~0.01 | ~0.01 | Fe, O, Si, Mg |

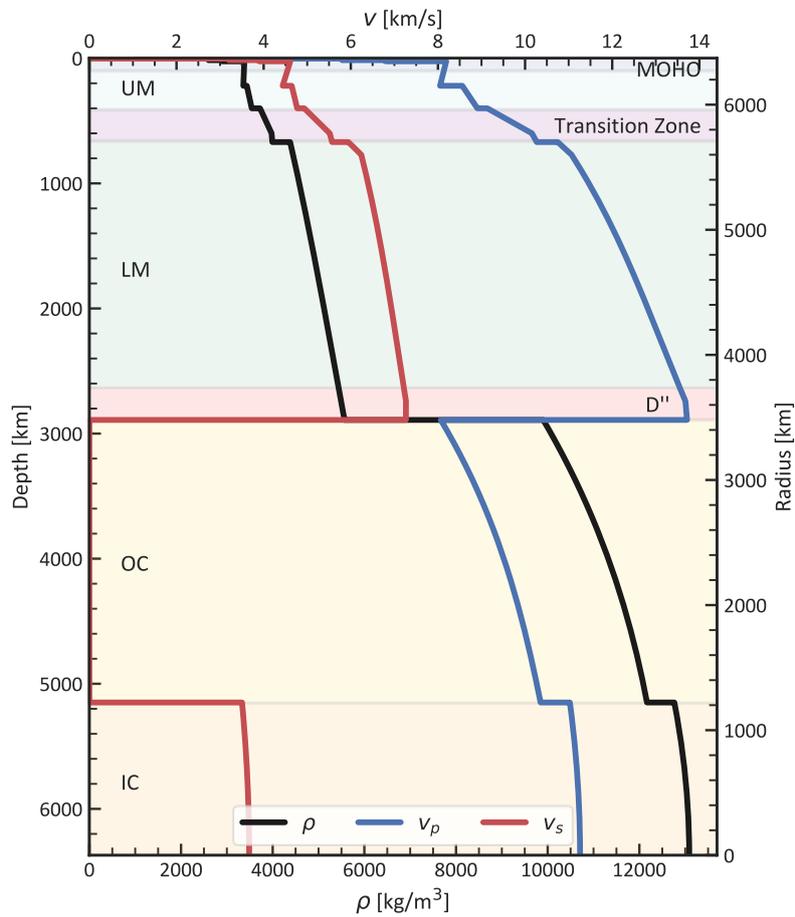

**Figure 17** - Profile of Earth's density (in black) and of primary and secondary seismic velocities (in blue and red, respectively) as a function of depth [61]. The density profile clearly highlights the main Earth's seismic discontinuities (MOHO, D'' and the Transition Zone) which delimit different reservoir of our planet: Upper Mantle (UM), Lower Mantle (LM), Outer Core (OC), Inner Core (IC).

## 5.2 The composition of the Earth

Despite Earth's internal structure is relatively well established, its deep interior remains inaccessible, making its bulk composition impossible to measure directly. However, there is broad agreement among Earth scientists in stating that Earth is mainly made out of Fe (32%wt), O (30%wt), Si (16%wt), Mg (15%wt), Ca (2%wt), Al (2%wt) and S (1%wt), which together account for ~97% of our planet's mass [64,65]. All other elements are present in smaller fractions and the assessment of their abundances requires the formulation of compositional models, which are still under great debate (Section 7).

The peculiar structure of the Earth is a consequence of the physical and chemical processes that occurred in Earth's early history. The fundamental result of planetary differentiation is that elements are not uniformly distributed in the Earth, but rather controlled by their combination of chemical and physical affinities. According to their preferred host phases (Figure 18), elements are generally grouped following the Goldsmith geochemical classification, into:

(i) lithophile elements which tend to occur with oxygen in oxides and silicates;
(ii) siderophile elements which tend to be metallic and readily dissolve in iron either as solid solutions or in molten state;
(iii) chalcophile elements which tend to concentrate as sulphides combining readily with sulfur and other chalcogen other than oxygen;
(iv) atmophile elements which tend to not form stable compounds (e.g. noble gases) and occur in liquids and/or gases (H, C, N) at temperatures and pressures found on the surface of the planet.

According to their condensation temperature[2] ($T_C$) elements are further categorized in refractory ($T_C$ > 1300 K) and volatile ($T_C$ < 1300 K).

**Figure 18** – Chemical affinities (denoted by colors) and tendency to volatize (described in the element property) for the different elements of the periodic table.

---

[2] The condensation temperatures are the temperatures at which 50% of the element will be in the form of a solid (rock) under a pressure of $10^{-4}$ bar.

In this complicated scenario, the only elements reliably accessible to Earth science are the refractory lithophile elements (RLEs). These elements followed the same behavior in the early Solar System, as demonstrated by the fact that they share the same abundance ratios in all types of chondritic meteorites known [66]. Considering that lithophile elements are not likely to be incorporated in the core, measuring the RLE content of M and LS, translates into assessing the RLE's content for the bulk Earth. Instead, since refractory siderophile elements are so concentrated in the core, these are known for their rarity in the Earth's crust. However, they are believed to be present in the bulk Earth according to their solar abundances.

On the other hand, volatiles appear heavily depleted on Earth when compared to meteorites and other undifferentiated bodies of the Solar System. The volatile lithophile elements show a coherent depletion pattern as a function of their $T_C$ [67-69], hinting to depletion mechanisms at the basis of our planet accretion and evolution [69,70]. For elements that are both volatile and siderophile, the complexities of the volatile-element depletion add to our lack of knowledge on how much of an element's loss is due to volatility and how much is due to partitioning into the core [67].

The present picture of the Earth sees a core mainly composed of Fe and a fraction of siderophile elements, which have sunk into the core because of their chemical affinity with iron. At inner core pressure and temperature conditions, it is predicted that a pure iron core should be solid, but its density would exceed the known density of the core by approximately 3%. This requires the presence of a light component in the IC (e.g. O, Si, S in the form of oxides or sulfides, accounting for 2-3%wt), in addition to the probable presence of Ni (up to 10%wt) [71]. The OC is instead liquid (as testified by the absence of S-waves propagation, Figure 17), it shares the same main composition of the IC, but it is expected to have about twice the fraction of light elements envisaged for the IC and to contain 8-13% of O [71,72]. Recent studies investigated the possibility of HPEs inclusion in the core, finding that only small amounts could be potentially included (up to 10 ng/g of U, 21 ng/g of Th, 250 μg/g of K) [73-78] (Section 6.2.1). Even so, this possibility cannot be completed ruled out yet and the debate is still open within the scientific community.

The remaining portion of the Earth, the so-called BSE, is instead rich in chalcophile and lithophile elements. These elements readily combined with chalcogens remaining close to the surface and not sinking into the Earth's core. Cooling and crystallization of mantle over timescales of millions of years resulted in its chemical differentiation according to density. This differentiation could have left most of the Earth's mantle different in composition from the uppermost sampled part of it [79], opening a debate as to whether the rest of the mantle has the same bulk composition. The outer portion of the mantle, slowly solidified in the now-called LS, which as a consequence ended up being highly enriched in incompatible elements (such as U and Th), unsuitable in size or charge to the cation sites of the surrounding minerals.

The M is the largest Earth's reservoir, and therefore it dominates any attempt to perform major-element mass balance calculations. Most estimates of its composition are based on rocks that sample only the uppermost mantle. The M is thought to be mainly made out of O, Mg, Si, Fe and other chalcophile elements.

Instead, the LS is the smallest solid Earth's subdivision (2% wt), but it contains a large fraction of the terrestrial inventory of many elements. The present surface crust represents 0.4% of the Earth's mass and 0.6% of the silicate Earth, but contains a very large proportion (20-70%, depending on the element) of incompatible elements, such as the two HPEs, U and Th. Thus, the crust factors prominently in any mass balance calculation for the Earth as a whole and in estimates of the thermal structure of the Earth.

### 5.3 A dynamic Mantle

Starting with the formulation of continental drift theory and plate tectonics, during the last century the idea of a dynamic Earth started to persuade geoscientists. Although initially it was thought that the mantle was too rigid to allow movements, later measurements (mantle viscosity by postglacial rebound) established that the mantle behaves as a fluid on long time scales. In the current understanding of our planet, continental motion and seafloor spreading are driven by convective motions of the mantle. Thanks to the advances in computing power over the 2000s decade, the field of mantle dynamics came a long way through the use of numerical simulations. The modern knowledge of mantle convection can rely on a multitude of inputs as (i) paleomagnetic studies which prove the relative continental motion and seafloor spreading, (ii) seismology which provides the delineation and locations of plate boundaries and subducting slabs along Wadati–Benioff zones (i.e. planar zones where the oceanic crust sinks under the continental lithosphere), and (iii) geodetic measurements of Earth's gravity field which ensure important constraints about the density structure of the mantle associated with convection [80]. Heat flow and bathymetry measurements show that lithospheric plates move from hot ridges to cold trenches, testifying that these structures are the expression of upwellings and downwellings. These plates spreading and subduction at ridges and slabs, demand vertical transfer of material from the mantle into the surface and vice versa. The material exchange is largely proven as shown by tomographic images of down-going slabs of oceanic lithosphere penetrating into the LM [81], leaving little doubts that the mantle convects. The current debate is about "how" it convects.

Indeed, an additional and independent source of information comes from mantle geochemistry. The disparity between the concentration of incompatible elements in composition of mid-ocean ridge basalts (MORBs) representative of the upper portion of the mantle and ocean island basalts (OIBs) thought to come from the LM was one of the driving motivations for supposing the preservation of isolated reservoirs and, thus, a layered mantle. Additional evidence come from geochemical arguments involving noble gas isotopes, volatile abundances and elemental ratios. However, these geochemical observations seem to conflict with geophysical evidence for whole mantle convection and this has engendered a long-standing debate about the details of mantle convection. High-pressure mineral physics experiments indicated that mantle discontinuities are most likely associated with solid–solid phase transitions, not compositional changes. The major UM component olivine was shown to undergo a change to a spinel structure called wadsleyite at 410 km depth; wadsleyite itself undergoes a less dramatic transition to a ringwoodite at around 510 km and then, at 660 km depth, ringwoodite changes to a combination of perovskite and magnesiowustite. Studies of convection in the presence of such phase changes indicated that they might impede convection temporarily but not indefinitely. Seismic tomographic studies using body waves showed that many slabs do indeed penetrate this boundary and sink well into the LM [82,83].

In the last years, various complexities were discovered in the deep LM that was previously considered as rather homogeneous. At small scale, a laterally intermittent layer at the base of D" ultralow-velocity zone (ULVZ) (Figure 19), with a maximum thickness near 40 km and a strong decrease of $v_p$, is most simply explained as the result of partial melt at this depth [84]. A pair of seismic discontinuities observed in some fast (cold) regions of D" could be the result of a double-crossing of the postperovskite phase boundary by the geotherm at two different depths [85]. Two deep slow velocity anomalies under West Pacific and Africa (roughly underneath the two maxima of the geoid) have unusual seismic properties. They have an anomalously large ratio of compressional to shear velocity ratio, $v_P/v_s$ [86], and an anticorrelation between ρ and seismic velocities [87] and between $v_P$ and $v_s$ [88]. These anomalous regions have very sharp boundaries [89] and depending on the authors have been named megaplumes, thermochemical piles, or large low-shear-velocity provinces (LLSVPs) [90]. These LLSVPs only cover part of the CMB surface (Figure 19), which is itself four times smaller than the Earth's surface, and as they only extend up to a few hundred kilometers, their total volume is three times larger that of the continental crustal volume [91]. These observations of the deep mantle heterogeneity cannot easily be explained by temperature variations. They seem to require lateral variations of Fe or Si contents in the mantle and, more in general, compositional inhomogeneities [92,93], although some authors interpret these observations in the framework of pure thermal models [94]. The LLSVP should be intrinsically denser to resist entrainment by convection. These compositional pyramids may anchor the hot spots [95]. The presence of a petrologically dense component of the source of hot spots also seems necessary to explain their excess surface temperature [96]. These abyssal heterogeneities help to bridge the gap between geochemical observations and convection modeling [97,98].

The current leading hypothesis for the LLSVPs is the accumulation of subducted oceanic slabs. This corresponds with the locations of known slab graveyards surrounding the Pacific LLSVP. These graveyards are thought to be the reason for the high velocity zone anomalies surrounding the Pacific LLSVP and are thought to have formed by subduction zones that were around long before the dispersion—some 750 million years ago—of the supercontinent Rodinia. Aided by the phase transformation, the temperature would partially melt the slabs, to form a dense heavy melt that pools and forms the ULVZ structures at the bottom of the CMB closer to the LLSVP than the slab graveyards (Figure 19). The rest of the material is then carried upwards due to chemical buoyancy and contributes to the high levels of basalt found at the mid-ocean ridge. The resulting motion forms clusters of small plumes right above the CMB that combine to form larger plumes and then contribute to "superplumes". The Pacific and African LLSVP, in this scenario, are originally created by a discharge of heat from the core (4000 K) to the much colder mantle (2000 K), the recycled LS is only fuel that helps to drive the superplume convection. Since it would be difficult for the Earth's core to maintain this high heat by itself, it gives support for the existence of radiogenic nuclides in the core, as well as the indication that if fertile subducted LS stops subducting in locations preferable for superplume consumption, it will mark the demise of that superplume.

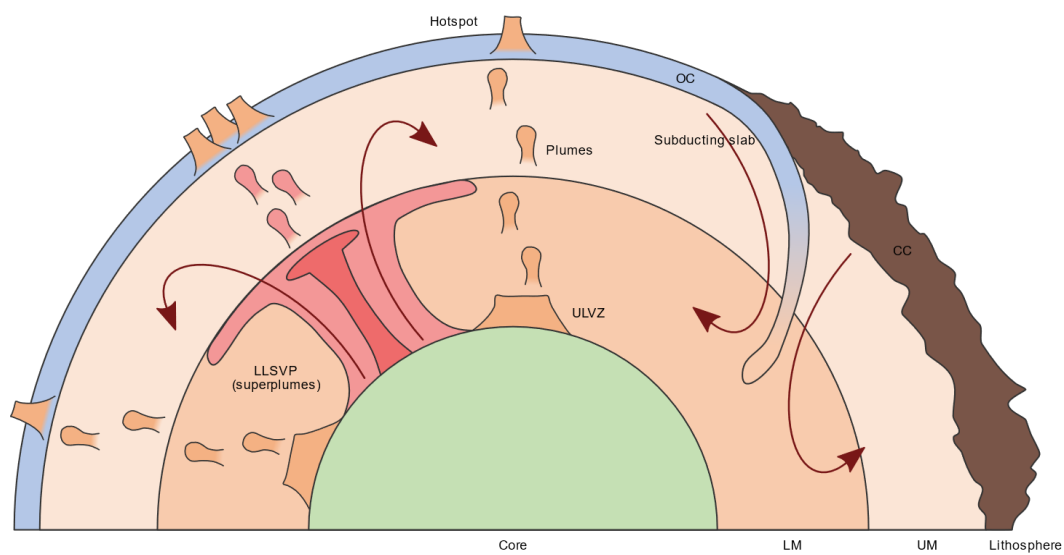

**Figure 19** – Schematic picture of main mantle features and the dynamic processes causing mantle inhomogeneities.

The continuous evolution of our knowledge on the mantle was reflected in the last decades in a multitude of different geochemical models proposed, predicting (i) an homogeneous model, with an UM and LM of similar compositions [65,72,99,100] and (ii) a layered model, with an UM and LM of distinctly different compositions [101-104]. Rather than using the usual seismic separation at 410 km, variants on these models envisage a compositional layering involving only parts of the LM, hence distinguishing the mantle in a so-called enriched mantle (EM) and a depleted mantle (DM). These concepts include basal mantle cumulate layers resultant from early Earth magma ocean conditions [105,106], or gravitationally sequestered layers of early-enriched crust [107].

## 6 Energetics of the Earth

The Earth has the peculiarity of having the highest surface heat flux among all the terrestrial planets of the Solar System. Its total heat loss (Q) is the combination of two distinct sources: (i) the radiogenic energy (H) produced by the radioactive decays of the HPEs contained therein and (ii) the energy released by the secular cooling (C) of our planet.

While decaying, the uranium, thorium and potassium radioisotopes contained in the Earth release geoneutrinos together with heat in a well-fixed ratio. Measuring the geoneutrino flux at surface hence translates in estimating H and in turn constraining C once that Q is known.

How well do we know Q? What are the present constraints on C? What is Earth's heat budget? This section aims at investigating and answering these questions.

### 6.1 The measured heat power (Q)

Heat flow measurements at Earth's surface tell us that our planet is cooling down: in other words, it is losing energy. After many years of research, the cooling of the Earth is still a central issue of the today's debate in solid Earth Sciences. In contrast with the thought of the last century, the conduction is not the only way of Earth's cooling. As a matter of fact, convective motions driving the oceanic plates and radioactive decays of HPEs are responsible for a large fraction of surface heat loss [108-110].

The Earth's heat flow brings to surface crucial information regarding the thermal conductivity and heat production of the Earth's interior. Heat flux measurements are characterized by strong variability on different spatial scale and are function of multiple variables, such as geologic age and geological settings. The spatial integration of individual measurements of heat flux over the surface represents the most direct method for calculating the heat loss rate of the Earth and for obtaining global maps of surface heat flux [111]. The weak points of this approach are mainly two: heat flow observations are (i) sparse and non-uniformly distributed across the globe and (ii) not reliable in the oceans. Oceanic heat flux measurements suffer by systematic errors due to the specific environment: the conventional measurement techniques only account for conductive heat transport (i.e. conduction through soil matrix in permeable rock and sediments). The quantitative assessment of heat transport by hydrothermal circulation (i.e. the water flow through pores and fracture into the sea) remains difficult and feasible only in small-scales studies. For this reason, the energy loss through the sea floor is generally estimated by means of models validated in selected environments which results are compared with the mantle temperature beneath mid-ocean ridge and the evolution of seafloor bathymetry (Jaupart et al., 2015). The half-space models or plate cooling models imply diverse boundary conditions, but all assume that the oceanic LS is hot at its formation at the mid-oceanic ridge, and it cools moving away from the spreading centers. The heat flux from the oceans can be calculated on the basis of temperature variation with depth, distance to the spreading center and different parameters such as thermal properties of the cooling LS, the age of the sea floor and temperature of the magma ascending [112]. For the continents, the situation appears less complicated since to date more than 50000 heat flux measurements [113] from the continents and their margins are available. Nevertheless, as previously mentioned, the irregular and biased spatial distribution motivates a careful statistical treatment of the raw heat-flow dataset including removal of obvious outliers, opportune weighted averaging and combining of statistical errors [114].

From the 1970s, comprehensive estimates of the global surface heat flux were undertaken by different authors adopting measurements of thermal conductivity of rocks and temperature gradients within bore holes for continents and energy loss-models for oceans. The measurements continue to increase and to be refined over time until they achieve Earth's surface heat flux estimates agreeing at around 44-47 TW [111]. The only exception is the lower limit estimation (31 TW) provided by [115] whose approach is based on direct heat flow measurements on sea floor. The remaining references report a global heat loss ranging from 41 to 47 TW of which the 62–77% is attributed to the energy loss occurring in oceans. The analysis reported in [115] were controversially commented by [116] who consider biased and misleading their understanding of heat flow of the Earth. The difference of ~10 TW respect to the previous estimates is attributable, according to [116], to the misconception of hydrothermal circulation which lead to a failed estimate of oceanic heat flow. This key issue was further discussed in [117] which on the opposite defines the half space cooling model as "failing paradigm" against the direct heat flux measurements adopted in the estimates previously published. Although [113] provide a new estimate (Q = 44 TW, see Table 12) based on about 70000 measurements and high resolution studies for hydrothermal calculation, we adopted the value Q = 47 ± 2 TW [114], which comes from a dataset with less measurements (~40000) but includes a comprehensive treatment of the uncertainties.

Table 12 - Integrated terrestrial surface heat fluxes (Q) estimated by different authors. If available, the contribution to the heat power from the continents ($Q_{CT}$) and from the oceans ($Q_{OCS}$) are reported together with the mean heat flux ($q_{CT}$ and $q_{OCS}$) and the surface areas ($A_{CT}$ and $A_{OC}$).

| | Continents | | | Oceans | | | Total |
|---|---|---|---|---|---|---|---|
| REFERENCE | $q_{CT}$ [mWm$^{-2}$] | $A_{CT}$ [10$^6$ km$^2$] | $Q_{CT}$ [TW] | $q_{OCS}$ [mWm$^{-2}$] | $A_{OC}$ [10$^6$km$^2$] | $Q_{OCS}$ (TW) | Q (TW) |
| Williams et al., 1974 [118] | 61 | 148 | 9 ± 1 | 92 | 362 | 33 | 43 ± 6 |
| Davies, 1980 [119] | 55 | 204 | 11 | 95 ± 10 | 306 | 30 | 41 ± 4 |
| Sclater et al., 1980 [120] | 57 | 202 | 12 | 99 | 309 | 30 | 42 |
| Pollack et al., 1993 [121] | 65 ± 2 | 201 | 13[a] | 101 ± 2 | 309 | 31[a] | 44 ± 1 |
| Hofmeister and Criss, 2005 [115] | 61[b] | - | - | 65[b] | - | - | 31 ± 1 |
| Jaupart et al., 2015 [109] | 65 | 210 | 14 ± 1 | 107 | 300 | 32 ± 2 | 46 ± 2 |
| Davies and Davies, 2010 [114] | 71 | 207 | 15 | 105 | 303 | 32 | 47 ± 2 |
| Davies, 2013 [111] | 65 | - | - | 96 | - | - | 45 |
| Lucazeau, 2019 [113] | 66.7 | 211.3 | 14.1 | 89.0 | 299.9 | 26.7 | 44[b] |
| **Adopted** | **71** | **207** | **15** | **105** | **303** | **32** | **47 ± 2** |

[a] Values inferred based on the percentage contribution of continents (30%) and of oceans (70%) reported in the reference.
[a] Median values of "All data" reported in Table 1 of the reference.
[b] Obtained forcing oceanic heat flow with a conductive model fitting subsidence and heat flow.

## 6.2 Earth's heat budget

The understanding of the Earth's present heat budget provides constraints on the internal processes characterizing the convective engine, on the ancient state and on the evolution through geological time of our planet. In this contest, the crux of the matter is represented by the study of the mantle convection which accounts for specific phenomena of present-day dynamics and it aims to the evaluation of past and active geological processes (Section 5.3). The mantle convection models must be defined in a time-dependent framework and must satisfy both the present-day energy budget and the distribution of heat flux at the surface. While the latter is well constrained by measurements on continents and plate cooling model for oceans, the balance of the main sources of the total energy remain uncertain [112]. On a global scale, the flux measured on surface can be seen as the results of the internal processes which occurs inside the Earth: radioactive heat production in the lithosphere ($H_{LS}$) and in the sublithospheric mantle ($H_M$), mantle cooling ($C_M$) and heat loss from the core ($C_C$) (Figure 20). Negligible contributions come from tidal dissipation (~ 0.1 TW) and gravitational potential energy released by the differentiation of crust from the mantle (~ 0.2 TW).

Subtracting the radiogenic energy production (H) from the total heat loss (Q), we can obtain the present Earth' secular cooling (C) in order to have insights on the thermal conditions of Earth's formation and on the dynamical processes in the mantle and core convection. If $H_{LS}$ can be envisioned as well constrained through direct observations, the determination of $H_M$ remains a tangled task. Geoneutrino detection comes into plays right here: taken for granted the accurate calculation of lithospheric flux, valuable insights can be derived about the mantle radioactivity and in turn on its relative contribution to the Earth's energy budget.

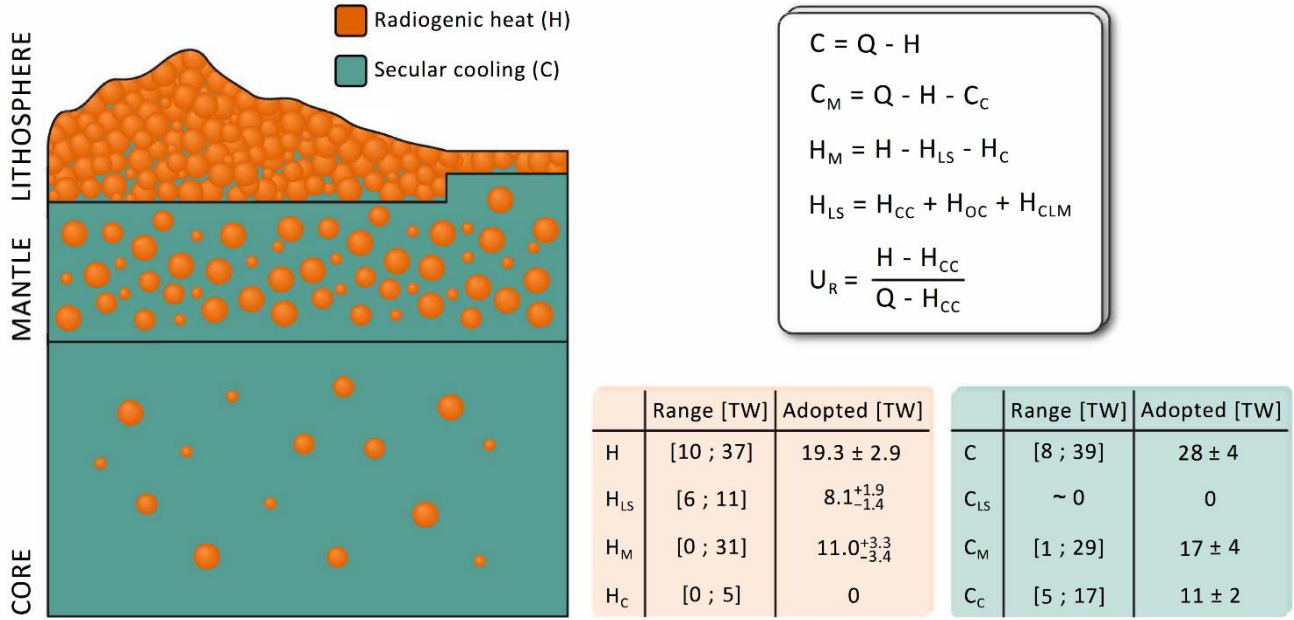

**Figure 20** – Schematic drawing (not in scale) of the Earth's profile illustrating the main contributions to the Earth's heat budget. The surface heat flux (Q) is the sum of secular cooling of the core ($C_C$) and mantle ($C_M$) and of the radiogenic heat from the core ($H_C$), mantle ($H_M$) and lithosphere ($H_{LS}$). Other negligible heat's sources (e.g. tidal dissipation) are ignored. The convective Urey ratio ($U_R$) is defined as the ratio between heat production and heat loss, leaving out $H_{CC}$ from both H and Q. For H, the central value is the average value of all models reported in Tables 18, 19 and 20, while the uncertainty is the average uncertainty with which every model value is estimated (i.e. 15%). For $H_{LS}$, following the arguments presented at the beginning of Section 9, we adopted the value reported in [25] (see Section 8). Due to the asymmetric uncertainty of $H_{LS}$, the value of $H_M = H - H_{LS} - H_C$ is calculated reconstructing the probability density function via Monte Carlo simulations. Still excluding the presence of HPEs in the core ($H_C$ = 0 TW), an unorthodox upper value is reported (see Section 6.2.1). The rounded range of H is given by the lowest-1σ (i.e. 11.4 - 1.6 = 9.8 TW) and the highest+1σ (i.e. 33.6 + 3.6 = 37.2 TW) values of the Low-H (Table 18) and Rich-H (Table 20) models, respectively. The rounded range of $H_{LS}$ is given by the lowest-1σ (i.e. 7.8 - 1.8 = 6.0 TW) and the highest+1σ (i.e. 8.2 + 2.6 = 10.8 TW) values reported in Table 21. The upper (lower) bound of the range of $H_M$ is obtained subtracting the lower (upper) bound of $H_{LS}$ from the upper (lower) bound of H. The adopted values of C and $C_M$ are calculated according to the equations reported in the figure considering the terms as linearly independent and Q = 47 ± 2 TW, while the adopted value $C_C$ is taken from [109], together with an estimated uncertainty equal to 1/6 of the range amplitude. The ranges of $C_M$ and $C_C$ are taken from the preferred interval estimated by [109] (Section 6.2.2). The extremes of the range of C are obtained by Q+1σ minus the lower value of H range (i.e. 47 + 2 - 10 = 39 TW) and Q-1σ minus the upper value of H range (i.e. 47 - 2 - 37 = 8 TW). We set $C_{LS}$ = 0 since the secular cooling of the lithosphere can be considered negligible (Section 6.2.2).

In the comprehensive understanding of the Earth thermal budget, a key parameter is represented by the Urey ratio that can be easily seen as the ratio of heat production over heat loss. In other words, it measures the efficiency of the Earth's convective engine in evacuating heat generated by radioactive decay [109].

It is worth mentioning that geophysicists and geochemists defined this nondimensional number in two different ways [108]: the Bulk Earth Urey ratio and the convective Urey ratio. Commonly, in the geochemical community, the Urey ratio denotes the Bulk Urey ratio calculated as follows:

$$U_R = \frac{H}{Q} \ (Bulk\ Earth) \tag{11}$$

where H is the radiogenic power of the entire Earth and Q is the total surface heat flux (see Section 5.2). The convective Urey ratio, extensively used in geophysical literature is given by:

$$U_R = \frac{H - H_{CC}}{Q - H_{CC}} \quad (Convective) \tag{12}$$

where the radiogenic power of the continental crust ($H_{CC}$, see Section 5.2.3) is leaved out from both heat loss and the heat production since the continental heat sources are not taken into account because they are not involved in mantle convection. The convective Urey ratio, hereafter Urey ratio ($U_R$), corresponds to the "original" Urey ratio appeared in [122]. According to its definition, $U_R$ assumes that the entire mantle convects as single layer (whole mantle convection) [108]. Given a total Earth heat flux (Table 12), $U_R$ is a BSE model-dependent parameter: [10] sets a range of present $U_R$ between 0.02 (low HPEs content) and of 0.75 (high HPEs content) while 0.29 and 0.23 are the best estimates proposed by [109] and by [108] respectively. According to thermal evolution models, the $U_R$ was close to 1 until ∼ 3 Ga ago, when it started to decrease with the emergence of plate tectonics [108].

### 6.2.1 Radiogenic heat production (H)

The radiogenic heat production inside the Earth is due to the energy released by the decays of radioactive nuclides which indeed play a starring role in the comprehension of geodynamical processes. Neglecting a fractional contribution coming from rare radionuclides ($^{87}$Rb, $^{138}$La, $^{147}$Sm, $^{176}$Lu, $^{187}$Re and $^{190}$Pt), the 99.5% of the present Earth's radiogenic heat production is due to the decay (or the decay chains) of $^{40}$K, $^{232}$Th, $^{235}$U and $^{238}$U, long-lived radionuclides ($T_{1/2} > 10^8$ years) created at the time of the Solar System formation and still extant now [123]. Due to their different half-lives, the relative amounts of heat-producing nuclides, and in turn their contribution to radiogenic budget, changed with time. At the early stages of Solar System, the concentration of $^{40}$K, $^{232}$Th, $^{235}$U and $^{238}$U were approximately 12, 1.25, 84 and 2 times higher respectively. In the first ∼10 Myr of the Solar System, the short-lived radionuclides $^{26}$Al ($T_{1/2}$ = 0.7 Myr) and $^{30}$Fe ($T_{1/2}$ = 1.2 Myr), now extinct in planetary bodies, were the dominant radiogenic sources [124]. Given the masses of K, U, Th in the present Earth and their decay properties, it is possible to trace the evolution of the Earth radiogenic power through the time together with the contribution of the different HPEs. In Figure 21a, the Earth radiogenic power is plotted respect to the time for the last 3.7 Gyr adopting a BSE composition of the medium-H model reported in [65] following the updated $a$(K)/$a$(U) from [125] (Section 7.3). The percentage contribution of the four long-lived radioisotopes changed with the time (Figure 21b) due to their different ratios. A decreasing trend is clearly observable of the $^{40}$K contribution which is now less than the half of that of the early stages of the Solar System; conversely the contribution of $^{238}$U and $^{232}$Th increases with the time up to reach respectively the 37% and 42% of the present radiogenic power. A negative trend is notable also for $^{235}$U contribution which reached a negligible present value of 2%.

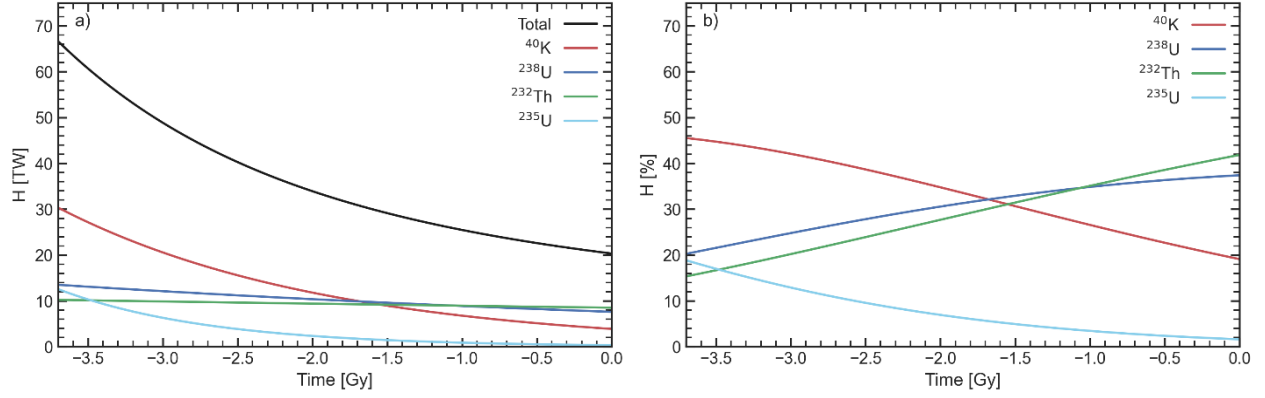

**Figure 21** – Earth's Radiogenic heat (H) in TW (a) and relative contributions in percentage (b) from $^{238}$U, $^{235}$U, $^{232}$Th and $^{40}$K over the last 3.7 Gyr. The compositional model is from [65] with the updated $a(K)/a(U)$ reported in [125]

For each radioactive decay, the heat production rate is strictly related to the energy released (Q-value), i.e. the difference between the mass of the parent ($m_p$) and the daughter nuclide(s) ($m_d$), multiplied by the square of the velocity of light vacuum, $c_0$. In the α decay, since the spectrum of decay energies to the different discrete energy levels of the daughter nuclides is discrete, all of the decay energy is transformed into heat; so, the Q-value is an optimal approximation of the heat production. For β⁻, β⁺ and electron capture, the heat production $E_H$ must be calculated by subtracting from the Q-value the energy carried away by the (anti)neutrino $E_{\bar{\nu}}$, which does not contribute to heat production [124]:

$$E_H = (m_p - m_d)c_0^2 - E_{\bar{\nu}} \tag{13}$$

The specific heat production (h) is given by:

$$h = E_H \lambda \tag{14}$$

where $\lambda = \frac{\ln 2}{T_{\frac{1}{2}}}$. The simplified assumption that the neutrino always carries away 2/3 of the decay energy made by different authors [15,126,127] may generate overestimation or underestimation of the real value; these could be canceled in long decay chains (e.g. U and Th decay chain) but not in short decay chains. In a recent work, [124] reevaluated and updated the radioactive heat production data of $^{26}$Al, $^{40}$K, $^{60}$Fe, $^{232}$Th, $^{235}$U and $^{238}$U, using newest available information (nuclear and atomic properties) and accounting for details of the decay processes. The results obtained for the four long-lived nuclides relevant for geosciences are listed in Table 13. For $^{40}$K, [124] observed a difference of 1-2% between the calculated values and the values reported by [128] and [127] that is mostly due to the difference in the mean β energy used for the calculation. An excellent agreement for all the radionuclides is instead highlighted with [16] who adopted a similar approach and actual decay spectra.

**Table 13** – Total (Q-value) and Heat-Effective Decay Energy per Atom ($E_H$), specific heat production (h) of $^{40}$K, $^{232}$Th, $^{235}$U, and $^{238}$U, elemental specific heat production (h') for K, U and Th. The decay energies of $^{40}$K are the weighted means of the two principal decay modes. For $^{238}$U are included the fraction/contribution of $^{234}$U. Adapted from [124].

|  | Q-value [MeV] | $E_H$ [MeV] | h [W/kg] | h' [W/kg] |
|---|---|---|---|---|
| $^{40}$K | 1.332 | 0.677 | $2.8761 \times 10^{-5}$ | $3.4302 \times 10^{-9}$ |
| $^{232}$Th | 42.646 | 40.418 | $2.6368 \times 10^{-5}$ | $2.6368 \times 10^{-5}$ |
| $^{235}$U | 46.397 | 44.380 | $5.6840 \times 10^{-4}$ | $9.8314 \times 10^{-5}$ |
| $^{238}$U | 51.694 | 47.650 | $9.4946 \times 10^{-5}$ | |

For an Earth reservoir (X), the radiogenic power ($H_X$) in TW is given by:

$$H_X = (M(K)_X \cdot h'_K + M(Th)_X \cdot h'_{Th} + M(U)_X \cdot h'_U) \times 10^{-12} \quad (15)$$

where M(K), M(Th) and M(U) are respectively the potassium, thorium and uranium masses in the reservoir. The radiogenic production of the Earth is attributable to the HPEs amount in the BSE. The presence of these elements, and in particular of K, in the Earth's core is envisaged by some authors but is still controversial and under debate. According to [73] a certain amount of K could alloyed with Fe at pressure > 26 GPa and incorporated into the core during the early core formation stage of the Earth. The K abundances proposed by different authors [73-78] ranges between 0.2 – 250 µg/g to which corresponds a heat generation of 0.01 – 1.7 TW. Although U and Th are considered essentially lithophile and resident only in the silicate phase, [129] suggested the a small fraction of U (10 ng/g) and Th (21 ng/g) may also partition into the metallic phase of the core and produce an additional heat source ($\sim$ 3 TW). Considering $H_C \sim 4.7$ TW the result of unorthodox models which hypothesize K, U and Th in the core, in this study we exclude the presence of HPEs therein, setting $H_C = 0$ (Figure 20).

The concentration of K, U and Th determines the present contribution of the radiogenic heat to the Earth's heat budget. The majority of compositional models of the BSE use the chondritic meteorites to describe the starting material of the Earth and in turn to determine the present concentration of HPEs. The Section 7 serves the purpose to illustrate and compare the classes of models which imply different estimates of heat generation in the bulk Earth. The lower and upper range of H can be defined on the basis of the lowest and the highest value for the Low-H and Rich-H models, respectively. The mean value of all available models is H = 19.3 ± 2.9, where the uncertainty is given by the average relative uncertainty (15%) with which every model is estimated (Figure 20).

The estimates provided by available models [25,109,130] (see Section 8) prove that $H_{Ls}$ varies between 7.8 and 8.2 TW (Table 21) with an average relative uncertainty of $\sim$25%. Considering the ± 1σ errors associated to each value, we can define for $H_{Ls}$ a full range of 6 – 10 TW. For the purpose of the extraction of the mantle radiogenic power from the geoneutrino signal, we adopted $H_{LS} = 8.1^{+1.9}_{-1.6}$ TW [10] obtained with the same lithospheric model [25] adopted for the geoneutrino signal estimation (Figure 20). Thanks to the increasing of availability of oceanic and subcrustal rock samples, the more recent models provide detailed information about the radiogenic heat from the LS components. While the earlier estimates took into account only the heat from the CC, nowadays we can quantify as $\sim$5% the contribution given by the HPEs in the OCC ($H_{OCC}$) and in the CLM ($H_{CLM}$) (Table 14).

**Table 14** – Radiogenic heat power of the continental crust ($H_{CC}$), oceanic crust ($H_{OCC}$) and continental lithospheric mantle ($H_{CLM}$) reported by different authors.

| REFERENCE | $H_{CC}$ [TW] | $H_{OCC}$ [TW] | $H_{CLM}$ [TW] |
|---|---|---|---|
| Taylor and McLennan, 1995 [131][a] | 5.6 | - | - |
| Rudnick and Fountain, 1995 [132][a] | 7.7 | - | - |
| Wedepohl, 1995 [133][a] | 8.5 | - | - |
| Mantovani et al., 2004 [22][b] | 8.4 | 0.2 | - |
| McLennan, 2001 [134][a] | 6.3 | - | - |
| Rudnick and Gao, 2003 [135][a] | 7.4 | - | - |
| Stacey and Davis, 2008 [136] | 8 | - | - |
| Hacker et al., 2011 [137][a] | 7.9 | - | - |
| Dye, 2010 [24] | 7.71 ± 1.5 | 0.24 ± 0.04 | - |
| Šrámek et al., 2013 [138] | 7.8 ± 0.9 | 0.22 ± 0.03 | - |
| Huang et al., 2013 [25][c] | $6.8^{+1.4}_{-1.1}$ | 0.3 ± 0.1 | $0.8^{+1.1}_{-0.6}$ |
| Wipperfurth, 2020 [130] (Litho 1.0)[d] | $7.1^{+2.1}_{-0.6}$ | 0.2 | $0.5^{+0.8}_{-0.3}$ |
| Wipperfurth, 2020 [130] (Crust 1.0)[d] | $6.7^{+2.1}_{-0.6}$ | 0.3 | $0.6^{+1.5}_{-.04}$ |
| Wipperfurth, 2020 [130] (Crust 2.0)[d] | $7.0^{+2.0}_{-1.6}$ | 0.2 | $0.6^{+1.6}_{-.04}$ |

[a] As appeared in [25].
[b] Obtained on the basis of radiogenic heat of the bulk crust and relative masses in continental and oceanic crust reported in the reference.
[c] The $H_{CLM}$ is taken as appeared from [10].
[d] The $H_{OCC}$ is obtained summing the heat of Sed and C of OC reported in Table S2, S3 and S4 of the reference.

Starting from these findings, one can infer the radiogenic heat of the mantle ($H_M$ = 11.3 ± 3.3 TW) by subtracting the relatively well constrained and independent contribution of $H_{LS}$ from H (Figure 20).

### 6.2.2 Secular Cooling (C)

Beyond the radiogenic heat production, the loss of internal energy of the Earth is balanced by the secular cooling (C), i.e. the gradual decrease of the primordial heat content. Nowadays, the cooling rate is estimated to exceed 100 K Gyr$^{-1}$ but it has not remained constant. As suggested by geological data and physical constraint on the thermal structure or the early Earth [109], the cooling rate increased with the time. Assuming a $U_R$ ranging from 0.08 to 0.38, [108] proposes a range 50 - 100 K Gyr$^{-1}$ for an average over the last 3 Gyr with a present value of 124 ± 22 K Gyr$^{-1}$ compatible with the value 106 K Gyr$^{-1}$ reported by [109].

The heat flow from the core ($C_c$) remains a controversial parameter of the thermal evolution models and, despite its not negligible value, was often ignored or embedded with the mantle heat flow. As a matter of fact, the core must have cooled by hundreds of degrees since its formation and sustained the operation of the geodynamo. The requirement of dynamo action in the core represents a constraint for the estimation of the heat across the CMB. In this puzzle, the thermal conductivity of high-pressure iron is the most important parameter [110]. Recent laboratory measurements and theoretical ab-initio calculation set its value at ~90 Wm$^{-1}$K$^{-1}$ [139], which is a factor 2-3 higher than the previous estimates [140].

Different authors estimate $C_C$ on the basis of different arguments. Using the entropy balance of the core and assuming a thermal conductivity increasing with depth, [141] estimates $C_c$ = 13.25 TW. The lowest values are proposed by [108] ($C_c$ = 4.5 ± 0.8 TW), by [136] (3.5 TW) and by [142] (5 TW) based on the heat carried by hot spots. Higher estimates come from studies of the thermal structure around the CMB (6 – 12 TW from [143]), of postperovskite phase diagrams (9 – 12 TW from [85]), numerical simulations (13 TW from [144]) and seismic tomography (10 – 30 TW from [145]). Our adopted value $C_c$ = 11 ± 2 TW (Figure 20) corresponds to the preferred estimation published by [109], who reviewed a range of values between 5 – 17 TW.

The cooling rate of the mantle can be estimated from its temperature calculated on the basis of MORB composition. A long-term average cooling rate of 50 K Gyr$^{-1}$ corresponding to $C_M \sim 7$ TW is suggested by petrological studies on Archean MORB rocks [112]. Higher estimates can be found calculating $C_M$ from the difference between the output (Q) and the input ranges ($H_{CC} + H_M + C_C$). Adopting this approach, the values of Q, $H_{CC}$, $H_M$ and $C_C$ proposed by [108] lead to an estimation of $C_M$ = 23 TW. In the same way, [109] found a preferred value of 16 TW and, considering all the uncertainties, a wide range of 1 – 29 TW (Figure 20). Neglecting the contribution of the continental lithosphere ($C_{LS} \sim 0$ TW), the adopted mantle secular cooling $C_M$ can be calculated as $C_M = Q - H - C_C = 17 \pm 4$ TW (Figure 20), where the uncertainties are propagated in quadrature.

The little consensus and the wide ranges proposed by the different authors for the secular cooling demand experimental confirmations. In this perspective, the insights gathered via geoneutrino measurements could prove useful: these estimations can be compared to the experimental results obtained by the combination of geoneutrino signals measured by KamLAND and Borexino (Section 10).

# 7 Bulk Silicate Earth models

## 7.1 Is the Earth compositionally similar to a primitive meteorite?

Although the Earth is the planet most familiar to us, direct probes provide a more uncertain geochemical bulk composition than the Sun's one [146]. The deepest hole that has ever been dug is about 12 km deep [147], while the deepest rock that has ever been recovered comes from ~700 km beneath the surface of our planet [148]. Hence, a coherent chemical description of our planet requires to embrace several indirect inputs.

It can be tempting to naively build an Earth in the image and likeness of a primitive meteorite. Indeed, ~$10^7$ kg of interplanetary solid material hit the Earth every year, with ~$10^4$ kg of meteorites falling to the ground [149]. Although chondritic meteorites are very common in the Solar System, they are the rarest to find on Earth's surface because of their resemblance to stones. On the other hand, they are also one of the most precious materials, since their thermal history partially preserved the initial chemical cocktail of our Solar System's formation. The most chemically primitive meteorites are carbonaceous chondrites that, together with enstatite chondrites, represent 5% and 2% of the stony meteorites fallen on Earth [150]. These two classes of meteorites are grouped according to their distinctive compositions (Table 15). Although their U and Th abundances can vary of a factor ~2 among the different groups, their Th/U ratio remains basically constant. The terrestrial radiogenic heat power (H) calculated adopting HPE abundances from the different chondritic groups is constrained in the range 21-26 TW.

**Table 15** – Uranium, thorium (in ng/g) and potassium (in μg/g) abundances ($a_{ch}$) [151] and their ratios for different chondrite groups belonging to carbonaceous and enstatite classes. The radiogenic heat (H) is calculated employing Eq. (15) by assuming a bulk earth mass (Table 11) with chondritic abundances ($a_{ch}$).

| Class | Group | $a_{ch}$(U) [ng/g] | $a_{ch}$(Th) [ng/g] | $a_{ch}$(K) [μg/g] | $a_{ch}$(Th)/ $a_{ch}$(U) | $a_{ch}$(K)/$a_{ch}$(U) [$10^4$] | H [TW] |
|---|---|---|---|---|---|---|---|
| Carbonaceous | CI | 8.2 | 29 | 560 | 3.5 | 6.8 | 20.8 |
| | CM | 11 | 40 | 400 | 3.6 | 3.6 | 20.9 |
| | CO | 13 | 45 | 345 | 3.5 | 2.7 | 21.8 |
| | CV | 17 | 60 | 310 | 3.5 | 1.8 | 25.8 |
| Enstatite | EH | 9 | 30 | 800 | 3.3 | 8.9 | 26.4 |
| | EL | 10 | 35 | 735 | 3.5 | 7.4 | 26.4 |

The chondrites with the chemical composition closest to the solar photosphere are the ones belonging to the CI carbonaceous group[3] [72], while those having the isotopic composition most similar to terrestrial samples are the enstatite chondrites [103,152]. The enstatite (carbonaceous) chondrites are characterized by the lowest (highest) oxidized and highest (lowest) metallic iron content. These peculiarities are relevant for two reasons: (i) a low (high) degree of oxidation proves a formation in an oxygen-poor (rich) environment, corresponding to inner (outer) portions of the solar nebula, (ii) a high (low) metallic iron content is a predisposing factor to the metallic core formation in planets. Indeed, a BSE compositional model employing carbonaceous or enstatite chondrites as its fundamental building blocks must comply with the essential constraint of having enough metallic iron to form Earth's core.

The gravitational segregation of metallic Fe and FeS melts in Earth's core started in the first 2-3 Myr and presumably lasted till 60-100 Myr since planetary formation [153]. The differentiation of a metallic core of mass $M_C$ from the BSE ($M_{BSE}$) brought incompatible elements, such as U and Th, to accumulate in the remaining silicate portion of the Earth, hence enriching their abundances of a factor:

$$f_C = \frac{M_{Earth}}{M_{Earth} - M_C} = \frac{M_{Earth}}{M_{BSE}} = 1.48 \tag{16}$$

---

[3] Even in these meteorites volatile elements have been depleted to various degrees, including the six most abundant elements (H, He, C, N, O and Ne) and lithium.

Therefore, the abundances of U and Th in the BSE appear ~50% higher than what observed in bulk carbonaceous and enstatite chondrites.

The differentiation of a metallic core was not the only process leading to an enrichment of U and Th in the silicate Earth. The loss of volatile elements in the planetary accretion stage is thought to have further increased the RLE abundances of the BSE [154], as suggested by several evidences: (i) the BSE is found to have ~2-3 times higher RLEs concentration than chondrites [154], (ii) the BSE abundances of the volatile lithophile elements show a coherent depletion pattern (when compared to chondrites) as a function of their $T_C$ [67,68], (iii) high temperatures in the early stages of Earth's formation (as testified by isotopic $\Delta^{17}O$ considerations) [155] and/or collisional erosions [67] were all predisposing factors for the loss of lighter elements. Without going into the details of the debate about the different volatilization mechanisms, we can parametrize the removal of part of the BSE with a depletion enrichment factor:

$$f_D = \frac{M_{BSE}}{M_{BSE} - M_V} \tag{17}$$

where $M_V$ is the mass of the material which left the Earth because of volatilization and/or collisional erosion. Therefore, following the two different mechanisms (core-mantle differentiation and mass depletion) leading to U and Th enrichment in the BSE, the abundances $a_{BSE}(U; Th)$ of these elements in the silicate Earth can be calculated from their chondritic abundances $a_{Ch}(U; Th)$ as:

$$a_{BSE}(U; Th) = f_C \cdot f_D \cdot a_{Ch}(U; Th) \tag{18}$$

Whilst there's wide agreement on $f_C$, which is known at the level of a few percent, the enrichment factor due to volatilization $f_D$ spans a wider range, between 0.8 and 3.3, according to different authors (Table 9). Among the proposed models, the one and only predicting $f_D$<1 is the collisional erosion model from [69], coherently with their hypothesis of preferential collisional erosion of the RLE-enriched crust (which hence removed part of the U and Th masses). All the BSE models starting from enstatite compositions predict $f_D$~1, reflecting the implicit conditions of enstatite material being virtually volatile-free [103] and thus not requiring any depletion correction. Models starting from carbonaceous chondrites predict $f_D$ values in the range 1.4-2.0 (1.7 on average), implicitly suggesting the removal of volatiles for as much as ~40% the mass of the silicate Earth. This is a consequence of the observation that the BSE is highly enriched in RLEs respect to carbonaceous chondrites. Models based on Earth's mantle dynamics predict $f_D$>2, independently by the chosen compositional building block, being it enstatitic or carbonaceous. For every model, the $f_D$ values obtained from Th abundances are higher than those obtained from U. This comes directly from the observation that the $a$(Th)/$a$(U) ratio of the BSE is expected to be bigger than what observed in chondrites (Table 15 and Table 19).

**Table 16** – Estimated volatilization enrichment factor ($f_D$) according to different models. $f_D$(U) and $f_D$(Th) are calculated by reversing Eq. (17), assuming a core-mantle differentiation enrichment factor of $f_c$=1.48 and chondritic abundances for $a_{Ch}$(U) and $a_{Ch}$(Th) following Table 15.

| Reference | Chondrite | $a_{BSE}$(U) [ng/g] | $f_D$(U) | $a_{BSE}$(Th) [ng/g] | $f_D$(Th) |
|---|---|---|---|---|---|
| Jackson and Jellinek, 2013 | CI-EH | 14 ± 3 | 1.2 | 55 ± 11 | 1.3 |
| O'Neill and Palme, 2008 | CI-EH | 10 | 0.8 | 40 | 0.9 |
| Javoy and Kaminski, 2014 | EH | 15 ± 2 | 1.2 | 51 ± 4 | 1.2 |
| Javoy et al., 2010 | EH | 12 ± 2 | 0.9 | 43 ± 4 | 1.0 |
| McDonough and Sun, 1995 | C | 20 ± 4 | 1.7 | 80 ± 12 | 1.9 |
| Lyubetskaya and Korenaga, 2007 | C | 17 ± 3 | 1.4 | 63 ± 11 | 1.5 |
| Palme and O'Neill, 2007 | C | 22 ± 3 | 1.8 | 83 ± 13 | 1.9 |
| Arevalo, 2010 | C | 20 ± 4 | 1.6 | 80 ± 13 | 1.9 |
| Wang et al., 2018 | C | 20 ± 2 | 1.6 | 75 ± 7 | 1.7 |
| Palme and O'Neill, 2014 | C | 23 ± 3 | 1.9 | 85 ± 13 | 2.0 |
| Turcotte, 2002* | CI-EH | 35 ± 4 | 2.8 | 140 ± 14 | 3.3 |
| Turcotte, 2014 | CI-EH | 31 | 2.5 | 124 | 2.9 |

[+]as reported in [138]

A compositional model attempting to accurately describe the Earth must consider several constraints, which are not verified simultaneously by any known class of meteorite. The cosmochemical inputs, which employ assumptions on chondrites and Solar System's compositions to describe the Earth, are not enough. Our understanding of the mantle has to rely on compositional models based on:
(i) geochemical information, which makes use of samples and observations of chemical processes occurring on Earth and on the uppermost part of the mantle;
(ii) geodynamical observations, which show whole-mantle convection and require a substantial energy input to justify the observed convective processes.

Indeed, regardless of Earth's composition's similarities with a class of chondrites rather than another, our planet is not a chondrite and hence has its peculiar and singular composition. For this reason, scientists proposed a wide variety of BSE compositional models (Table 18-20), which can be grouped on the basis of their expected radiogenic heat production (H) [130] in (i) poor-H models, (ii) medium-H models and (iii) rich-H models (Table 17).

**Table 17** – Masses of U, Th and K in the BSE, Th/U and K/U mass ratios and the radiogenic heat (H) expected from poor-H, medium-H and rich-H BSE models together with the corresponding Urey ratio ($U_R$) (See Section 6.2). For each class, the masses of the HPEs are obtained multiplying the adopted abundances reported in Table 18, Table 19 and Table 20 by the BSE mass (Table 11). The standard deviations on the masses are obtained by propagating the errors on the abundances, here considered as the only source of uncertainty. For each model class, H is calculated multiplying the masses of U, Th and K by the specific heat production coefficients h' reported in Table 13. The obtained values are then summed to obtain H(U+Th) and H(U+Th+K). Their uncertainties are propagated from the masses' standard deviations, considering U, Th and K as fully correlated. The uncertainty on the coefficients h' is here neglected.

| Classes | $M_{BSE}$(U) [$10^{16}$ kg] | $M_{BSE}$(Th) [$10^{16}$ kg] | $M_{BSE}$(K) [$10^{19}$ kg] | $M_{BSE}$(Th)/$M_{BSE}$(U) | $M_{BSE}$(K)/$M_{BSE}$(U) [$10^4$] | $H_{BSE}$(U+Th) [TW] | $H_{BSE}$(U+Th+K) [TW] | $U_R$ |
|---|---|---|---|---|---|---|---|---|
| Poor-H | 5.2 ± 0.9 | 19.1 ± 2.4 | 67.3 ± 11.5 | 3.7 | 1.3 | 10.1 ± 1.5 | 12.4 ± 1.9 | $0.14^{+0.06}_{-0.05}$ |
| Medium-H | 8.2 ± 1.3 | 31.3 ± 4.6 | 98.7 ± 16.9 | 3.8 | 1.2 | 16.3 ± 2.5 | 19.7 ± 3.1 | 0.32 ± 0.08 |
| Rich-H | 13.3 ± 1.5 | 53.3 ± 5.3 | 133.2 ± 13.3 | 4.0 | 1.0 | 27.1 ± 2.9 | 31.7 ± 3.4 | 0.62 ± 0.09 |

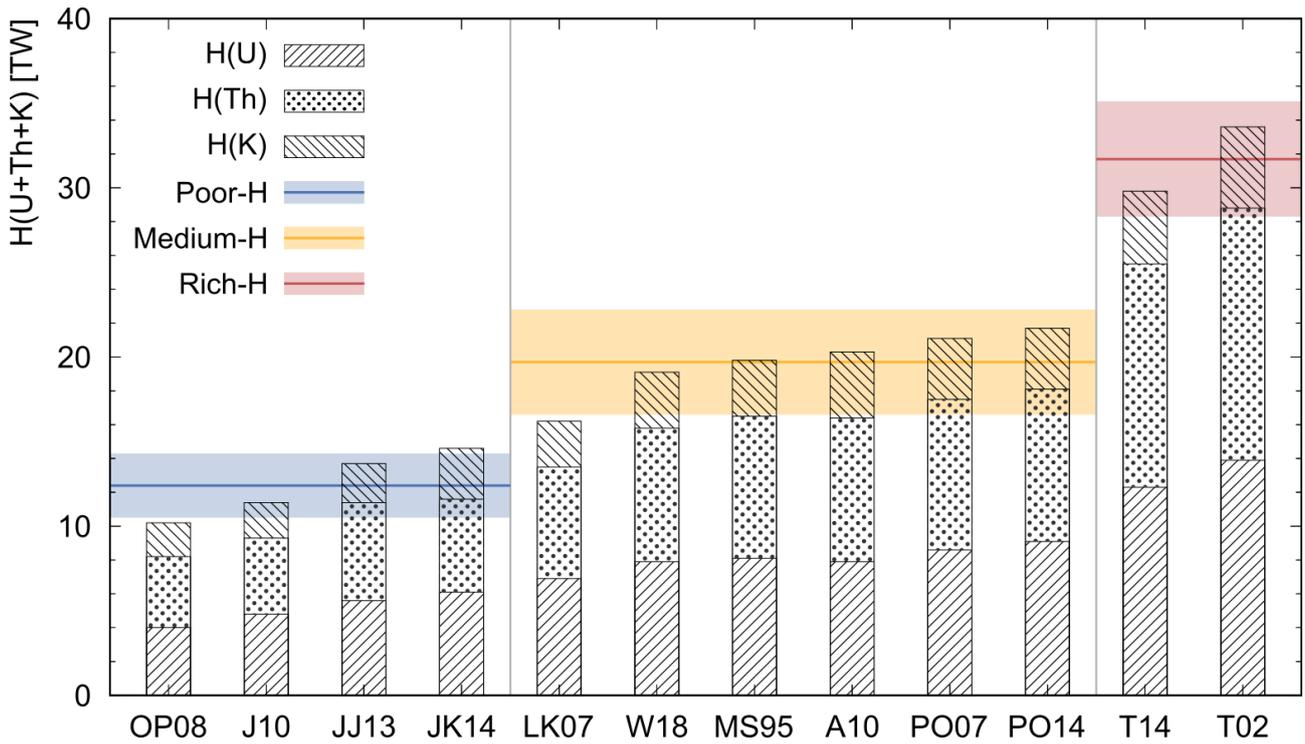

**Figure 22** – Histogram of the predicted radiogenic heat production H(U+Th+K) (in TW) according to the different BSE compositional models reported in Table 18, Table 19 and Table 20. Each bar is subdivided according to the relative contribution of U, Th and K to the overall radiogenic power. The horizontal bands represent the H(U+Th+K) (together with its standard deviation) expected from the adopted poor-H (in blue), medium-H (in yellow) and rich-H (in red) BSE classes reported in Table 17.

### 7.2 Poor-H BSE models

The BSE enstatite models [103,156,157] are based on a mixture of enstatite chondrites composition (~2/3 of EH chondrites and ~1/3 of EL chondrites) since this class of meteorites (i) are the chondrite group isotopically most similar to the Earth [103], (ii) share a common oxygen reservoir with our planet [158], (iii) are largely degassed, so that the total mass reduction to match the depleted Earth is estimated to 4.7 ± 1.6% [157], (iv) have sufficiently high iron content to explain the metallic core and the observed oxidized iron in the mantle. All these observations seem to suggest a common origin between Earth and enstatite chondrites, thus making this meteorite type a good candidate as the building block of our planet. However, since the enstatite Mg/Si ratio is 35% lower than that of the measured in the UM [159], these models implicitly require compositional layering of the mantle [103]. Because of the small mass correction due to volatilization ($f_D$~1), the predicted BSE composition ends up being low in U and Th, constraining H in the range [10, 16] TW assuming a $a(K)/a(U)$ ratio of 14000 [125].

Non-chondritic and collisional erosion models consider the Earth to be either an unsampled material or a combination of chondritic materials altered by collisional erosion processes [69]. In particular, these models observe that (i) terrestrial mantle rocks' isotopic ratios of $^{142}Nd/^{144}Nd$, $^{187}Os/^{188}Os$, the atmospheric $^{84}Kr/^{130}Xe$ and the isotopic anomalies in $\Delta^{17}O$, $\varepsilon^{48}Ca$, $\varepsilon^{50}Ti$, $\varepsilon^{54}Cr$, $\varepsilon^{64}Ni$, $\varepsilon^{92}Mo$, $\varepsilon^{100}Ru$, and $\mu^{142}Nd$ [152] are not matched by any known meteorite and cannot be explained by the nucleosynthetic variability among chondrites [67,158,160], (ii) Moon isotopic composition seems to match Earth's one, hinting to a collisional origin caused by large impact after Earth differentiation, (iii) the Earth's volatility pattern shows a depletion dependence on incompatibility, suggesting volatilization from early-formed crust during the latter stages of accretion. This assumption would explain the apparent paradox of the missing $^{40}Ar$ planetary budget, otherwise requiring degassed hidden reservoirs [67,160]. As a consequence of preferential collisional erosion, these models suppose that the BSE (which initially could have accreted in chondritic proportions) was largely deprived of its HPEs by the removing of 10% of the RLE-enriched crust during intense collisions with large impacts. Therefore, these models predict low HPEs abundances and constrain H at ~10 TW.

**Table 18** – Uranium, thorium (in ng/g) and potassium (in µg/g) abundances, masses, and predicted radiogenic heat (H) for different poor-H BSE compositional models. The masses of U, Th and K are obtained multiplying the HPEs abundances by the BSE mass (Table 11). The standard deviations on the masses are obtained by propagating the errors on the abundances, here considered as the only source of uncertainty. H is calculated multiplying the masses of U, Th and K by the specific heat production coefficients h' reported in Table 13. The obtained values are then summed to obtain H(U+Th) and H(U+Th+K). Their uncertainties are propagated from the masses' standard deviations, considering U, Th and K as fully correlated. The uncertainty on the coefficients h' is here neglected. The last row reports the HPEs abundances, masses and H (together with their standard deviations) adopted in this study to represent poor-H models. The abundances central values are calculated by averaging all the values proposed by the different models considered. The errors on the abundances are calculated by considering the average relative uncertainty of the different models and multiplying it by the central value. The adopted abundances are then employed to obtain the HPEs masses and predicted radiogenic heat as explained above.

| | | | $a_{BSE}$(U) [ng/g] | $a_{BSE}$(Th) [ng/g] | $a_{BSE}$(K) [µg/g] | $M_{BSE}$(U) [$10^{16}$ kg] | $M_{BSE}$(Th) [$10^{16}$ kg] | $M_{BSE}$(K) [$10^{19}$ kg] | $M_{BSE}$(Th)/$M_{BSE}$(U) | $M_{BSE}$(K)/$M_{BSE}$(U) [$10^4$] | H(U+Th) [TW] | H(U+Th+K) [TW] |
|---|---|---|---|---|---|---|---|---|---|---|---|---|
| Poor-H | Jackson and Jellinek, 2013 | JJ13 | 14 ± 3 | 55 ± 11 | 166 ± 33 | 5.6 ± 1.2 | 22.2 ± 4.4 | 67.0 ± 13.3 | 3.9 | 1.2 | 11.4 ± 2.4 | 13.7 ± 2.8 |
| | O'Neill and Palme, 2008 | OP08 | 10 | 40 | 140 | 4.0 | 16.1 | 56.5 | 4.0 | 1.4 | 8.2 | 10.2 |
| | Javoy and Kaminski, 2014 | JK14 | 15 ± 2 | 51 ± 4 | 216 ± 25 | 6.2 ± 0.7 | 20.7 ± 1.8 | 87.0 ± 10.2 | 3.3 | 1.4 | 11.6 ± 1.2 | 14.6 ± 1.5 |
| | Javoy et al., 2010[†] | J10 | 12 ± 2 | 43 ± 4 | 146 ± 29 | 4.8 ± 0.8 | 17.4 ± 1.6 | 58.9 ± 11.7 | 3.6 | 1.2 | 9.3 ± 1.2 | 11.4 ± 1.6 |
| | **Adopted values** | | **13 ± 2** | **47 ± 6** | **167 ± 29** | **5.2 ± 0.9** | **19.1 ± 2.4** | **67.3 ± 11.5** | **3.7** | **1.3** | **10.1 ± 1.5** | **12.4 ± 1.9** |

[†]as reported in [138]

## 7.3 Medium-H BSE models

Geochemical models recognize that there is no group of meteorites that has a bulk composition matching that of the Earth, but combine observations from chondrites and the residuum-melt relationship between peridotites and basalts to estimate the composition of the BSE [65,161,162]. Earth is assumed to have a bulk major-element composition matching that of CI chondrites since this class of meteorites (i) are chemically the most primitive and not differentiated known meteorites, (ii) they perfectly match the photosphere composition, (iii) they correctly set the Mg/Si ratio and the absolute abundances of the refractory element abundances [72]. However, these models do not explain the isotopic anomalies of our planet and require large volatilization corrections to match Earth's depletion pattern [68]. As a consequence of the ~40% degassing correction, these models appear 2-3 times enriched in HPEs when compared to CI chondrites and predict H to be in the range [13.3, 25.0] TW.

**Table 19** – Uranium, thorium (in ng/g) and potassium (in µg/g) abundances, masses, and predicted radiogenic heat (H) for different medium-H BSE compositional models. The masses of U, Th and K are obtained multiplying the HPEs abundances by the BSE mass (Table 11). The standard deviations on the masses are obtained by propagating the errors on the abundances, here considered as the only source of uncertainty. The radiogenic heat H is calculated multiplying the masses of U, Th and K by the specific heat production coefficients h' reported in Table 13. The obtained values are then summed to obtain H(U+Th) and H(U+Th+K). Their uncertainties are propagated from the masses' standard deviations, considering U, Th and K as fully correlated. The uncertainty on the coefficients h' is here neglected. The last row reports the HPEs abundances, masses and radiogenic heat (together with their standard deviations) adopted in this study to represent medium-H models. The abundances central values are calculated by averaging all the values proposed by the different models considered. The errors on the abundances are calculated by considering the average relative uncertainty of the different models and multiplying it by the central value. The adopted abundances are then employed to obtain the HPEs masses and predicted radiogenic heat as explained above.

| | | | $a_{BSE}(U)$ [ng/g] | $a_{BSE}(Th)$ [ng/g] | $a_{BSE}(K)$ [µg/g] | $M_{BSE}(U)$ [$10^{16}$ kg] | $M_{BSE}(Th)$ [$10^{16}$ kg] | $M_{BSE}(K)$ [$10^{19}$ kg] | $M_{BSE}(Th)/M_{BSE}(U)$ | $M_{BSE}(K)/M_{BSE}(U)$ [$10^4$] | H(U+Th) [TW] | H(U+Th+K) [TW] |
|---|---|---|---|---|---|---|---|---|---|---|---|---|
| Medium - H | McDonough and Sun, 1995^ | MS95 | 20 ± 4 | 80 ± 12 | 240 ± 48 | 8.2 ± 1.6 | 32.1 ± 4.8 | 96.8 ± 19.4 | 3.9 | 1.2 | 16.5 ± 2.9 | 19.8 ± 3.5 |
| | Lyubetskaya and Korenaga, 2007 | LK07 | 17 ± 3 | 63 ± 11 | 190 ± 40 | 7.0 ± 1.2 | 25.3 ± 4.3 | 76.7 ± 16.1 | 3.6 | 1.1 | 13.5 ± 2.3 | 16.2 ± 2.9 |
| | Palme and O'Neill, 2007 | PO07 | 22 ± 3 | 83 ± 13 | 260 ± 39 | 8.8 ± 1.3 | 33.7 ± 5.0 | 104.9 ± 15.7 | 3.8 | 1.2 | 17.5 ± 2.6 | 21.1 ± 3.2 |
| | Arevalo, 2010° | A10 | 20 ± 4 | 80 ± 13 | 280 ± 60 | 8.1 ± 1.6 | 32.3 ± 5.2 | 113.0 ± 24.2 | 4.0 | 1.4 | 16.4 ± 3.0 | 20.3 ± 3.8 |
| | Wang et al., 2018 | W18 | 20 ± 2 | 75 ± 7 | 237 ± 25 | 8.0 ± 0.8 | 30.1 ± 2.7 | 95.6 ± 10.1 | 3.8 | 1.2 | 15.8 ± 1.5 | 19.1 ± 1.9 |
| | Palme and O'Neill, 2014 | PO14 | 23 ± 3 | 85 ± 13 | 260 ± 39 | 9.2 ± 1.4 | 34.3 ± 5.1 | 104.9 ± 15.7 | 3.7 | 1.1 | 18.1 ± 2.7 | 21.7 ± 3.3 |
| | **Adopted values** | | **20 ± 3** | **78 ± 11** | **245 ± 42** | **8.2 ± 1.3** | **31.3 ± 4.6** | **98.7 ± 16.9** | **3.8** | **1.2** | **16.3 ± 2.5** | **19.7 ± 3.1** |

° derived from [65] following the updated a(K)/a(U) from [125]
^ uncertainties on a(U) and a(Th) (20% and 15% respectively) should be updated to 10% following the suggestion of [163]

### 7.4 Rich-H BSE models

Geodynamical models try to estimate the abundances of the HPEs on the basis of the energetic constraints dictated by past and active geological processes, mantle dynamics and surface heat flow [164]. They seek to solve the balance of mantle forces between thermal/momentum diffusivity versus viscosity and buoyancy examining the time evolution of the secular cooling and radiogenic contributions. These parameterized thermal evolution models require a significant fraction of the present-day mantle energy source to be contributed by radiogenic heating to prevent extremely high temperatures in Earth's early history [165]. Most of these models predict $U_R$ of 0.6-0.8, thus requiring high HPEs abundances to justify the high energy demand. These models do not treat or explain Earth's isotopic anomalies and Earth's elemental ratios and thus implicitly or explicitly require layered mantle convection to explain Earth's Mg/Si ratio and UM composition [164,166]. These models predict H to be in the range [29.8, 37.2] TW, but trade-offs in assigned values of thermal conductivity, core-mantle heat exchange or viscosity can result in alternative solutions ranging from poor-H to rich-H compositional models.

Fully radiogenic models assume that the terrestrial heat flow is fully accounted by radiogenic production [10,22]. This can be obtained by keeping the BSE abundance ratios fixed at chondritic values and scaling the HPEs bulk abundances to match the expected radiogenic production of 47 TW (Section 6.1). These models represent maximal scenarios and do not account for any chemical or physical evidence of our planet.

**Table 20** – Uranium, thorium (in ng/g) and potassium (in µg/g) abundances, masses, and predicted radiogenic heat (H) for different rich-H BSE compositional models. The masses of U, Th and K are obtained multiplying the HPEs abundances by the BSE mass (Table 11). The standard deviations on the masses are obtained by propagating the errors on the abundances, here considered as the only source of uncertainty. The radiogenic heat H is calculated multiplying the masses of U, Th and K by the specific heat production coefficients h' reported in Table 13. The obtained values are then summed to obtain H(U+Th) and H(U+Th+K). Their uncertainties are propagated from the masses' standard deviations, considering U, Th and K as fully correlated. The uncertainty on the coefficients h' is here neglected. The last row reports the HPEs abundances, masses and radiogenic heat (together with their standard deviations) adopted in this study to represent rich-H models. The abundances central values are calculated by averaging all the values proposed by the different models considered. The errors on the abundances are calculated by considering the average relative uncertainty of the different models and multiplying it by the central value. The adopted abundances are then employed to obtain the HPEs masses and predicted radiogenic heat as explained above.

| | | | $a_{BSE}$(U) [ng/g] | $a_{BSE}$(Th) [ng/g] | $a_{BSE}$(K) [µg/g] | $M_{BSE}$(U) [$10^{16}$ kg] | $M_{BSE}$(Th) [$10^{16}$ kg] | $M_{BSE}$(K) [$10^{19}$ kg] | $M_{BSE}$(Th)/$M_{BSE}$(U) | $M_{BSE}$(K)/$M_{BSE}$(U) [$10^4$] | H(U+Th) [TW] | H(U+Th+K) [TW] |
|---|---|---|---|---|---|---|---|---|---|---|---|---|
| Rich - H | Turcotte, 2002[†] | T02 | 35 ± 4 | 140 ± 14 | 350 ± 35 | 14.1 ± 1.6 | 56.5 ± 5.6 | 141.2 ± 14.1 | 4.0 | 1.0 | 28.8 ± 3.1 | 33.6 ± 3.6 |
| | Turcotte, 2014 | T14 | 31 | 124 | 310 | 12.5 | 50.0 | 125.1 | 4.0 | 1.0 | 25.5 | 29.8 |
| | **Adopted values** | | **33 ± 4** | **132 ± 13** | **330 ± 33** | **13.3 ± 1.5** | **53.3 ± 5.3** | **133.2 ± 13.3** | **4.0** | **1.0** | **27.1 ± 2.9** | **31.7 ± 3.4** |

[†]as reported in [138]

## 8 Review of lithospheric models and their uncertainties

From 25% to 85% of the geoneutrino signal detected at surface comes from the closest Earth reservoir, the lithosphere [25,167].

The use of geoneutrinos as probes for Earth's interior (i.e. the mantle) requires a deep knowledge of the lithospheric signal, so as to remove its contribution and to isolate the signal coming from the most interior reservoirs. Since the differentiation of the different signal contributions has proven difficult from the experimental point of view (Section 11.5), the current evaluation of the lithospheric signal has to rely on geophysical and geochemical 3D models of the outer parts of our planet.

The main geophysical reservoirs composing the LS (Figure 23) are the crust, subdivided in CC and OCC and covered by a sedimentary layer (SED), and the CLM (Section 5). In Table 21 are reported masses, mass ratios and radiogenic heat of U, Th e K in the LS estimated according to the model reported in [25] (H13) and in [130]; the latter includes three different geophysical models, i.e. CRUST 2.0 (W20 – C2), CRUST 1.0 (W20 – C1) and LITHO 1.0 (W20 – L1).

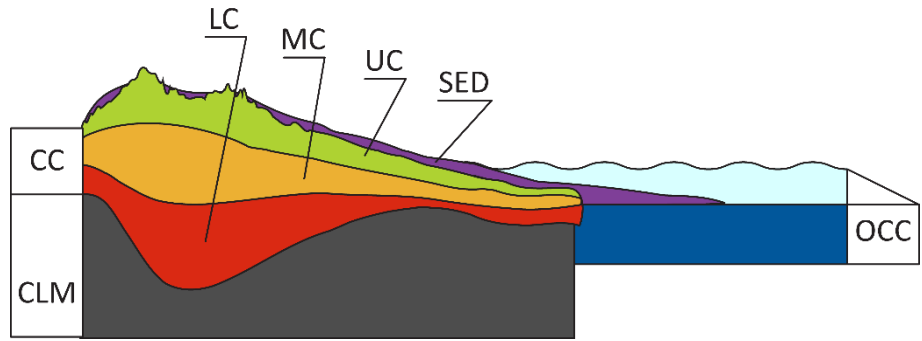

**Figure 23** - Schematic drawing of the structure of the lithosphere (not to scale). The continental crust (CC) is covered by the sedimentary layer (SED) and subdivided in the seismically defined upper crust (UC), middle crust (MC) and lower crust (LC). Under the CC we can distinguish the continental lithospheric mantle (CLM), not present in correspondence of the oceanic crust (OCC).

**Table 21** – HPEs' masses (M), mass ratios, radiogenic heat of the lithosphere ($H_{LS}$) considering the global lithospheric models: H13 based on [25], W20-C2, W20-C1 and W20-L1 based on [130]. The H13 model from [25] has been adopted on the basis of the rationale described in Section 9.

|  | $M_{LS}$(U) [$10^{16}$ kg] | $M_{LS}$(Th) [$10^{16}$ kg] | $M_{LS}$(K) [$10^{19}$ kg] | $M_{LS}$(Th)/ $M_{LS}$(U) | $M_{LS}$(K)/ $M_{LS}$(U) [$10^4$] | $H_{LS}$(U+Th) [TW] | $H_{LS}$(U+Th+K) [TW] |
|---|---|---|---|---|---|---|---|
| H13[a] [25] | $3.3^{+0.8}_{-0.7}$ | $14.3^{+4.8}_{-2.8}$ | $36.9^{+8.4}_{-6.0}$ | 4.3 | 1.2 | $6.9^{+1.6}_{-1.2}$ | $8.1^{+1.9}_{-1.4}$ |
| W20 – C2[b] [130] | $3.3^{+1.0}_{-0.7}$ | $14.5^{+5.3}_{-3.6}$ | $36.3^{+11.0}_{-8.0}$ | 4.4 | 1.1 | $6.9^{+2.6}_{-1.9}$ | $8.1^{+2.3}_{-1.8}$ |
| W20 – C1[b] [130] | $3.1^{+1.0}_{-0.7}$ | $14.1^{+5.5}_{-3.6}$ | $35.8^{+11.4}_{-7.1}$ | 4.5 | 1.2 | $6.5^{+2.5}_{-1.8}$ | $7.8^{+2.5}_{-1.8}$ |
| W20 – L1[b] [130] | $3.2^{+0.9}_{-0.7}$ | $14.2^{+4.6}_{-3.4}$ | $36.6^{+10.2}_{-7.9}$ | 4.4 | 1.1 | $6.9^{+2.3}_{-1.8}$ | $8.2^{+2.6}_{-1.9}$ |
| **Adopted** | **$3.3^{+0.8}_{-0.7}$** | **$14.3^{+4.8}_{-2.8}$** | **$36.9^{+8.4}_{-6.0}$** | **4.3** | **1.2** | **$6.9^{+1.6}_{-1.2}$** | **$8.1^{+1.9}_{-1.4}$** |

[a] As appeared in [10].
[b] Calculated based on data of Bulk CC, LM, OC (Sed and C) reported in Supplementary materials of the reference.

### 8.1.1 The Bulk Crust (BC)

The crust can be categorized in a dense and thinner (5–20 km thick) oceanic crust (OCC) and a lighter and thicker (20–70 km thick) continental crust (CC) (Figure 23). While the latter is generally acid and in turn characterized by higher HPE abundances ($a$(U)~1 ppm, $a$(Th)~10 ppm and $a$(K)~1%), the OCC is basic and depleted in HPEs ($a$(U)~0.1 ppm, $a$(Th)~0.1 ppm and $a$(K)~0.1%) [25]. Present antineutrino experiments such as KamLAND and Borexino (and future planned experiments, Section 10) sit on top of the CC, where the crustal contribution to the geoneutrino signal reaches ~65% and ~70%, respectively [25]. Along the years, the proposal of an "oceanic detector" (Section 11.4) placed on top of the OCC has been repeatedly formulated, but not realized until now due to the technical difficulties foreseen for its construction. Such a detector would permit to minimize the impact of the crustal contribution, which would represent only ~20% of the detected signal, with the remaining ~80% being representative of the mantle [25].

The CC is not homogeneous, it exhibits a fine structure marked by the seismic Conrad discontinuity at 15-20 km deep [168]. The structure of the CC is seismically defined to consist of upper, middle, and lower crustal layers (UC, MC and LC, respectively) (Figure 23), which show increasing densities and seismic velocities ($\rho_{UC} = 2.72 \pm 0.05 \frac{g}{cm^3}$, $\rho_{MC} = 2.81 \pm 0.05 \frac{g}{cm^3}$, $\rho_{LC} = 2.95 \pm 0.06 \frac{g}{cm^3}$ and $V_{UC} = 6.0 \pm 0.2 \frac{km}{s}$, $V_{MC} = 6.4 \pm 0.2 \frac{km}{s}$, $V_{LC} = 6.9 \pm 0.2 \frac{km}{s}$) [169]. This gradual shift in the geophysical properties is accompanied by a smooth change in composition. Given the observed anticorrelation between seismic velocities and the $SiO_2$ content of rocks [25], deeper crust layers exhibit mafic/basic characteristics, while shallower layers show greater felsic/acidic components. As a result, the observed abundance of HPEs in the crust decreases with depth, with UC, MC and LC typically showing ~3 ppm, ~1 ppm and ~0.2 ppm of U and ~10 ppm, ~5 ppm and ~1 ppm of Th, respectively [25].

Along the past few decades, several geophysical models of increasing complexity and spatial resolution were proposed [170]. The first global geophysical model, 3SMAC [171], appeared in 1996. It included a tomographic model reporting the thickness and physical properties of all ice layers, sediment accumulations and oceanic and continental crust from Earth's surface to the upper mantle. In 1998, the 3SMAC model has been replaced by CRUST 5.1 [172], a model having a 5° x 5° resolution grid reporting the thickness and physical properties of lithospheric layers and the depth profile of seismic velocities. This model was further refined along the years, integrating an increasing amount of geophysical and geochemical data including new reflection and refraction seismic data, until the release of more resolute updates called CRUST 2.0 [173], having a 2° x 2° spatial resolution, and the 1° x 1° resolution Crust 1.0 [169]. Finally, the geophysical information included in CRUST 1.0 was used as starting model and perturbed to fit high-resolution surface wave dispersion to obtain LITHO 1.0 [174], a tessellated model of the crust and uppermost mantle of the Earth which interpolates the data over a ~1° triangular tin providing a continuous database of lithospheric properties on the Earth's surface.

Combining the geophysical information provided by the abovementioned global models and the chemical inputs obtained from geochemistry, authors produced in turn a variety of different crustal compositional models for predicting the expected geoneutrino signal at surface (Table 22).

**Table 22** – Abundances of uranium and thorium (in µg/g) for the different reservoirs of the continental (CC) and oceanic crust (OCC) according to different authors.

| | a(U) [µg/g] | | | | | | a(Th) [µg/g] | | | | | |
|---|---|---|---|---|---|---|---|---|---|---|---|---|
| | CC | | | | | OCC | CC | | | | | OCC |
| Reservoir | Sed | UC | MC | LC | Sed | OCC | Sed | UC | MC | LC | Sed | OCC |
| Mantovani et al., 2004 [22] | 1.68 | 2.5 | 1.6 | 0.62 | 1.68 | 0.1 | 6.9 | 9.8 | 6.1 | 3.7 | 6.9 | 0.22 |
| Enomoto et al., 2007 [3] | 2.8 | 2.8 | 1.6 | 0.2 | 1.7 | 0.10 | 10.7 | 10.7 | 6.1 | 1.2 | 6.9 | 0.22 |
| Dye, 2010 [24] | 2.70 ± 0.57 | 2.70 ± 0.57 | 1.30 ± 0.40 | 0.20 ± 0.16 | 1.70 ± 0.34 | 0.10 ± 0.02 | 10.5 ± 1.1 | 10.5 ± 1.1 | 6.50 ± 0.52 | 1.20 ± 0.96 | 6.90 ± 0.69 | 0.22 ± 0.02 |
| H13° [25] | 1.73 ± 0.09 | 2.7 ± 0.6 | $0.97^{+0.58}_{-0.36}$ | $0.16^{+0.14}_{-0.07}$ | 1.73 ± 0.09 | 0.07 ± 0.02 | 8.10 ± 0.59 | 10.5 ± 1.0 | $4.86^{+4.30}_{-2.25}$ | $0.96^{+0.18}_{-0.51}$ | 8.10 ± 0.59 | 0.21 ± 0.06 |
| W20 – L1 [130] | 2.7 ± 0.4 | 2.7 ± 0.6 | $0.78^{+0.76}_{-0.38}$ | $0.16^{+0.23}_{-0.09}$ | 1.73 ± 0.07 | 0.07 ± 0.02 | 10.49 ± 0.64 | 10.5 ± 1.0 | $3.55^{+5.22}_{-2.11}$ | $0.16^{+0.23}_{-0.09}$ | 8.10 ± 0.43 | 0.21 ± 0.05 |
| W20 – C1 [130] | 2.7 ± 0.4 | 2.7 ± 0.6 | $0.87^{+0.84}_{-0.43}$ | $0.18^{+0.26}_{-0.11}$ | 1.73 ± 0.07 | 0.07 ± 0.02 | 10.49 ± 0.65 | 10.5 ± 1.0 | $4.22^{+6.21}_{-2.51}$ | $1.10^{+2.75}_{-0.78}$ | 8.09 ± 0.44 | 0.21 ± 0.05 |
| W20 – C2 [130] | 2.7 ± 0.4 | 2.7 ± 0.6 | $0.76^{+0.78}_{-0.38}$ | $0.15^{+0.21}_{-0.09}$ | 1.73 ± 0.07 | 0.07 ± 0.01 | 10.49 ± 0.71 | 10.41 ± 0.99 | $3.31^{+5.50}_{-2.07}$ | $0.80^{+1.87}_{-0.56}$ | 8.10 ± 0.44 | 0.21 ± 0.04 |

By employing the abundances reported in Table 22 with difference geophysical models, authors produced a variety of different crustal models predicting different HPEs masses and in turn different radiogenic heat production for the bulk crust ($H_{BC}$) (Table 23). Studies focused on the impact of geophysics variability showed that geoneutrino signal estimates obtained through the use of different geophysical models (Litho1.0, Crust 1.0 and Crust 2.0) yield similar results [130]. Hence, the main factor affecting the variability in the predicted signal at surface proves to be the HPEs abundances employed for the different crustal layers. The crustal signal rates at KamLAND and Borexino experimental sites calculated according to the different crustal models of Table 23 are reported in Table 24. Other authors produced crustal models reporting the expected geoneutrino flux for different experimental sites [26], while others produced interactive websites and maps predicting the geoneutrino flux at surface [175,176].

**Table 23** – Masses of uranium, thorium (in $10^{16}$ kg) and potassium (in $10^{19}$ kg) together with the associated $M_{BC}(Th)/M_{BC}(U)$, $M_{BC}(K)/M_{BC}(U)$ ratios and the predicted radiogenic heat production coming from U+Th and U+Th+K (in TW) for the Bulk Crust (BC) reported by different authors. The recalculation of the radiogenic heats based on the updated h' coefficients (Table 13) is estimated to be less than 5%. The H13 model from [25] has been adopted on the basis of the rationale described in Section 9.

| Model | $M_{BC}(U)$ [$10^{16}$ kg] | $M_{BC}(Th)$ [$10^{16}$ kg] | $M_{BC}(K)$ [$10^{19}$ kg] | $M_{BC}(Th)/M_{BC}(U)$ | $M_{BC}(K)/M_{BC}(U)$ [$10^4$] | $H_{BC}(U+Th)$ [TW] | $H_{BC}(U+Th+K)$ [TW] |
|---|---|---|---|---|---|---|---|
| Mantovani et al., 2004 [22] | 3.53 ± 0.35 | 14.5 ± 1.5 | 36.7 ± 3.9 | 4.1 | 1.0 | 7.3 ± 0.7 | 8.6 ± 0.9 |
| Enomoto et al., 2007 [3] | 3.4 | 12.9 | / | 3.8 | / | 7.0 | / |
| Dye, 2010 [24] | 3.1 ± 0.8 | 13.9 ± 1.8 | 36.1 ± 6.1[+] | 4.3 | 1.1 | / | 7.8 ± 1.5[*] |
| H13 [25] | $2.8^{+0.6}_{-0.5}$ | $12.3^{+3.2}_{-2.1}$ | $33.0^{+6.5}_{-4.9}$ | 4.4 | 1.2 | $5.9^{+1.4}_{-1.2}$ | $7.0^{+1.4}_{-1.1}$ |
| W20 – C2 [130] | $2.9^{+0.8}_{-0.6}$ | $12.5^{+0.8}_{-0.6}$ | $33.9^{+9.6}_{-7.5}$ | 4.3 | 1.1 | $5.9^{+2.0}_{-1.6}$ | $7.2^{+2.0}_{-1.6}$ |
| W20 – C1 [130] | $2.7^{+0.7}_{-0.6}$ | $12.3^{+4.1}_{-3.0}$ | $32.2^{+9.2}_{-7.1}$ | 4.5 | 1.2 | $5.9^{+2.2}_{-1.7}$ | $7.0^{+2.1}_{-1.6}$ |
| W20 – L1 [130] | $2.9^{+0.8}_{-0.6}$ | $12.8^{+3.1}_{-1.4}$ | $32.4^{+8.9}_{-7.0}$ | 4.4 | 1.2 | $6.2^{+2.1}_{-1.6}$ | $7.4^{+2.1}_{-1.6}$ |
| **Adopted** | $\mathbf{2.8^{+0.6}_{-0.5}}$ | $\mathbf{12.3^{+3.2}_{-2.1}}$ | $\mathbf{33.0^{+6.5}_{-4.9}}$ | **4.4** | **1.2** | $\mathbf{5.9^{+1.4}_{-1.2}}$ | $\mathbf{7.0^{+1.4}_{-1.1}}$ |

[*]As appeared in [10].
[+]Calculated considering a $^{40}$K isotopic abundance from Table 1.

**Table 24** – Geoneutrino signals (in TNU) of the bulk crust expected at Borexino and KamLAND experimental sites according to different authors. In order to obtain comparable signal estimates, models based on out-of-date oscillation parameters [3,22,24] were scaled according to the updated $<P_{ee}>$ = 0.55.

| Model | Borexino $S_{BC}(U+Th)$ [TNU] | KamLAND $S_{BC}(U+Th)$ [TNU] |
|---|---|---|
| Mantovani et al., 2004 [22][*] | 29.6 ± 2.9 | 24.7 ± 2.4 |
| Enomoto et al., 2007 [3][o] | / | 24.2 ± 2.2 |
| Dye, 2010 [24][o] | 29.5 ± 7.1 | 23.5 ± 5.6 |
| H13 [25] | $29.0^{+6.0}_{-5.0}$ | $20.6^{+4.0}_{-3.5}$ |
| W20 – C2 [130] | $31.6^{+8.1}_{-6.4}$ | $23.5^{+5.9}_{-4.7}$ |
| W20 – C1 [130] | $31.1^{+8.0}_{-6.4}$ | $24.6^{+6.6}_{-5.2}$ |
| W20 – L1 [130] | $33.0^{+7.9}_{-6.4}$ | $27.0^{+7.1}_{-5.6}$ |

[*]Obtained applying Eq. (7) to the reported fluxes of U and Th with $<P_{ee}>$ = 0.55, summed as fully correlated.
[o]Scaled according to the approach illustrated in Appendix A.3.

### 8.1.2 Continental Lithospheric Mantle (CLM)

The CLM is a portion of the mantle underlying the CC included between the MOHO and the LAB (Section 5). Previous models of geoneutrino flux [3,22-24] relied on the density profile of the mantle as given by the Preliminary Reference Earth Model (PREM) [61]. In these models, the crust and the mantle were treated as two separate geophysical and geochemical reservoirs. In particular, the mantle was conventionally described as a shell between the crust and the core and considered compositionally homogeneous [3,24]. These models did not consider the heterogeneous topography of the base of the crust, or the likely differences in composition of the lithospheric mantle underlying the oceanic and continental crusts. In more recent models [25,26,130], the CLM beneath the continents is instead treated as a distinct geophysical and geochemical reservoir that is coupled to the crust in the reference Earth model, forming the LS. The assumption is that the lithospheric mantle beneath the oceans is compositionally identical to the DM and is therefore usually incorporated in this reservoir during modeling: accurately describing its thickness is not crucial in the estimation o the geoneutrino signal. On the other hand, the lithospheric mantle under the CC (i.e. CLM) is compositionally different from both the CC and the DM, with its top starting from the MOHO surface and the bottom being difficult to constrain.

It is only recently that authors started to include the CLM in the geochemical and geophysical modelling of the LS. Despite the adoption of different geophysical models, both the models proposed by H13 and W20 employed the same geochemical abundances $a_{CLM}(U) = 0.03^{+0.05}_{-0.02}$ µg/g , $a_{CLM}(Th) = 0.15^{+0.28}_{-0.10}$ µg/g and $a_{CLM}(K) = 0.03^{+0.04}_{-0.02}$ %, which in turn produced consistent predicted masses and radiogenic heat for the LS (Table 25).

By employing the reported HPE masses, authors estimated the expected geoneutrino signals $S_{CLM}(U + Th)$ at KamLAND and Borexino experimental sites, listed in Table 26 together with the values adopted for the mantle signal extraction from experimental results. Coherently with the strategy followed for the estimation of the signal produced by the BC (Section 8.2), the adopted CLM signal is taken from [25] both for KamLAND and Borexino experiments.

**Table 25** - Masses of uranium, thorium (in $10^{16}$ kg) and potassium (in $10^{19}$ kg) together with the associated $M_{CLM}(Th)/M_{CLM}(U)$, $M_{CLM}(K)/M_{CLM}(U)$ ratios and the predicted radiogenic heat production coming from U+Th and U+Th+K (in TW) for the CLM. The H13 model from [25] has been adopted on the basis of the rationale described in Section 9.

| Model | $M_{CLM}(U)$ [$10^{16}$ kg] | $M_{CLM}(Th)$ [$10^{16}$ kg] | $M_{CLM}(K)$ [$10^{19}$ kg] | $M_{CLM}(Th)/M_{CLM}(U)$ | $M_{CLM}(K)/M_{CLM}(U)$ [$10^4$] | $H_{CLM}(U+Th)$ [TW] | $H_{CLM}(U+Th+K)$ [TW] |
|---|---|---|---|---|---|---|---|
| H13 [25] | $0.29^{+0.54}_{-0.20}$ | $1.45^{+2.94}_{-0.94}$ | $3.1^{+4.7}_{-1.8}$ | 5.0 | 1.1 | $0.7^{+1.1}_{-0.5}$ | $0.8^{+1.1}_{-0.6}$ |
| W20 – C2 [130] | $0.24^{+0.61}_{-0.17}$ | $1.06^{+3.06}_{-0.79}$ | $2.3^{+5.6}_{-1.6}$ | 4.4 | 1.0 | $0.4^{+1.1}_{-0.3}$ | $0.5^{+0.8}_{-0.3}$ |
| W20 – C1 [130] | $0.20^{+0.60}_{-0.15}$ | $0.91^{+2.97}_{-0.70}$ | $1.9^{+5.5}_{-1.4}$ | 4.6 | 1.0 | $0.4^{+1.1}_{-0.3}$ | $0.5^{+1.6}_{-0.4}$ |
| W20 – L1 [130] | $0.19^{+0.28}_{-0.12}$ | $0.88^{+1.54}_{-0.56}$ | $1.9^{+2.5}_{-1.1}$ | 4.6 | 1.0 | $0.5^{+1.4}_{-0.4}$ | $0.6^{+1.6}_{-0.4}$ |
| **Adopted** | $\mathbf{0.29^{+0.54}_{-0.20}}$ | $\mathbf{1.45^{+2.94}_{-0.94}}$ | $\mathbf{3.1^{+4.7}_{-1.8}}$ | $\mathbf{5.0}$ | $\mathbf{1.1}$ | $\mathbf{0.7^{+1.1}_{-0.5}}$ | $\mathbf{0.8^{+1.1}_{-0.6}}$ |

**Table 26** – Continental lithospheric mantle geoneutrino signals (in TNU) expected at Borexino and KamLAND experimental sites according to different authors together with the values adopted in this study (taken from H13). The H13 model from [25] has been adopted on the basis of the rationale described in Section 9.

| Model | Borexino $S_{CLM}(U+Th)$ [TNU] | KamLAND $S_{CLM}(U+Th)$ [TNU] |
|---|---|---|
| H13 [25] | $2.2^{+3.1}_{-1.3}$ | $1.6^{+2.2}_{-1.0}$ |
| W20 – C2 [130] | $1.8^{+3.0}_{-1.1}$ | $1.3^{+2.2}_{-0.8}$ |
| W20 – C1 [130] | $1.7^{+2.9}_{-1.1}$ | $1.4^{+2.4}_{-0.9}$ |
| W20 – L1 [130] | $0.8^{+1.2}_{-0.5}$ | $0.8^{+1.2}_{-0.5}$ |
| **Adopted** | $\mathbf{2.2^{+3.1}_{-1.3}}$ | $\mathbf{1.6^{+2.2}_{-1.0}}$ |

## 8.2 Geoneutrinos from the region near the detectors

In geoneutrino science, multiple sites geoneutrino studies appear as the most reliable tool for disentangling the unresolved riddles about the Earth's heat budget, BSE compositional paradigms and mantle convection. Given the predicted signal of the accessible lithosphere, the mantle component is assumed to be the same for diverse geoneutrino detectors and it can be indirectly inferred from a combined treatment of experimental signals [20,138] (Section 9). A correct discrimination of the mantle geoneutrino signal must be grounded on a solid prediction of the crustal component and its uncertainty. In this puzzle, the geophysical and geochemical modelling of the crust, especially of the portion near to the detector, is certainly the most compelling task and, at the same time, the thorniest charge for scientists of the geoneutrino community.

Borexino and KamLAND - but also the forthcoming SNO+ and JUNO (Section 11) - measure a geoneutrino signal which mantle contribution is about one quarter [25,26].
The site-dependent crustal component represents the dominant contribution due to the concomitance of diverse factors affecting the geoneutrino production, propagation, and detection.

The ~30 km thick CC hosting and standing above the underground detectors is characterized by U and Th abundances which are globally at less one order of magnitude higher respect to OCC or CLM (Table 22).

The distance dependency of the antineutrino survival probability makes the geoneutrino signal from the Near-Field Crust (NFC) even more influential for the indirect study of the mantle radioactivity. While the average survival probability (<$P_{ee}$> = 0.55) could be a reasonable approximation for describing the oscillation during the propagation at long distances (~10000 km), a particular attention must be paid if dealing with small distances (~100 km) (Figure 3). The result is an amplification/reduction effect of the geological peculiarities of the NFC which can be translated in a not negligible (1 - 2 TNU) difference between the signals calculated with an average or a precise oscillation probability [177].

Additionally, the isotropic $1/4\pi r^2$ spherical scaling factor (Section 2) ensures that the geoneutrino flux reaching a detector is dominated by the natural radioactivity surrounding the detector. This contribution can be, in a first attempt, estimated exploiting the global models of the LS illustrated in Section 8 (Table 27). H13, W20 – C2, W20 – C1 and W20 – L1. The geophysical structure of the 9° × 9° area centered in the Borexino and KamLAND detectors depicted in the four global models is represented in Figure 26.

For the KamLAND region (Figure 24), all the global models predict a SED layer with a rather homogenous thickness with an average value of 0.8 km and the higher values (> 1.2 km) recorded in the eastern area of the Japan sea. For the continental area, the average SED thickness is ~ 0.4 km (Figure 24a) but a clear disagreement among the four models is observed, particularly in the region at southwest of the detector (Figure 24d, f, h), and in turn for the BC (Figure 24j), an opposite trend is observed: the global models substantially agree on the values for the continental area while this does not happen for the Japan sea where the percent relative range is always higher than 100%. In the continental area the average thickness of UC and LC is 12 km (Figure 24c, g), while the MC appears thinner (~ 10 km) (Figure 24e), resulting in a total BC thickness ranging from 28 km to 38 km (Figure 24i). In the oceanic region thickness values ranging from 3 – 6 km are observed for the three crustal layers (c, e, g), while the average value for the BC varies between 8 and 18 km (Figure 24i).

According to the global crustal models the thickness of the SED layers in the area hosting Borexino is characterized by a high variability (Figure 25a), with an average value ranging from 0.7 km in the proximity of the detector to 7 km in the Adriatic Sea (eastern area). Moreover, the four global crustal model present values that differ greatly from each other with a percentage relative range that is ~ 200% in the 3° × 3° tiles surrounding Borexino (Figure 25b). For the UC, a percentage relative range higher (~ 100%) is observed only in the Tyrrhenian sea area (Figure 25d) while the global models substantially agree on a homogeneous average value of 4 km in the rest of the region (Figure 25c). The same trend is observed for the MC and the LC (Figure 25f - h) which have an average thickness respectively of 5 km and 8 km in the Tyrrhenian sea area and of 12 km in the rest of region (Figure 25e - g). The BC thickness is characterized by an increasing trend from southwest to northeast with average values ranging from 15 to 48 km (Figure 25i). The high discrepancy among the global models in the predicted thickness of the BC is observed in the sea region and in the area at south of the detector where the percentage relative range is ~ 50% (Figure 25j).

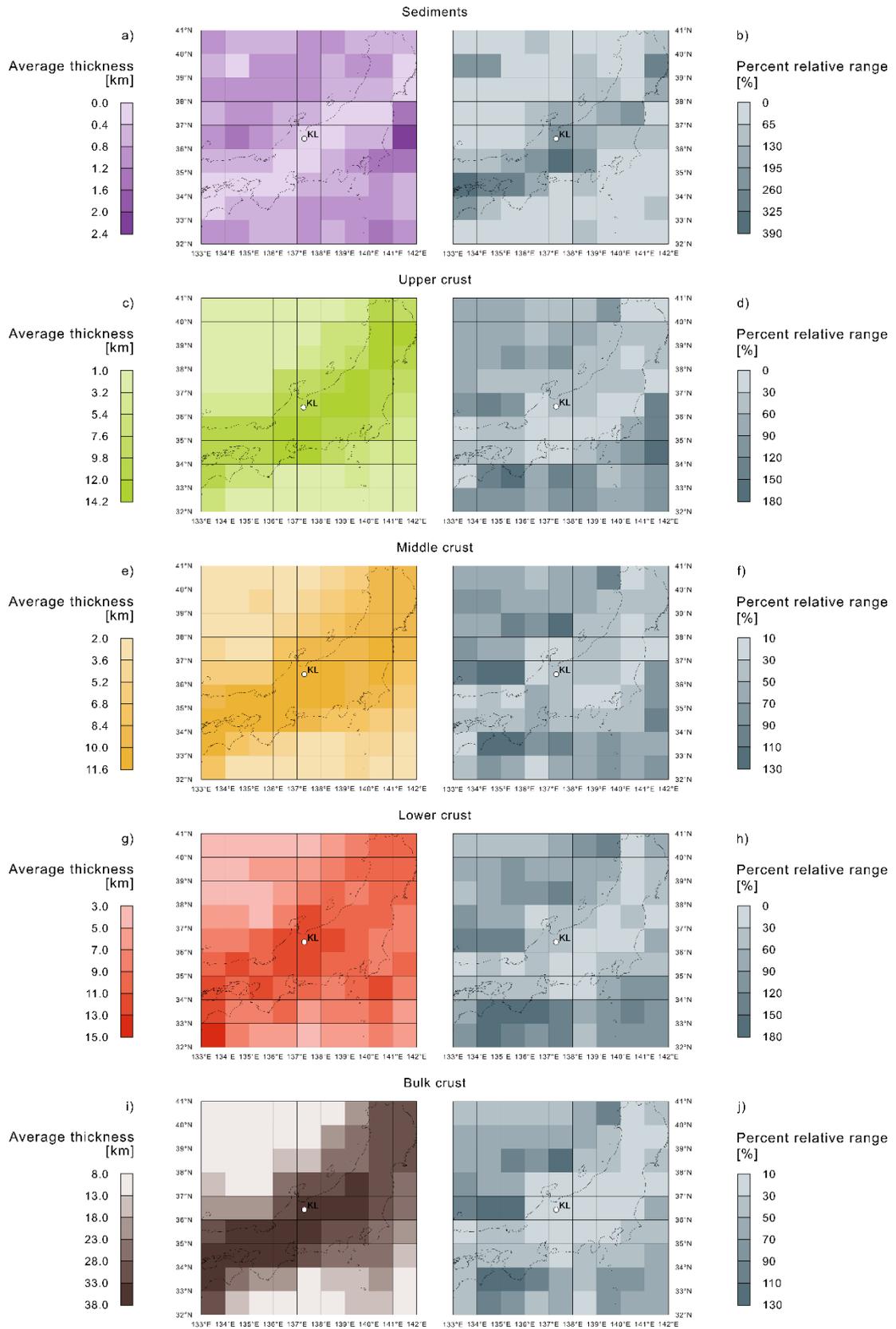

**Figure 24** – Left panels: thickness of the Sediments (SED), Upper Crust (UC), Middle Crust (MC), Lower Crust (LC) and Bulk Crust (BC) of the 9° × 9° area centered in the KamLAND detector. The value of each 1° × 1° tile is the average obtained considering the model, CRUST 2.0, CRUST 1.0, Litho 1.0 and H13 models. Right panels: percent relative range [(maximum value – minimum value)/average value *100] of the models obtained for each tile for SED, UC, MC, LC and BC.

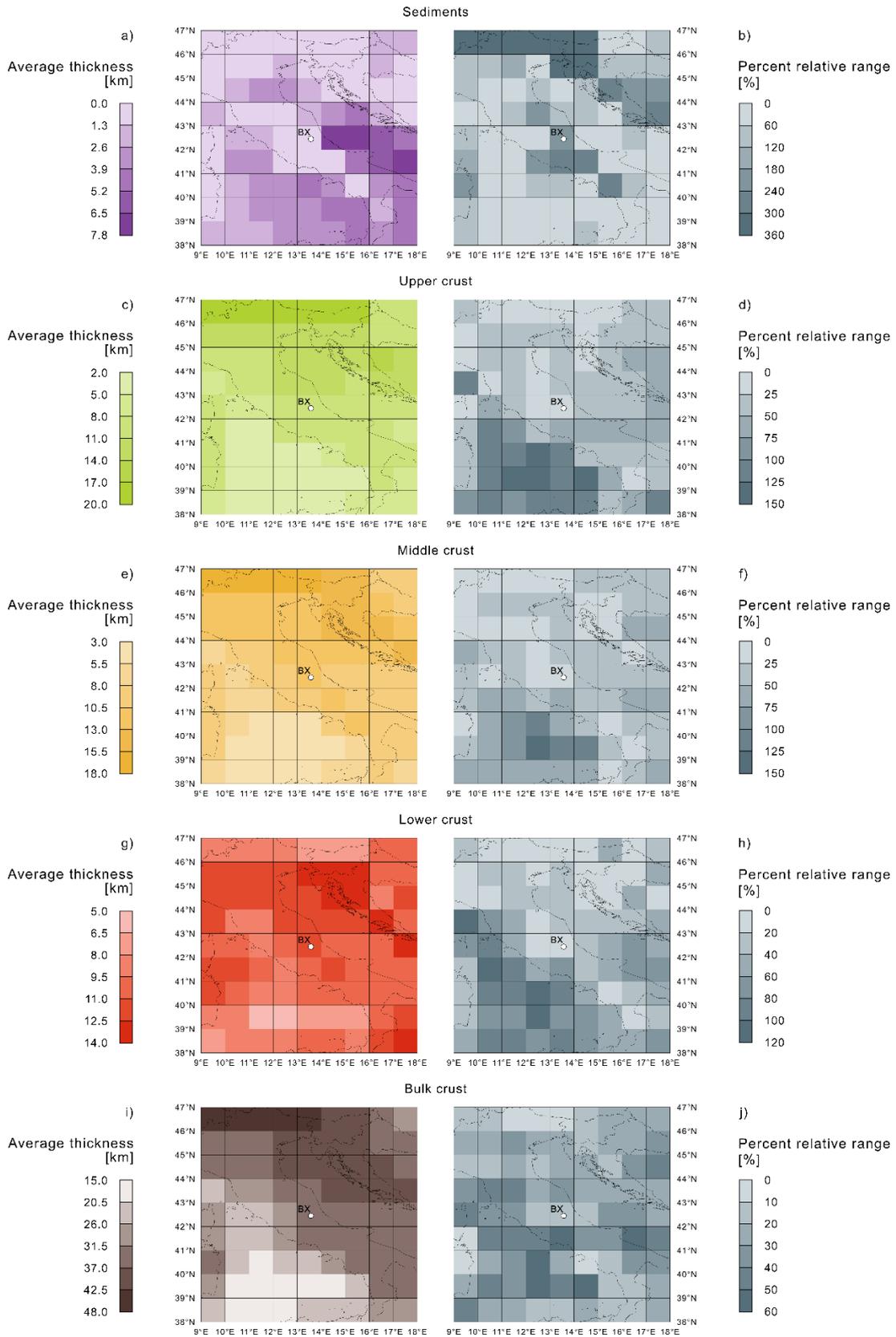

**Figure 25 -** Left panels: thickness of the Sediments (SED), Upper Crust (UC), Middle Crust (MC), Lower Crust (LC) and Bulk Crust (BC) of the 9° × 9° area centered in the Borexino detector. The value of each 1° × 1° tile is the average obtained considering the model, CRUST 2.0, CRUST 1.0, Litho 1.0 and H13 models. Right panels: percent relative range [(maximum value – minimum value)/average value *100] of the models obtained for each tile for SED, UC, MC, LC and BC.

The geochemical modeling performed by [25] and [130] have some common ground: the SED and the UC are considered compositionally homogenous while the uranium and thorium abundances in MC and LC in each voxel depend on seismic arguments (Section 8.1.1). Looking into details, the abundances assigned to UC, SED and crust of OCC are identical, but the continental sediments are treated with two different approaches (Table 22). [130] assume that, due to the weathering effect, the sediments have the same U and Th abundances of UC, higher than those assigned by [25] on the basis of the Global Subducting Sediments II model [178]. Due to the proximity of this layer to the detectors, this divergence is the main origin of the increasing trend of the geoneutrino signals reported in [130] (Table 27).

Despite these differences, the mentioned global models obtained the similar ratio between the geoneutrino signal from the NFC, i.e. the 24 voxels close to the detectors (Figure 26), and the BC geoneutrino signal that is $S_{NFC}/S_{BC}$ ~ 0.67 for KamLAND[4] and $S_{NFC}/S_{BC}$ ~ 0.55 for Borexino. These results prove that such small regions accounts for the most relevant contribution to the geoneutrino signal. For this reason, local refined models based on specific geophysical and geochemical data were developed to provide a more accurate and reliable predictions. For KamLAND experiment, [179] and [3] proposed a site-specific geophysical and geochemical modeling of the crust near the detector while a 3D geophysical and geochemical model of the crust surrounding the Borexino detector was proposed by [180].

The Japan island arc, hosting the KamLAND detector, is part of a continental shelf located close to the eastern margin of the Eurasian plate. The Philippine plate and the Pacific plate are moving towards the Eurasian plate and are subducting respectively beneath the southern and the northern part of Japan. The submarine trenches are thus formed with parallel uplifted areas and intense igneous activity. The KamLAND site is sited in a typical continental crust of Island Arc and Forearc environment. The Sea of Japan, situated between the Japan island arc and the Asian continent, is classified as marginal sea and it is bordered by islands and expanded basins on the back-arc side (back arc basin). [20] and [3] adopted the same geophysical and geochemical inputs and additionally studied the effects on geoneutrino signal of the peculiarities characterizing the subducting slab and the Japan Sea crust (Appendix A.1).

The Gran Sasso range, where the Borexino experiment is located, is a massif of the Central sector of the Apennines, a peri-Mediterranean chain part of the Adria plate. The actual geological structure of the Apennine chain is the result of the geodynamical processes occurred during its orogenesis began in the early Neogene (20 million years ago). A refined reference model for the Gran Sasso area was developed by [180] in which local and specific geophysical and geochemical information were used to provide an estimate of the geoneutrino signal originated from the NFC (Figure 26)[5]. The model subdivides the study area in two zones, the central tile (CT) and the rest of the region (RR), which are described with different degree of resolution (Appendix A.2).

---

[4] Note that the NFC defined by Wipperfurth et al. 2020 is not coincident with the NFC defined by Huang et al. 2013 (Figure 26).
[5] The 3D geophysical model is available at https://www.fe.infn.it/radioactivity/Borexino/

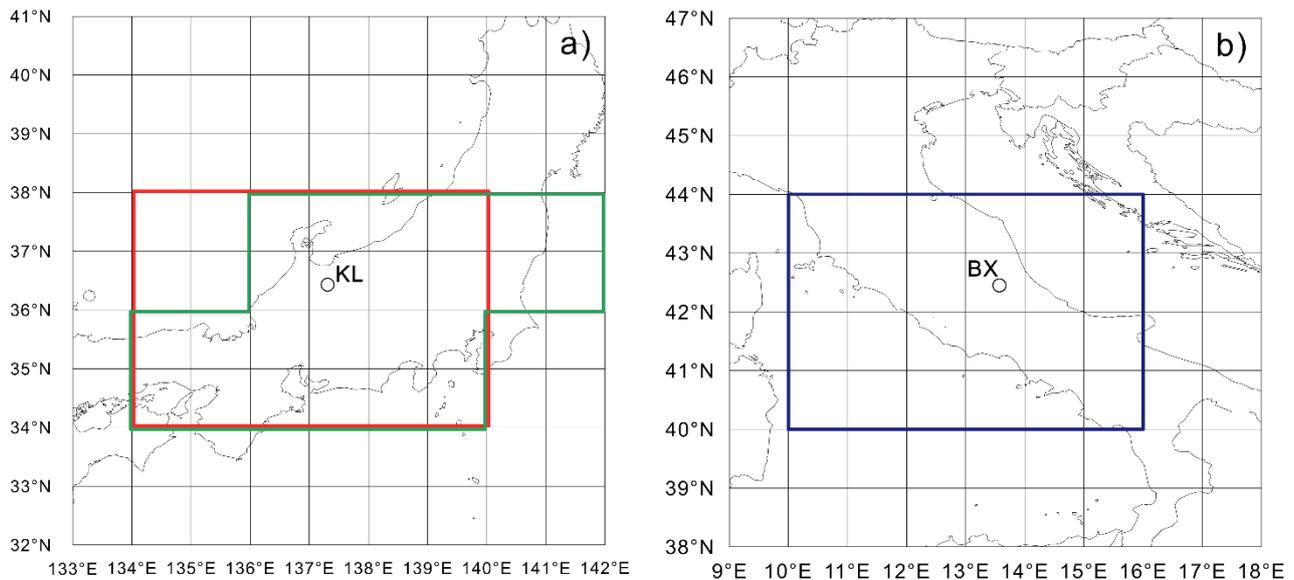

**Figure 26** – a) Near Field Crust (NFC) identified by [130] (red box) and by [25] and [20] (green box) for KamLAND; b) NFC identified for Borexino: all authors chose the same region (blue box).

The geoneutrino signals expected at KamLAND and Borexino calculated adopting the above mentioned global and refined models are reported in Table 26 for the NFC, the Far Field Crust (FFC) and the BC (BC = NFC + FFC) together with the adopted values for the purpose of mantle signal extraction from experimental results (Section 9). Since for Borexino the crustal portion of NFC, and as a consequence for the FFC, is the same for all the references reported (Figure 23b), the BC signal can be calculated integrating the refined estimate of NFC [180] with the FFC contribution given by any global model. Conversely, for KamLAND, due to the spatial differences of the NFC (Figure 23a) attention may be paid in the calculation of the BC signal: starting from the refined estimate [20] for the NFC, only the FFC signal reported in [25] can be adopted. With the aim to have a coherent mantle signal extraction, the same model is used for the FFC contribution at Borexino together with the most updated estimates for the NFC given by [10].

As already highlighted for global models, for KamLAND the NFC contribution estimated by the refined model is higher respect to the Borexino case. Note that for the BC estimates reported in [3], the NFC and FFC signals with their corresponding uncertainties could not be inferred since only percentage contribution are reported.

For Borexino all the refined models provide NFC and FFC signals estimates with central values that are approximately the same; for the NFC the uncertainties are different because in [180] the maximal and minimal excursions of various input values and uncertainties are taken as the ± 3σ error range. The discrepancy recorded in FFC signals uncertainties, for [180] and [20] has to be ascribed to the different geochemical and geophysical datasets considered.

**Table 27** – The geoneutrino signals expected at Borexino and KamLAND according to global and refined models for the Near Field Crust (NFC), the Far Field Crust (FFC) and the corresponding adopted value for the mantle signal calculation (see Section 9). For details on calculation method and uncertainties treatment see Appendix A.3. The models from [20] and [10] were adopted for describing the NFC of KamLAND and Borexino, respectively, on the basis of the rationale described in Section 9. Based on the same arguments, the global model from [25] was adopted to describe the FFC of the two experiments.

| Experiment | Type of model | Reference | $S_{NFC}$ (U+Th) [TNU] | $S_{FFC}$ (U+Th) [TNU] |
|---|---|---|---|---|
| KamLAND | Global | Huang et al., 2013 [25] | $13.4^{+2.3}_{-2.1}$ | $7.3^{+1.5}_{-1.2}$ |
| | | Wipperfurth et al., 2020 [130] – CRUST 2.0 | $15.8^{+3.9}_{-3.1}$ | $7.8^{+2.1}_{-1.7}$ |
| | | Wipperfurth et al., 2020 [130] – CRUST 1.0 | $16.7^{+4.5}_{-3.5}$ | $7.8^{+2.2}_{-1.7}$ |
| | | Wipperfurth et al., 2020 [130] – LITHO 1.0 | $18.2^{+4.7}_{-3.7}$ | $8.8^{+2.5}_{-1.9}$ |
| | Refined | Enomoto et al., 2007 [3] | $26.0 \pm 2.4$ | |
| | | Fiorentini et al., 2012 [20] | $17.7 \pm 1.4$ | $8.8 \pm 1.4$ |
| | Adopted | | $17.7 \pm 1.4$ | $7.3^{+1.5}_{-1.2}$ |
| Borexino | Global | Huang et al., 2013 [25] | $15.3^{+2.8}_{-2.3}$ | $13.7^{+2.8}_{-2.3}$ |
| | | Wipperfurth et al., 2020 [130] – CRUST 2.0 | $17.4^{+4.3}_{-3.5}$ | $14.1^{+3.8}_{-3.0}$ |
| | | Wipperfurth et al., 2020 [130] – CRUST 1.0 | $17.5^{+4.3}_{-3.5}$ | $13.6^{+3.8}_{-2.9}$ |
| | | Wipperfurth et al., 2020 [130] – LITHO 1.0 | $18.2^{+3.9}_{-3.2}$ | $14.8^{+4.0}_{-3.2}$ |
| | Refined | Coltorti et al., 2011 [180] | $9.67 \pm 3.82$ | $15.65 \pm 1.50$ |
| | | Fiorentini et al., 2012 [20] | $9.67 \pm 1.26$ | $15.67 \pm 2.43$ |
| | | Agostini et al., 2020 [10] | $9.2 \pm 1.2$ | $13.7^{+2.8}_{-2.3}$ |
| | Adopted | | $9.2 \pm 1.2$ | $13.7^{+2.8}_{-2.3}$ |

## 9 Extracting the mantle signal

The KamLAND (KL) and Borexino (BX) experiments observed at >5σ level signals of U and Th geoneutrinos coming from the whole Earth. In absence of an experimental way (e.g. directionality analysis, Section 11.5) to disentangle the contribution from the lithosphere and the mantle, the employment of geological models is required to estimate the mantle geoneutrino component.

The correct subtraction of the lithospheric component from the experimental signals of KL and BX must comply with the following constraints: (i) the global crustal model (Section 8.2) employed for the FFC needs to be unique for the two experiments for avoiding systematic biases, (ii) the local models (Section 8.2) of the NFC should be built with geochemical and/or geophysical information typical of the regions surrounding the detectors for substituting global features with local geological data, (iii) they must be geometrically complementary to the FFC area and (iv) all geoneutrino signal contributions should be separately reported.

Under these assumptions, the mantle signals $S_M^{BX}(U+Th)$ and $S_M^{KL}(U+Th)$ can be inferred by subtracting the estimated lithospheric components from the experimental total signals, $S_{Exp}^{BX}(U+Th)$ for BX and $S_{Exp}^{KL}(U+Th)$ for KL (Section 3, 4):

$$\begin{aligned} S_M^{BX}(U+Th) &= S_{Exp}^{BX}(U+Th) - S_{NFC}^{BX}(U+Th) - S_{FFC}^{BX}(U+Th) - S_{CLM}^{BX}(U+Th) \\ S_M^{KL}(U+Th) &= S_{Exp}^{KL}(U+Th) - S_{NFC}^{KL}(U+Th) - S_{FFC}^{KL}(U+Th) - S_{CLM}^{KL}(U+Th) \end{aligned} \tag{19}$$

where the lithospheric signals are modelled in three independent components: (i) the NFC ($S_{NFC}^{BX}(U+Th)$ and $S_{NFC}^{KL}(U+Th)$), (ii) the FFC ($S_{FFC}^{BX}(U+Th)$ and $S_{FFC}^{KL}(U+Th)$) (Section 8.2) and (iii) the CLM ($S_{CLM}^{BX}(U+Th)$ and $S_{CLM}^{KL}(U+Th)$) (Section 8.1.2).

Respect to the adopted KL local model from [20], the study proposed by [3] did not explicitly report the spatial extension of NFC (Figure 26). Analogously local model surrounding BX [10] was chosen as an update of [180], which reported geophysical and geochemical data from the Italian crust. Finally, the global model from [25] was adopted since it identified separately the contributions from the FFC and the CLM for the two experiments and it was the only model geometrically complementary to the adopted KL NFC (Figure 26).

It has to be noted that even if the different components are considered uncorrelated for KL and BX separately, the FFC and the CLM signals of KL and BX are fully correlated ($S_{FFC}^{KL}(U+Th) \propto S_{FFC}^{BX}(U+Th)$ and $S_{CLM}^{KL}(U+Th) \propto S_{CLM}^{BX}(U+Th)$) since they are derived from the same geophysical and geochemical model of [25].

Using only the experimental signals published by BX and KL collaborations without any spectral information, the PDFs of $S_{Exp}^{BX}(U+Th) = 47.0_{-8.1}^{+8.6}$ TNU and $S_{Exp}^{KL}(U+Th) = 32.1 \pm 5.0$ TNU are reconstructed for inferring the following mantle signals at KL and BX: $S_M^{KL}(U+Th) = 4.8_{-5.9}^{+5.6}$ TNU and $S_M^{BX}(U+Th) = 20.8_{-9.2}^{+9.4}$ TNU (Table 28).

Note that the KL mantle signal $6.0_{-5.7}^{+5.6}$ TNU preliminarily published in [7] includes also the CLM contribution, since it is obtained by subtracting the BC contribution according to [3]. Moreover, it is worth to highlight that the BX mantle signal of $21.2_{-9.0}^{+9.6}$ TNU provided in [10] derives from a comprehensive fitting procedure accounting for the spectral information.

Since both KL and BX adopt the same assumption of chondritic ratio $M_{BSE}(Th)/M_{BSE}(U) = 3.9$ [163,181] in the extraction of the geoneutrino signal (Sections 3.3 and 4.3), the above results can be properly combined in the estimation of a joint bivariate PDF ($S_M^{KL+BX}(U+Th)$) under the assumption of site-independent mantle signal. In principle, because of the different depth of the Lithosphere-Asthenosphere boundary under the two experimental sites, a slight difference exists in the mantle signal detected by KL and BX. However, this difference is expected to account for less than 2% [25] and it can be here neglected in view of present experimental uncertainties.

The joint distribution $S_M^{KL+BX}(U+Th)$ can be inferred from the PDFs $S_M^{BX}(U+Th)$ and $S_M^{KL}(U+Th)$ by requiring that $S_M^{BX}(U+Th) = S_M^{KL}(U+Th)$ (Figure 27). Under these hypotheses, the combination of KL and BX constrains the mantle geoneutrino signal to $S_M^{KL+BX}(U+Th) = 8.9_{-5.5}^{+5.1}$ TNU. This result can be inserted in a trend of combined mantle geoneutrino signals published in the last 10 years (Figure 28) providing a valuable indication for testing the mantle signals expected from BSE compositional models (Section 7).

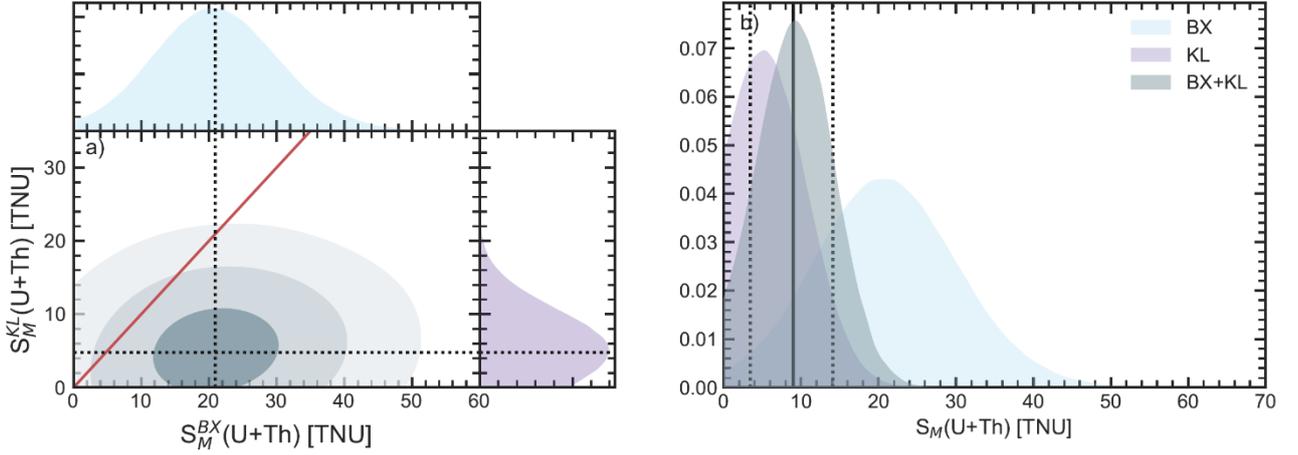

**Figure 27 – a)** Probability density functions of the extracted mantle signals (in TNU) for KL ($S_M^{KL}(U+Th)$, in violet) and BX ($S_M^{BX}(U+Th)$, in blue) together with the joint bivariate distribution ($S_M^{KL+BX}(U+Th)$, in grey). The dashed vertical and horizontal black lines represent the median values obtained for BX and KL respectively. The grey contours mark the areas corresponding to the 2D coverage of 39.3%, 86.5% and 98.9%. The red line represents the constraint $S_M^{BX}(U+Th) = S_M^{KL}(U+Th)$. **b)** Probability density functions of the extracted mantle signals (in TNU) for KL (in violet) and BX (in blue) together with the joint distribution $S_M^{KL+BX}(U+Th)$ (in grey) resulting from their combination. The black vertical line represents the median value of the distribution, while the vertical dashed lines report the 1σ interval.

**Table 28 –** Experimental signals ($S_{Exp}$) and adopted values for the modelled signals of the NFC ($S_{NFC}$), FFC ($S_{FFC}$) and CLM ($S_{CLM}$) for Borexino (BX) and KamLAND (KL), together with their derived mantle signals ($S_M$). The last row reports the mantle signal resulting from the combination of KL and BX observations.

|  | $S_{Exp}(U+Th)$ [TNU] | $S_{NFC}(U+Th)$ [TNU] | $S_{FFC}(U+Th)$ [TNU] | $S_{CLM}(U+Th)$ [TNU] | $S_M(U+Th)$ [TNU] |
|---|---|---|---|---|---|
| KL | $32.1 \pm 5.0$ | $17.7 \pm 1.4$ | $7.3_{-1.2}^{+1.5}$ | $1.6_{-1.0}^{+2.2}$ | $4.8_{-5.9}^{+5.6}$ |
| BX | $47.0_{-8.1}^{+8.6}$ | $9.2 \pm 1.2$ | $13.7_{-2.3}^{+2.8}$ | $2.2_{-1.3}^{+3.1}$ | $20.8_{-9.2}^{+9.4}$ |
| KL+BX | - | - | - | - | $8.9_{-5.5}^{+5.1}$ |

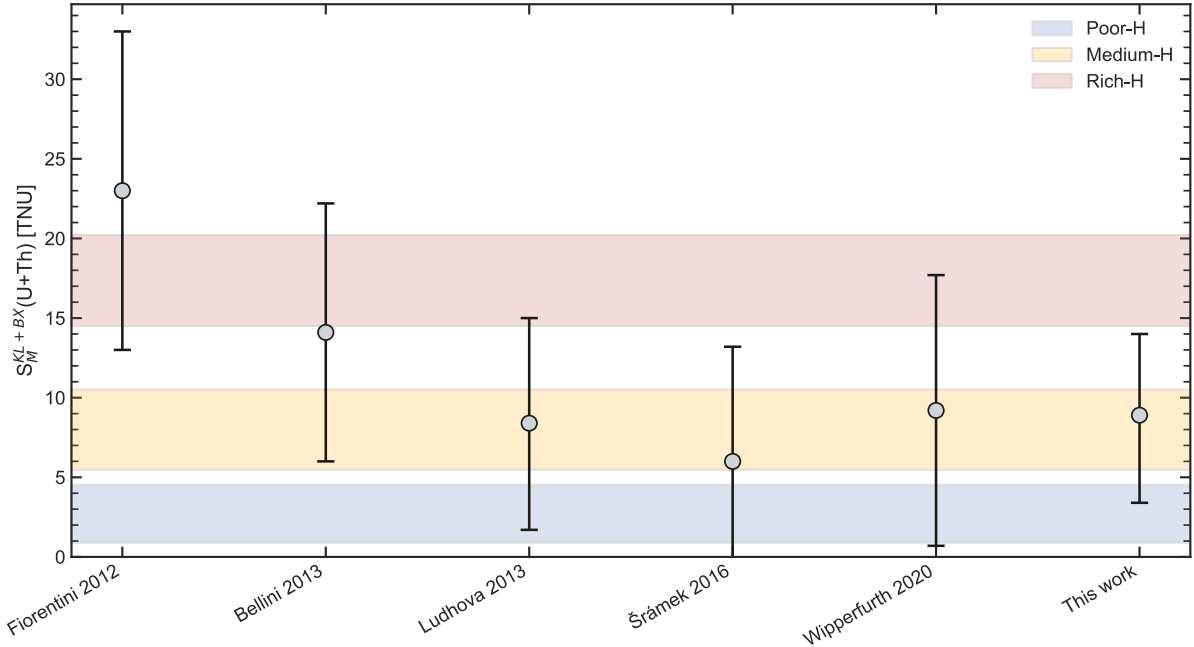

**Figure 28** - Collection of the published mantle signals ($S_M^{KL+BX}(U+Th)$) obtained from the combination of Borexino and KamLAND experimental results, together with the predictions (horizontal bands) of Poor-H, Medium-H and Rich-H models presented in Table 17. The reported model-dependent signals for the mantle are taken from [8,17,20,26,182] and this study. The horizontal blue, red and yellow bands correspond to the 68% coverage interval for the mantle signal predicted by Poor-H, Medium-H and Rich-H models (Table 29), calculated by substituting predicted mantle heat ($H_M(U+Th)$) in Eq. (23).

In the perspective of multi-site mantle investigation, a joint effort among the experimental collaborations in building a common analysis framework could improve the impact in geoscience. The formalization of a global χ² (or likelihood) function incorporating correlations, embedding U and Th experimental event rates, neutrino oscillation parameters and experimental statistical/systematic uncertainties would greatly boost the robustness of the extracted geoneutrino mantle signal, as detailed in [20,183].

## 10 What can we learn from geoneutrinos?

### 10.1 Mantle radiogenic power and composition

The mantle geoneutrino signal potentially brings to the surface valuable information about the unexplored Earth and particularly on the mantle radioactivity and the Earth's energetics. Since the adopted radiogenic heat power of the LS ($H_{LS}$ (U+Th)= $6.9^{+1.6}_{-1.2}$ TW, Table 21) is independent from the BSE model, the discrimination capability of the combined geoneutrino measurement among the different BSE models can be studied in the space $S_M(U + Th)$ vs $H_M(U + Th)$. The mantle signal ($S_M(U + Th)$) can be expressed as a linear function of the mantle radiogenic heat ($H_M(U + Th)$):

$$S_M(U + Th) = \beta \cdot H_M(U + Th) \tag{20}$$

where the $\beta$ coefficient depends only on U and Th distribution in the mantle and ranges between $\beta_{low}$ = 0.75 TNU/TW and $\beta_{high}$ = 0.98 TNU/TW. The lower and upper values are obtained assuming that the HPEs' masses are placed in a layer just above the CMB (low scenario) (Figure 29a) and assuming that they are homogeneously distributed in the mantle (high scenario)[6] (Figure 29b) [10].

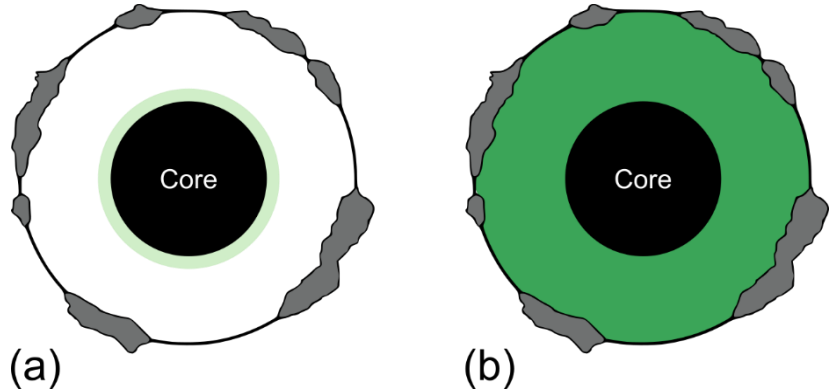

**Figure 29** – Cartoons of the distribution of HPEs' masses in the mantle predicted according to two different scenarios. (a) Low scenario: the HPEs are placed in a thin layer (in light green) above the CMB (b) High scenario: the HPEs are distributed homogenously (in dark green) in the mantle (modified after [10]).

The radiogenic power of the mantle $H_M(U + Th)$ can be further expressed as function of the U mass in the mantle $M_M(U)$:

$$H_M(U + Th) = h'(U) \cdot M_M(U) + h'(Th) \cdot M_M(Th) = [h'(U) + 3.7\, h'(Th)] \cdot M_M(U) \tag{21}$$

where the Th/U mass ratio $\frac{M_M(Th)}{M_M(U)} = 3.7$ is constrained by the $\frac{M_{LS}(Th)}{M_{LS}(U)} = 4.3$ in the LS according to the H13 model (Table 21) and the $\frac{M_{BSE}(Th)}{M_{BSE}(U)} = 3.9$ adopted in the extraction of the geoneutrino signal at KL and BX (Sections 3.3 and 4.3). Considering the equations above, it follows:

---

[6] The possibility of a layer enriched in HPEs in the upper part of the Mantle is disproved by several geochemical arguments and observations.

$$S_M(U+Th) = \beta \cdot [h'(U) + 3.7\, h'(Th)] \cdot M_M(U) \tag{22}$$

The linear relation between $S_M(U+Th)$ and $H_M(U+Th)$ is plotted in Figure 30. Assuming that the U and Th abundances in the mantle are radial, non-decreasing function of the depth and in a fixed ratio, the area between the two extreme lines (green lines) depicts the region allowed by all possible distributions of the U and Th in this reservoir. The maximal and minimal excursions of mantle geoneutrino signal is taken as a proxy for the 3σ error range. The solid black horizontal line in Figure 30 traces the combined signal $S_M^{KL+BX}(U+Th) = 8.9_{-5.5}^{+5.1}$ TNU, which falls within the prediction of the Medium-H models. The 68% coverage interval falls outside the prediction of the Rich-H models and results compatible with Poor-H models.

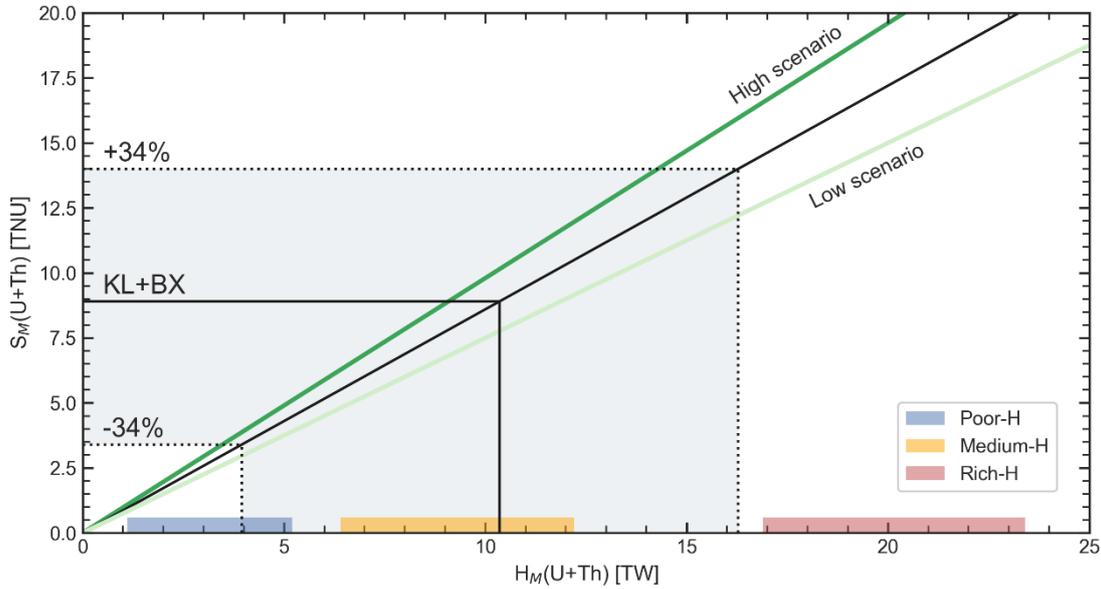

**Figure 30** – Mantle geoneutrino signal ($S_M(U+Th)$) as a function of U and Th mantle radiogenic heat $H_M(U+Th)$: the area between the green lines denotes the full range allowed between a homogenous mantle (high scenario, Figure 29b) and unique rich layer just above the CMB (low scenario, Figure 29a). The slope of the central inclined black line ($\beta_{centr}$ = 0.86 TNU/TW) is the average of $\beta_{low}$ and $\beta_{high}$. The blue, red and yellow bands on the X-axis correspond to the 68% coverage interval of the mantle radiogenic heat ($H_M(U+Th)$) of Poor-H, Medium-H and Rich-H models (Table 29), respectively. The black solid horizontal line represents the median mantle signal obtained by the combination of experimental signals from KL and BX $S_M^{KL+BX}(U+Th)$ (Table 28). The dashed horizontal black lines represent the 68% coverage interval.

The combined mantle signal ($S_M^{KL+BX}(U+Th)$) can be converted to the corresponding radiogenic heat by inverting the Eq. (20). Since the experimental error on the mantle signal is much larger than the systematic variability associated to the U and Th distribution in the mantle, the radiogenic power from U and Th in the mantle $H_M^{KL+BX}(U+Th)$ inferred from the combined mantle signal $S_M^{KL+BX}(U+Th)$ can be obtained with:

$$H_M^{KL+BX}(U+Th) = (1/\beta_{centr}) \cdot S_M^{KL+BX}(U+Th) = 1.16 \cdot S_M^{KL+BX}(U+Th) \tag{23}$$

Starting from $S_M^{KL+BX}(U+Th) = 8.9_{-5.5}^{+5.1}\, TNU$ it can be derived $H_M^{KL+BX}(U+Th) = 10.3_{-6.4}^{+5.9}\, TW$. The implications of this estimate in terms of total radiogenic heat (H), HPEs abundances ($a_M(U)$, $a_M(Th)$ and $a_M(K)$) and masses ($M_M(U)$, $M_M(Th)$ and $M_M(K)$) in the mantle can be studied with the comparison with the estimates provided by the Poor-H, Medium-H and Rich-H models (Table 29). The combined mantle geoneutrino measurement constrains at 68% C.L. the mantle composition to $a_M(U)$ > 5 ng/g, $a_M(Th)$ > 19 ng/g and $a_M(K) > 60$ μg/g and the mantle radiogenic heat power to $H_M(U + Th) > 4.0\ TW$ and $H_M(U + Th + K) > 4.8\ TW$.

**Table 29** – HPEs' abundances and masses, radiogenic heat in the mantle according to the classes of BSE models and obtained based on the combined geoneutrino measurement. The HPEs abundances are derived from the HPEs masses adopting the mass of the sublithospheric mantle $\boldsymbol{M_M} = \boldsymbol{3.911 \times 10^{24}\ kg}$ from the H13 model (Table 11). The masses and the radiogenic heat for the Poor-H, Medium-H and Rich-H models are obtained subtracting from the corresponding value of the BSE (Table 15) the adopted estimates of the LS according to the H13 model (Table 21) and considering them linearly independent. For the combined measurement the HPEs masses constrained by the combined measurement are obtained adopting $\boldsymbol{M_M}(Th)/\boldsymbol{M_M}(U) = 3.7$ and $\boldsymbol{M_M}(K)/\boldsymbol{M_M}(U) = 1.2 \times 10^4$; the $\boldsymbol{H_M}(U+Th)$ and $\boldsymbol{H_M}(U+Th+K)$ are obtained using the Eq. (15) and assuming that the K contribution to the total radiogenic heat is 17%.

| Model | $a_M(U)$ [ng/g] | $a_M(Th)$ [ng/g] | $a_M(K)$ [μg/g] | $M_M(U)$ [$10^{16}$ kg] | $M_M(Th)$ [$10^{16}$ kg] | $M_M(K)$ [$10^{19}$ kg] | $H_M(U+Th)$ [TW] | $H_M(U+Th+K)$ [TW] |
|---|---|---|---|---|---|---|---|---|
| Poor-H | $5 \pm 3$ | $12^{+10}_{-12}$ | $76^{+34}_{-36}$ | $1.9^{+1.1}_{-1.2}$ | $4.6^{+3.8}_{-4.6}$ | $29.6^{+13.4}_{-14.2}$ | $3.2^{+2.0}_{-2.1}$ | $4.2^{+2.4}_{-2.6}$ |
| Medium-H | $13 \pm 4$ | $42^{+14}_{-17}$ | $155^{+47}_{-48}$ | $4.9^{+1.5}_{-1.6}$ | $16.5^{+5.6}_{-6.5}$ | $60.7^{+18.4}_{-18.8}$ | $9.3 \pm 2.9$ | $11.4^{+3.5}_{-3.6}$ |
| Rich-H | $26^{+4}_{-5}$ | $98^{+16}_{-18}$ | $244^{+38}_{-40}$ | $10.0^{+1.7}_{-1.8}$ | $38.4^{+6.4}_{-7.0}$ | $95.4^{+15.0}_{-15.7}$ | $20.2^{+3.2}_{-3.3}$ | $23.4^{+3.8}_{-3.9}$ |
| Combined KL+BX | $13 \pm 8$ | $50^{+29}_{-31}$ | $157^{+90}_{-97}$ | $5.3^{+3.0}_{-3.3}$ | $19.5^{+11.2}_{-12.1}$ | $61.5^{+35.3}_{-38.0}$ | $10.3^{+5.9}_{-6.4}$ | $12.5^{+7.1}_{-7.7}$ |

## 10.2 Studying Earth's energetics with geoneutrinos

The mantle radiogenic heat constrained by the combined geoneutrino measurement of KL and BX allows for making the first step towards the understanding of the Earth's present heat budget, the evolution through geological time of our planet and the ratio of heat production over heat loss (Section 6).

In Table 30 are reported the results in terms of contributions to the Earth's heat budget obtained on the basis of the combined mantle geoneutrino signal and compared to the adopted value presented in Figure 20. Summing to $H_M^{KL+BX}(U + Th + K)$ the radiogenic power of the lithosphere $H_{LS}(U + Th + K) = 8.1^{+1.9}_{-1.4}$ TW, the inferred Earth's radiogenic power is $H^{KL+BX}(U + Th + K) = 20.8^{+7.3}_{-7.9}\ TW$ which falls in the 68% coverage range of the Medium-H models and it is compatible at 1σ level with the Poor-H models (Figure 31a).

**Table 30** – Comparison between the contributions to the Earth's heat budget based on the constraints set by the combined geoneutrino measurement of KL and BX and the adopted values described in Section 6 (see Figure 20). The value of the total heat power (Q), radiogenic heat of LS ($H_{LS}$(U+Th+K)) and secular cooling of the core ($C_C$) are the same for both the cases. The Earth's radiogenic heat (H(U+Th+K)) is obtained summing the independent components of $H_{LS}$(U+Th+K) and $H_M$(U+Th+K). The secular cooling of the Earth (C) and of the mantle ($C_M$) are obtained according to the equations reported in Figure 20 considering the terms as linearly independent.

|  | Adopted | Combined KL + BX |
|---|---|---|
| **Q [TW]** | \multicolumn{2}{c}{$47 \pm 2$} | |
| $H_{LS}$ (U+Th+K)[TW] | \multicolumn{2}{c}{$8.1^{+1.9}_{-1.4}$} | |
| $H_M$ (U+Th+K)[ TW] | $11.3^{+3.3}_{-3.4}$ | $12.5^{+7.1}_{-7.7}$ |
| **H (U+Th+K) [TW]** | $19.3 \pm 2.9$ | $20.8^{+7.3}_{-7.9}$ |
| $C_M$ [TW] | $17 \pm 4$ | $15 \pm 8$ |
| $C_C$ [TW] | \multicolumn{2}{c}{$11 \pm 2$} | |
| **C [TW]** | $28 \pm 4$ | $26 \pm 8$ |

The total radiogenic heat $H^{KL+BX}(U + Th + K)$ can be used also to extract the convective Urey ratio $U_R$ (Section 6.2) according to Eq. (12), taking into account the adopted Q and radiogenic heat of the continental crust ($H_{CC}$ (U + Th + K) = $6.8^{+1.4}_{-1.1}$ TW, Table 14). The 68% C.L. of the resulting value $U_R^{KL+BX} = 0.35^{+0.19}_{-0.20}$, if compared to the $U_R$ predicted by the three classes of BSE models (Figure 31b), shows an agreement with the corresponding range of the Medium-H models and a slight compatibility with both Poor-H and High-H models.

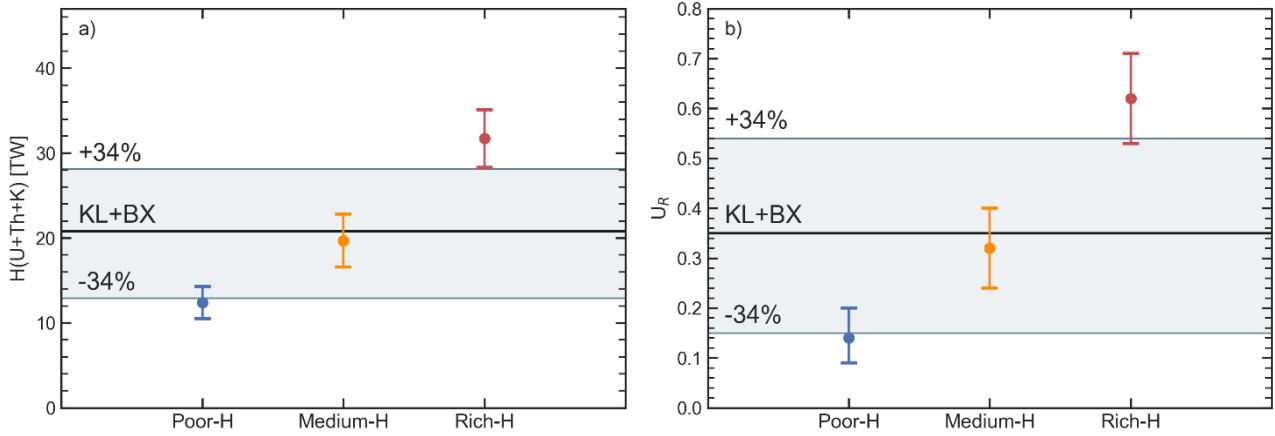

**Figure 31 – (a)** Comparison of the Earth's radiogenic heat (H(U+Th+K) constrained by the combined geoneutrino measurement of KL and BX with the estimates of the Poor-H, Medium-H and Rich-H models (Table 17). The lithospheric component is the same for all BSE models and for the combined measurement and it corresponds to the adopted value of the H13 model $H_{LS}$(U + Th + K) = $8.1^{+1.9}_{-1.4}$ TW (Table 21). **(b)** Comparison of the convective Urey ratio ($U_R$) constrained by the combined geoneutrino measurement of KL and BX with the estimates (68% coverage interval) of the Poor-H, Medium-H and Rich-H models. The $U_R$ is calculated according to Eq. (12), assuming that the total heat power Q = 47 ± 2 TW (Table 12) and the radiogenic heat of the continental crust $H_{CC}$ (U+Th+K) = $6.8^{+1.4}_{-1.1}$ TW (Table 14). The grey band represents the 68% coverage interval of $U_R^{KL+BX} = 0.35^{+0.19}_{-0.20}$.

Subtracting the total radiogenic heat $H^{KL+BX}(U + Th + K)$ from the adopted total heat power $Q = 47 \pm 2\,TW$ (Table 12), it follows that Earth's secular cooling is $C = 26 \pm 8\,TW$. Moreover, taking into account the adopted value of secular cooling of the core $C_C = 11 \pm 2\,TW$, the secular cooling of the mantle is $C_M = 15 \pm 8\,TW$. These results led to estimate the percentage contributions of the radiogenic heat and of the secular cooling to the Earth's heat budget (Figure 32). More than half (C~56%) of the total heat power is given by the heat loss of the mantle ($C_M$~32%) and from the core ($C_C$~24%). The remaining heat is attributable to the total radiogenic heat (H) which is due mainly to the contribution of the mantle ($H_M$~27%).

The results of the combined geoneutrino measurement of KL and BX in terms of Earth's energetics agree with the estimates of the Medium-H models which assume that the Earth has a bulk major-element composition matching that of CI chondrites.

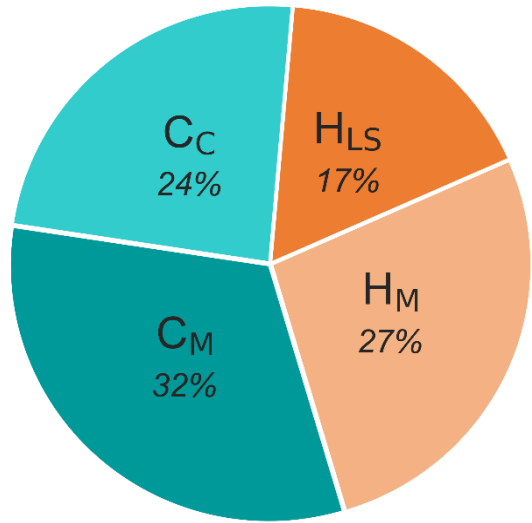

**Figure 32 –** Percentage contributions, constrained by the combined measurement of KL and BX, to the total heat power Q = 47 ± 2 TW of the secular cooling from the mantle ($C_M$) and from the core ($C_C$) and of the radiogenic heat from the lithosphere ($H_{LS}$) and from the mantle ($H_M$).

## 11 What next?

In the near future, the BX and KL experiments will not be the only ones able to detect geoneutrinos. The SNO+ experiment in Canada will soon start its "pure scintillator phase" which will include geoneutrinos detection; in China, the Jiangmen Underground Neutrino Observatory (JUNO) is currently under construction in the Guangdong province and in the Sichuan province is proposed the realization of a new detector in the China Jinping Underground Laboratory. These detectors are characterized by peculiar features in terms of depth and entity of background sources (reactor antineutrino and cosmic muon flux) (Table 31) from which arise different responses to the geoneutrinos signal. The plots reported in Figure 33 highlight that SNO+ and the proposed Jinping have the lower cosmic-ray muon fluxes while at JUNO is expected the highest muon flux together with the highest reactor antineutrino signal.

As depicted in Figure 34, while the above-mentioned experiments are located on continental crust, the pioneering proposal of the Ocean Bottom Detector (OBD) aims to realize the first detector sited in oceanic crust providing a strong sensitivity to geoneutrinos originating from Earth's mantle.

In this section, the main features of the future detectors are discussed together with the potential advances in geoneutrino science represented by the detection of directionality antineutrino and of potassium geoneutrinos not yet permitted by the present technology.

**Table 31** - Location, depth and water equivalent depth (in km), expected muon flux (in $cm^{-2} s^{-1}$), active mass (in ktons) and expected antineutrino signals (in TNU) for SNO+, JUNO and Jinping detectors. Detectors depths are taken from [184], [185] and [186], while reported muon fluxes come from [184], [187] and [186]. For each detector, the total expected geoneutrino signal S(U+Th), the lithospheric signal $S_{LS}$(U+Th) and the mantle contribution $S_M$(U+Th) are calculated according to H13. For SNO+, the obtained lithospheric signal is updated adopting the refined crustal model of [188] for the NFC. The expected reactor antineutrino ($S_{Rea}$) signals in the Geoneutrino Energy Region (GER, from 1.8 MeV to 3.3 MeV) and in the Full Energy Region (FER, 1.8 MeV to 10 MeV)) at detector site were calculated from 2019 PRIS data following [54]. The ratios between S(U+Th)/$S_{Rea}$(GER) and $S_M$(U+Th)/S(U+Th) are also reported.

|  | SNO+ (46.4667°N, 81.1703°W) | JUNO (22.1181°N, 112.5181°E) | Jinping (28.1532°N, 101.7114°E) |
|---|---|---|---|
| Depth [km] (Water equivalent [km.w.e.]) | 2.1 (5.9) | 0.7 (1.8) | 2.4 (6.7) |
| Mass [kton] | 0.78 | 20.0 | 4.0 |
| Muon flux [$cm^{-2} s^{-1}$] | $3.3 \cdot 10^{-10}$ | $4 \cdot 10^{-7}$ | $3.53 \cdot 10^{-10}$ |
| S(U+Th) [TNU] | $42.9^{+9.2}_{-5.3}$ | $39.7^{+6.5}_{-5.2}$ | $54.6^{+10.7}_{-8.9}$ |
| $S_{LS}$(U+Th) [TNU] | $34.2^{+9.2}_{-5.3}$ | $30.9^{+6.5}_{-5.2}$ | $46.0^{+10.7}_{-8.9}$ |
| $S_M$(U+Th) [TNU] | 8.7 | 8.8 | 8.6 |
| $S_{Rea}$(GER) [TNU] | $47.1^{+1.7}_{-1.4}$ | $282.3^{+35.4}_{-32.3}$ | $4.7^{+0.1}_{-0.1}$ |
| $S_{Rea}$(FER) [TNU] | $189.7^{+4.6}_{-4.2}$ | $631.9^{+44.8}_{-40.4}$ | $17.9^{+0.4}_{-0.4}$ |
| S(U+Th)/$S_{Rea}$(GER) | 0.91 | 0.14 | 11.6 |
| $S_M$(U+Th)/S(U+Th) | 0.203 | 0.222 | 0.158 |

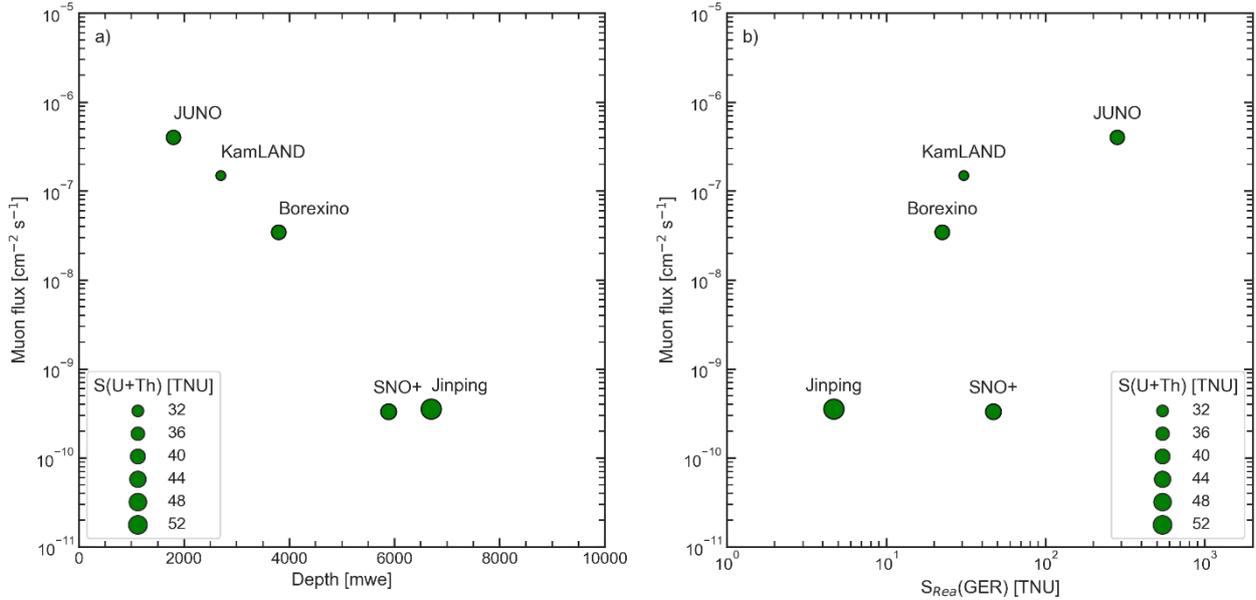

**Figure 33** – Muon flux expected at detector site vs water equivalent depth of the detectors **(a)** and vs **(b)** reactor antineutrino signal in the GER ($S_{Rea}$(GER)) expected at detector site (Table 31).

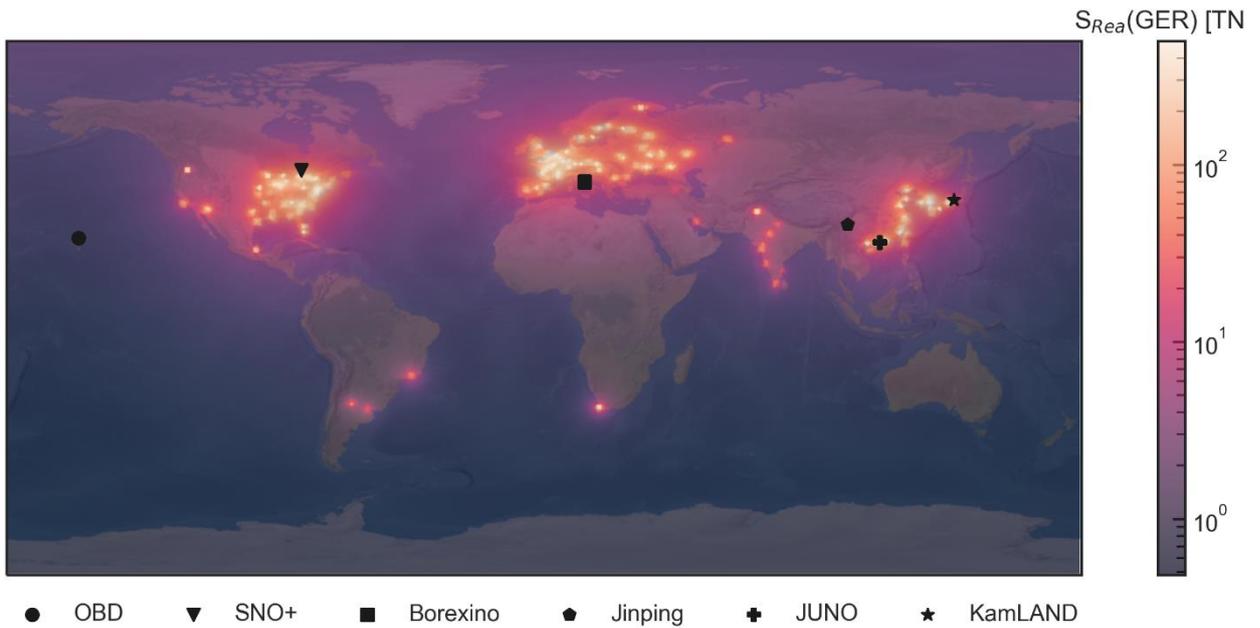

**Figure 34** – Map of the worldwide predicted antineutrino signals from nuclear power plants in the GER ($S_{Rea}$(GER)) expressed in TNU. For each 1°x 1° cell, the signal is yearly calculated from 2019 PRIS data following [54]. The adopted numerical data are available at www.fe.infn.it/radioactivity/antineutrino/index.html. The locations of current and future liquid scintillator experiments are superimposed with different marker symbols.

## 11.1  SNO+

SNO+ is a 780 tons liquid scintillator experiment located at the underground SNOLAB facility in Vale's Creighton mine near Sudbury, Canada (46.47°N, 81.17°W). The detector is covered by a 2092 ± 6 m rock overburden, corresponding to 5890 ± 94 mwe, which provides an effective shield against cosmic muons (Figure 33), making SNOLAB the laboratory having the lowest cosmic flux in the world [189]. SNO+ reuses much of the existing infrastructure of SNO, consisting of a 6 m radius acrylic vessel surrounded by almost 9300 PMTs. Due to the incompatibility between acrylic and existing widely used scintillators, SNO+ will employ a newly developed liquid scintillator of linear alkyl benzene and the fluor 2.5-diphenyloxazole (PPO) at 2 g/L [190]. Besides being compatible with acrylic, linear alkyl benzene exhibits a competitive light yield while enabling longer attenuation lengths, superior safety characteristics, chemical simplicity, ease of handling and for these reasons it will be employed in future neutrino experiments including JUNO.

Although the main goal of SNO+ is the search for the neutrinoless double-beta decay of $^{130}$Te, the experiment has also the potential to observe solar neutrinos, reactor antineutrinos, geoneutrinos, supernova neutrinos and invisible nucleon decays [191]. Data taking will consist of three different phases aiming at different physics goals [192]: (i) the water phase, in which the detector will act as a pure water Cherenkov detector, allowing to measure events occurring both inside and outside the acrylic vessel with directional information, and to characterize the optical properties of the outer water and PMT response; (ii) the pure scintillator phase, in which the vessel will be filled with linear alkyl benzene, allowing to characterize the optical properties and backgrounds of the scintillator; (iii) the Tellurium phase, in which linear alkyl benzene will be loaded with tellurium in the search for $^{130}$Te neutrinoless double-beta decay. After completing an extensive physics campaign operating as an ultrapure water Cherenkov detector, linear alkyl benzene has now been deployed in the acrylic vessel [190]. During this phase (as well as during phase ii), SNO+ will be able to exploit IBD to detect geoneutrinos and reactor antineutrinos, expected to account for $S(U+Th) = 42.9^{+9.2}_{-5.3}$ TNU and $S_{Rea}(GER) = 47.1^{+1.7}_{-1.4}$ TNU at detector site, respectively (Table 30).

The SNOLAB facility is located in a geothermally anomalous region [193], the Sudbury Structure, whose bulk crustal radioactivity was in the past estimated through inversion of heat flux measurements [194], yielding though nonunique constraints for modeling the geoneutrino flux. By making use of compiled geological, geophysical, and geochemical information [195] provided a detailed 3D model of the 6° x 4° regional crust centered at SNO+[7]. Crustal cross sections obtained from refraction and reflection seismic surveys were used to characterize the crust and assign uncertainties to its geophysical structure. This crustal model was further refined by [188], which accompanied compiled geological observations and geophysical surveys with a set of 112 rock samples representative of the geological formations collected with an ad-hoc sampling in the 50 km x 50 km upper crust region surrounding SNO+. Spectroscopic analyses on these samples conducted through HPGe gamma detector and ICPMS techniques permitted a detailed study of the PDFs of U and Th abundances and of their correlation, enabling a bivariate analysis used for a robust treatment of geochemical uncertainties. This study found that the lithospheric contribution to the geoneutrino signal at SNO+ will account for $S_{LS}(U+Th) = 34.2^{+9.2}_{-5.3}$ TNU, which shall be subtracted from the measured experimental signal in order to recover information on the mantle, expected to account for $S_M(U+Th) = 8.7\ TNU$ [25] (Table 31).

This *ad-hoc* 3D modelling of the Sudbury region put on view how the SNO+ detector lies in the contact zone between two distinct and peculiar geological units (Huronian Supergroup and Norite-gabbro of Sudbury Igneous Complex), making it necessary an additional detailed characterization of the surrounding rock overburden, which [196] estimated to contribute with a signal increase of $1.4^{+1.8}_{-0.9}$ TNU.

## 11.2 JUNO

The Jiangmen Underground Neutrino Observatory (JUNO) is a 20 ktons multipurpose underground liquid scintillator detector currently under construction at a depth of 700 m (1800 mwe) in the Guangdong Province of South China (N 22.12°, E 112.52°). By featuring a 75.2% photodiode coverage achieved via a primary calorimetry system consisting of 20000 20-inch PMTs and an additional calorimetry system of 25600 3-inch PMTs, JUNO will permit to further push the boundaries of neutrino physics [197]. Besides the determination of the neutrino mass hierarchy, the primary physics goal of this linear alkyl benzene based experiment, the excellent energy resolution (3%/√E[MeV]) and the unprecedently large fiducial mass (~17 ktons) planned for the JUNO detector will offer exciting opportunities for addressing many important topics in the field of geoneutrinos.

---

[7] The 3D geophysical model is available at https://www.fe.infn.it/radioactivity/SNO+/

The detector will be placed at 53 km from the Yangjiang and Taishan nuclear power plants, which will contribute to an overall reactor antineutrino signal of $S_{Rea}(GER) = 282.3^{+35.4}_{-32.3}$ TNU (Table 31). This contribution will need to be properly characterized to recover the 7 times-lower geoneutrino flux. For this reason, the antineutrino spectrum coming from one of Taishan's reactor cores is planned to be closely monitored by the Taishan Antineutrino Observatory (TAO) [198], which is expected to have a 2-times better resolution and a 30-times higher statistics than JUNO.

According to the different modeling of the local lithospheric signal, the total geoneutrino signal at JUNO is expected to range between $S(U + Th) = 39.7^{+6.5}_{-5.2}$ TNU [199] and $S(U + Th) = 49.1^{+5.6}_{-5.0}$ TNU [200] (Table 31). While the signal expected from U and Th is comparable to what observed in BX and KL [199], thanks to its unprecedented size and sensitivity JUNO will allow for the recording of nearly 300–500 geoneutrino interactions per year, significantly improving the statistics of existing geoneutrino event samples. In approximately nine months of data taking, JUNO will match the present world sample of recorded geoneutrino interactions, currently accounting for less than 220 events (Section 3.3 and Section 4.3).

JUNO will sit on top of a ~30 km thick continental crust, which is expected to represent the major contribution to the detected geoneutrino signal; modeling of the mantle envisage a $S_M(U + Th) = 8.8\,TNU$ contribution to the geoneutrino signal, representing ~22% of the total [199] (Table 31). The near crust within ~100 km from the detector is expected to generate itself a geoneutrino signal equivalent to that of the whole mantle. For this reason, a refined geological model of the region around the detector is of critical importance in view of gathering insights on mantle radiogenic power and composition from a future geoneutrino measurement at JUNO. The scientific community already produced JULOC [200] and GIGJ [201], two refined crustal models based respectively on Bayesian inversion of gravimetric data and seismic ambient noise tomography. GIGJ is a 3D numerical model having 50 × 50 × 0.1 km resolution, built by inverting GOCE (Gravity field and steady-state Ocean Circulation Explorer) gravimetric data over the 6° × 4° area centered at JUNO[8]. The model incorporates the *a-priori* knowledge derived by deep seismic sounding profiles, receiver functions, teleseismic P wave velocity models, and MOHO depth maps, in order to provide a site-specific subdivision of crustal layers' mass, thickness and density together with the associated geophysical uncertainties. JULOC is a 3D crustal model covering a 10° × 10° area centered in JUNO. This model recovers the thickness and the uncertainties of crustal layers by inverting shear P and S wave velocities measured via seismic ambient noise tomography and it derives layers' density by employing the relation between P wave velocity and rock density. Thanks to ICPMS and wet-chemistry analyses on 3000 surface rock samples collected in the research area, JULOC accompanies its geophysical model with geochemical information on U and Th abundances, predicting a geoneutrino signal contribution from the local 10° × 10° crust of 28.5 ± 4.5 TNU [200].

**11.3   Jinping Underground Laboratory**

The Jinping neutrino experiment is proposed at the China Jinping Underground Laboratory (CJPL) in Sichuan, China (28.15°N, 101.71°E). Thanks to its 2400 m depth (corresponding to 6700 mwe), this laboratory has the worldwide lowest cosmic-ray muon flux after SNO+ and the lowest reactor antineutrino flux of any other laboratory (Figure 33), and it is therefore ideal to carry out low-energy neutrino experiments [202].

---

[8] The 3D geophysical model is available at https://www.fe.infn.it/radioactivity/GIGJ/

The collaboration plans to build two cylindrical caverns each of 20 m in diameter and 24 m in height, containing two distinct cylindrical neutrino detectors [203]. With a total mass of 4 ktons (of which 3 ktons fiducial for geoneutrinos), the Jinping experiment could improve present solar neutrinos measurements, investigate the matter oscillation effect in solar neutrino oscillation and acquire the geoneutrino signals mainly produced from the Asian continental crust.

The CJPL facility lies on the slopes of Himalaya, the Earth's region having the thickest continental crust (~70 km). The total geoneutrino signal is hence expected to reach $S(U+Th) = 54.6^{+10.7}_{-8.9}$ TNU, with the lithosphere and the mantle accounting for $S_{LS} = 45.97^{+10.7}_{-8.9}$ TNU and $S_M = 8.6$ TNU, respectively (Table 31). Thanks to the unprecedently low muon flux ($2 \cdot 10^{-10}$ cm$^{-2}$ s$^{-1}$) and reactor background ($S_{Rea}(GER) = 4.7^{+0.1}_{-0.1}$ TNU) (Table 31), the Jinping experiment is expected to measure in 1500 days the total geoneutrino flux and the Th/U ratio with 4% and 27% precision respectively [203]. Due to predominance of the lithospheric contribution over the mantle signal, which accounts for only 15% of the total geoneutrino signal, Jimping will not constrain the mantle's radiogenic heat production itself, but it will represent an unvaluable resource in view of a multi-site analysis combining observations from different experiments [26].

**11.4 Ocean Bottom Detector**

KamLAND and Borexino experiments show that geoneutrinos are useful to observe geoscientific insights, although leaving a question of the mantle's contribution to the global signal. Distinguishing the contribution from the mantle by current and future detectors, which are all located on continents, is challenging, since the crustal contribution is about 70% of the total flux. The thin 7 km (cf., continental crust ~35 km thick) and simple oceanic crust, having an order of magnitude lower Th and U abundance, make a detector sited in the middle of the ocean ideally sensitive to identifying geoneutrinos originating from the mantle. In case of the oceanic detector the contribution of mantle geoneutrinos is about 70% of the total expected anti-neutrino flux while the land-based detectors have <20 % contributions. Leaving far from the coast keep a distance from nuclear reactors which are unseparated background, reactor neutrinos. Given the detector can be moved and placed into the ocean, multi-site measurement offers a possibility to tighten the constraint on mantle radioactivity abundance and distribution.

The idea of placing a neutrino detector in the ocean was firstly proposed as "Hanohano" experiment [204]. The group at University of Hawaii, together with Makai Ocean Engineering, reported on technological developments and a detailed detector design [205]. Recently in Japan an ongoing collaboration between Tohoku University and JAMSTEC (Japan Agency for Marine-Earth Science and Technology) has reinvigorated this idea with Ocean Bottom Detector (OBD). OBD collaboration is planning to deploy a small size liquid scintillator detector (~20 kg) in the ocean at 1 km depth in 2022. The sensitivity of ~1.5 kt detector for mantle geoneutrino is estimated to be 3.4σ with 3-year measurement by detector simulation. Figure 35 shows the expected energy spectrum of 1.5 kt OBD. OBD project broadens our perspective and works across the disciplinary boundaries of particle physics, geoscience and ocean engineering. The kt scale detector will be a breakthrough in the interdisciplinary community.

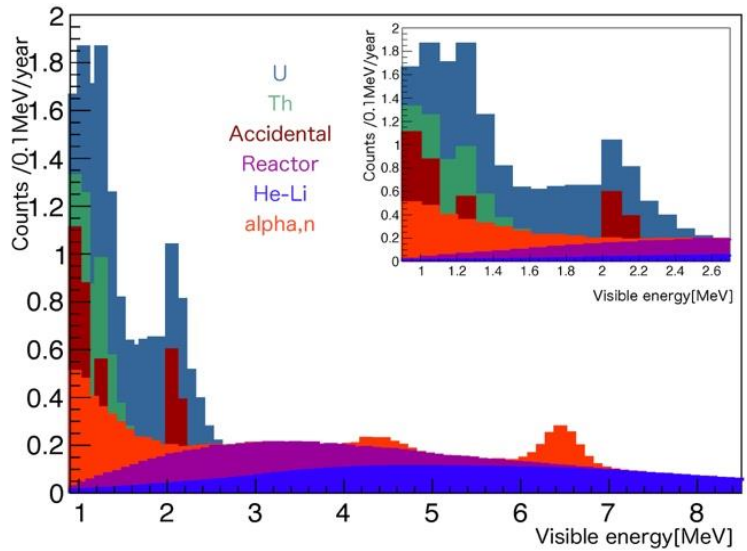

**Figure 35 -** Expected energy spectrum of 1.5 kt at OBD estimated by Geant4 simulation. Right top panel focuses on geoneutrino energy range (< 2.6 MeV). The detector location is assumed to be off coast of Hawaii in the depth of 2.7 km. A 70 cm fiducial volume cut is applied to reduce backgrounds from the radioactive contamination in the detector materials.

### 11.5     Directionality

Liquid scintillator detectors have the sensitivity to measure the total amount of geoneutrinos from the Earth's crust and mantle. Furthermore, the detectors successfully measured various kinds of neutrinos, such as solar, reactor, and extraterrestrial neutrinos. However, we do not have the technology to track the direction of incoming neutrinos at present due to the high misidentification in the neutrino's track reconstruction [206,207]. A direction-sensitive detector could map out the U and Th distribution inside the Earth and could be able resolve crust versus mantle contributions.

Low energy antineutrinos are detected by the IBD reaction. This reaction is tagged by the delayed coincidence based on the prompt positron signal and delayed neutron capture signal. In the inverse beta decay, a relationship between the positron's kinetic energy and scattering angle has been developed by [208]. By momentum conservation law, the neutron's kinetic energy and scattering angle can be calculated. Since the neutron scattering angle is < 35 degrees in geoneutrino energy region, the neutron retains the directional information of the incoming neutrinos. The team at Tohoku University in Japan have developed $^6$Li-loaded liquid scintillator to (i) minimize thermal diffusion of the neutron before it is captured by other nuclei and (ii) get point-like delayed signal for precise vertexes. Optical discrimination of energy deposit point by high resolution imaging devices is required to separate prompt and delayed signals. The team is now working on prototype detector which has 30L of $^6$Li-loaded liquid scintillator and two of the imaging detectors to precisely measure vertexes with reconstructed 3D images. The directional sensitive detector can be applied to not only measure geoneutrinos but also monitor nuclear reactor and track astrophysical sources of neutrinos.

### 11.6     Geoneutrinos from $^{40}$K

While the abundances of RLEs such as U and Th are well constrained by observations in chondrites, the silicate Earth seems strongly depleted in volatile elements such as K (Section 5.2). The abundance of this element envisaged for the BSE spans a factor 2 among the proposed compositional models (Section 7), with predictions exhibiting a terrestrial $a_{BSE}$(K)/$a_{BSE}$(U) ratio from ~1/3 to ~1/8 of what observed in chondrites. Different theories attribute the origin of this "*missing K*" to either loss to space during accretion [38] or segregation into the core [209], but no experimental measurement was able to solve this riddle so far. Estimating the K content of the Earth with a direct measurement would help the comprehension of Earth's origin and composition, providing key tests of bulk Earth compositional paradigms and providing critical information about the behavior of volatile elements during Earth's early-stage formation.

With a luminosity of ~$10^{25}$ decays/s, $^{40}$K currently accounts for ~20% of present radiogenic heat production inside the Earth and is thought to have accounted for nearly 50% of the radiogenic power at the first stages of Earth formation (Section 6.2.1). Although it is the most abundant among the main HPEs, K has never been directly investigated through the use of geoneutrinos.

The radioisotope $^{40}$K is both a neutrino and an antineutrino emitter (Figure 36). With an 89.28% Branching Ratio (B.R.), $^{40}$K undergoes β⁻ decay to $^{40}$Ca ground state, emitting an electron and an antineutrino ($E_{max}$=1.3 MeV) [14]. On the other hand, 10.72% of the times, $^{40}$K emits a neutrino while decaying to $^{40}$Ar in one of the following processes: i) electron capture to $^{40}$Ar ground state (0.046% B.R.) with the emission of a 1.5 MeV monoenergetic neutrino, ii) electron capture to $^{40}$Ar excited state (10.67% B.R.) with the emission of a 43.6 keV monoenergetic neutrino and iii) β⁺ decay to $^{40}$Ar ground state (0.001% B.R.) with the emission of a positron and a neutrino ($E_{max}$ = 0.5 MeV) (Figure 36). Therefore, on the basis of Medium-H compositional BSE models, the resulting $^{40}$K electron neutrino and antineutrino fluxes expected at surface are ~$10^5 \frac{\nu_e}{cm^2 \cdot s}$ and ~$10^6 \frac{\bar{\nu}_e}{cm^2 \cdot s}$ respectively.

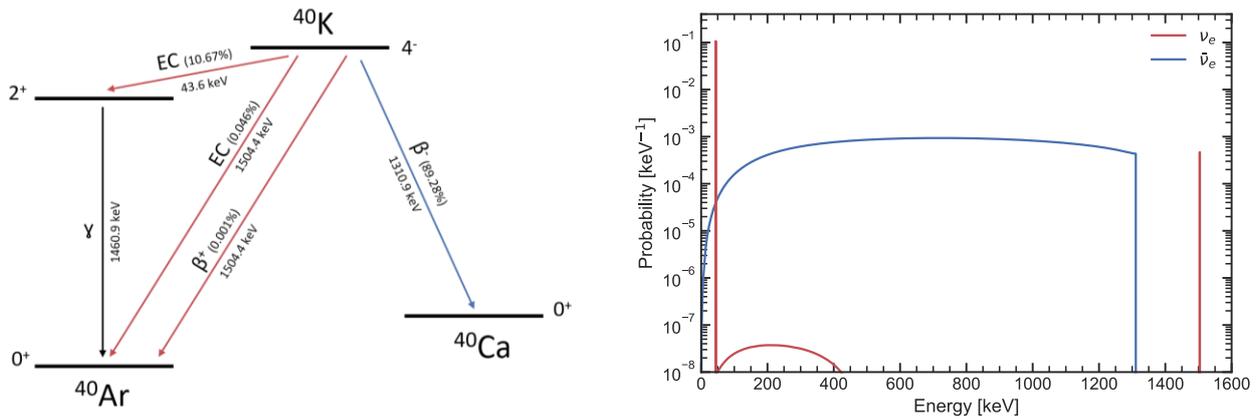

**Figure 36 – (a)** Decay scheme of $^{40}$K from [14]. For each decay branch, the Branching Ratio (B.R.) (in brackets), the decay mode and the associated energy are reported. Decay lines are red and blue when associated to the emission of a neutrino or an antineutrino, respectively. **(b)** Neutrino (in red) and antineutrino (in blue) emission energy spectra for the decay scheme reported in panel (a) obtained from Betashape [210].

Ongoing geoneutrino experiments make use of the IBD reaction on free protons to detect antineutrinos. Unfortunately, the 1.311 MeV $^{40}$K antineutrino endpoint falls below the IBD kinematic threshold of 1.806 MeV, making these geoneutrinos currently undetectable. In the past decades, several approaches were suggested to exploit both charged current and neutral current interactions to detect $^{40}$K geoneutrinos.

Proposed charged current interactions foresee the use of targets heavier than hydrogen to lower the current IBD detection threshold for $^{40}$K antineutrinos' [211]. However, in the absence of a delayed neutron capture signal following positron annihilation (Figure 4), an alternative technique must be found in order to effectively tag IBD reactions. Since the nuclei resulting from IBD reactions are usually β⁻ emitters, two possible approaches involve the tagging of the delayed β⁻ coincidence (for $T_{1/2}$<1 day) or the radiochemically counting the resulting nuclei (for $T_{1/2}$>1 day), depending on the product's $T_{1/2}$. Several isotopes suitable for radiochemical counting (e.g. $^3$He, $^{35}$Cl, $^{14}$N) have been proposed by [212], while among easily taggable IBD targets, $^{106}$Cd has been suggested by [213] because of its peculiar double positron signature. Other possible targets sensitive to $^{40}$K geoneutrinos include $^{63}$Cu, $^{79}$Br and $^{151}$Eu [214], but the single positron produced by these reactions cannot be currently discerned from background in current scintillator detectors. The antineutrino cross-section envisaged for these target isotopes is several orders of magnitude lower than the one for IBD on free protons. Reaching acceptable statistical significance for $^{40}$K, would require non-viable detector sizes (~$10^5$ tons) and high loading fractions, therefore compromising scintillator transparency and making their employment effectively challenging in current-generation liquid scintillator detectors. Innovative approaches enabling the detection of single positrons through the employment of opaque scintillators [215] promise to enable high loading in liquid scintillators and to possibly permit the employment of such IBD targets in view of $^{40}$K geoneutrino detection.

Proposed neutral current approaches employ elastic scattering on electrons to tag $^{40}$K neutrinos and antineutrinos events. Not having the possibility to distinguish neutrinos from antineutrinos, the main disadvantage of this approach is represented by the huge solar flux background (~$10^{11} \frac{\nu_e}{cm^2 \cdot s}$), which is several orders of magnitude higher than the expected $^{40}$K flux. The only viable way to discern $^{40}$K geoneutrino signals over the background is the experimental separation of the events on the basis of the recoil direction of scattered electrons [206]. However, the short recoil tracks and the direction of the incoming particle are not effectively recoverable via current liquid scintillator techniques. Possible solutions foresee the use of direction-sensitive gas-filled detectors (e.g., CF$_4$) fulfilling the detection requirements posed by $^{40}$K geoneutrinos [206] or the employment of Cherenkov neutrino detectors [216]. Other innovative neutral current approaches exploit the coherent scattering on nucleon, whose cross section is several orders of magnitude higher than IBD and elastic scattering on electrons. However, the detection of the low kinetic energies of the outgoing nucleons and their direction are impossible to appreciate with current experimental technologies and require the employment of innovative approaches based on nanoscale explosives [217] or semiconductors such as Ge or Si [218].

## Acknowledgements

KI and HW gratefully thank the KamLAND Collaboration for continues supports and contributions. The KamLAND experiment has been supported by the Japan Society for the Promotion of Science (JSPS), the Japanese Ministry of Education, Culture, Sports, Science and Technology (MEXT), and the United States Department of Energy (DOE). The Kamioka Mining and Smelting Company has provided services for activities in the mine.

GB, FM, VG, AS thank the members of the collaboration who in various aspects contributed to the Borexino geoneutrino results in addition to the Gran Sasso Laboratory and the Italian INFN, the US NSF as well as the German BMBF, DFG, HGF, MPG, the Russian RFBR, the Polish RSF, for their support.
GB thanks Livia Ludhova for the usefull discussions and Sandra Zavatarelli for her important help. FM, VS and AS would like to thank Giovanni Fiorentini, Eligio Lisi, William McDonough, Scott Wipperfurth and Ondřej Šrámek for the fruitful discussions and Kassandra Raptis for her precious support.

**Appendixes**

**Appendix A.1 - The crust near KamLAND**

The Japan island arc, hosting the KamLAND detector, is part of a continental shelf located close to the eastern margin of the Eurasian plate. The Philippine plate and the Pacific plate are moving toward the Eurasian plate and are subducting respectively beneath the southern and the northern part of Japan. The submarine trenches are thus formed with parallel uplifted areas and intense igneous activity. The KamLAND detector is sited in a typical continental crust of Island Arc and Forearc environment. The Japan Sea (JS), situated between the Japan island arc and the Asian continent, is classified as marginal sea and it is bordered by islands and expanded basins on the back-arc side (back arc basin).

[20] and [3] proposed a site-specific geophysical and geochemical modeling of the crust near the KamLAND detector and additionally studied the effects on geoneutrino signal of the peculiarities characterizing the subducting slab and the JS crust. The geophysical structure of the Japanese crust depicted, based on [219], envisages the presence of two layers, the UC and the LC, separated by the Conrad discontinuity and doesn't account for the presence of SED and MC. The BC thickness ranges between 32 and 40 km with both the UC and LC accounting for the half of the total thickness. From the geochemical point of view, the LC is treated as a homogenous layer with $a_{LC}$(U) = 0.85 ± 0.23 µg/g and $a_{LC}$(Th) = 5.19 ± 2.08 µg/g based on the model of the reported in [220]. A more refined modeling is dedicated to the UC, which U and Th abundances are distributed with a 0.25° × 0.25° resolution grid adopting the chemical composition estimated by [221]. The measurements on 166 samples, collected on the exposed crust and associated to 37 geological groups, are adopted to infer the geochemical abundances for the whole UC. The surface exposure weighted average abundances are estimated to be $a_{UC}$(U) = 2.32 µg/g and $a_{UC}$(Th) = 8.3 µg/g, slightly lower than the typical continental crust abundances. It is worth highlighting that although the analyzed dataset does not include only rocks from the crystalline basement rocks, this approach ignores the presence of a distinct SED layer.

The subducting slabs of the Philippine and Pacific plates could represent a radionuclides enrichment factor for the LC of the Japan Arc. [20] modeled a single slab penetrating below Japan with an average velocity $v$ = 60 mm/year on a time scale T $\sim 10^8$ year and encompasses two extreme scenarios for the evaluation of the impact of the subduction processes on the prediction of geoneutrino signal, i.e. (i) the slab keeps its trace elements during the subduction and supposing (ii) all the U from the subducting crust is dissolved in fluids and transported to the base of the LC of Japan arc. The corresponding enrichment factor are 1.06 and 2.57 translating in an estimation of the total signal of the subducting slab of $S_{Slab}$(U+Th) = 2.92 ± 0.88 TNU. According to [3], the subducting slab is a oceanic crust layer with a thickness of 10 km that, with the same composition of the OCC, originate an increase on the total geoneutrino flux of 0.21% for U and 0.11 % for Th. Note that, based on seismic arguments, [3] set also the presence of a "cold" slab accumulated at the boundary between the UM and the LM ($\sim$ 670 km). The U and Th abundances assigned to the slab ($a_{Slab}$(U) = 0.021 µg/g and $a_{Slab}$(Th) = 0.065 µg/g) are assumed to originate an increase of the total flux of 2.1% and 1.0%.

An additional peculiarity of KL consists in the controversial nature of the crust beneath the JS. Although global models classify this portion of crust as a typical OCC, its higher thickness and the presence of fragments of CC make it unique and different. [20] estimated the minimal and maximal geoneutrino production assuming for the JS crust two extreme scenario: (i) a typical OCC with a thickness of 7 km and a overlaying 1 km SED layer; (ii) a typical CC characterized by a thickness of 19 km and an overlaying 4 km SED layer. The contribution to the signal from the JS $S_{JS}$(U+Th)= 0.43 ± 0.13 TNU is thus defined as the central value of these two extremes with uncertainties encompassing the extreme values with 3σ. [3] studied that the effect of the JS crust can produce an increase of the total geoneutrino flux ranging between the 0.36% and 2%, assigning a continental or oceanic composition respectively.

The $S_{NFC}$(U+Th) estimated by [20] (Table 27) includes the signal produced by the six tiles (Figure 26a), the subducting slab and the JS crust. For [3], only the $S_{BC}$(U+Th) is reported since the data does not permit to infer the signal from the NFC and from the FFC.

**Appendix A.2 - The crust near Borexino**

The Gran Sasso range, where the Borexino experiment is located, is a massif of the Central sector of the Apennines, a peri-mediterranean chain part of the Adria plate. The actual geological structure of the Apennine chain is the result of the geodynamical processes occurred during its orogenesis began in the early Neogene (20 million years ago).

A refined reference model for the Gran Sasso area was developed by [180] in which local and specific geophysical and geochemical information are used to provide an estimate of the geoneutrino signal originated from the 6° × 4° (492 × 444 km) portion of the crust surrounding the LNGS (Figure 26). The model subdivides the study area in two zones, the central tile (CT) and the rest of the region (RR), which are described with different degree of resolution. The CT, i.e. the crustal portion within ~100 km from the Borexino detector, is described with a simplified tectonic model characterized by a typical resolution of (2.0 km × 2.0 km × 0.5 km).

The crust has a layered structure typical of Central Apennines, characterized by a SED cover thicker than that reported for the same area in any global crustal model (~1 km, see Figure 25). The deep structure of the Central Apennines was investigated analyzing data from the eastern part of CROP 11 deep reflection seismic profile that cuts across the whole chain. The interpretation of this profile, coupled with detailed information coming from deep (~4 km) exploration wells, assures around the Gran Sasso area the existence of a thick (>10 km) sedimentary sequence overlying the crystalline crust, detailed in Figure 8 of [222]. Excluding the rare and shallow volcanic deposits, the sedimentary pile includes different sequences of carbonate and terrigenous sediments from Late Triassic to Pleistocene which reflect diverse depositional environments (carbonate platform and silicoclastic depositional systems). The U and Th mass abundances were obtained by ICP-MS and gamma spectroscopy measurements of the rock samples representative of the sedimentary succession and collected within 200 km from the LNGS. Considering the relative volume of the different reservoirs estimated on the basis of the 3D geological model, the weighted average abundance obtained for U ($a_{SED}$(U) = 0.8 ± 0.2 µg/g) and Th ($a_{SED}$(Th) = 2.0 ± 0.5 µg/g) are incompatible at more than 5σ level with global estimates (Table 21).

The overall thickness of the crust (~35 km) modeled by [180] is in agreement with the global reference models (~34 km, Figure 25) and it is confirmed by the studies reported in [223] and [224]. The local seismic sections do not highlight any evidence of MC and as result the crystalline basement is subdivided into UC (~13 km) and LC (~9 km). The U and Th mass abundances are obtained by ICP-MS and gamma spectroscopy measurements of the rock samples collected from the closest representative outcrops of UC and LC of the South Alpine basement, located in Ivrea-Verbano Zone and in Valsugana. The U and Th abundances adopted for the UC and LC are compatible at 1σ level with the estimates provided by the global models (Table 21).

The geoneutrino signal of the NFC is $S_{NFC}$(U + Th) = 9.2 ± 1.2 TNU where 77% of the signal originates from U and Th distributed in the CT. The maximal and minimal excursions of various input values and uncertainties reported in [180] are taken as the ± 3σ error range. The U and Th signal errors are conservatively considered fully positively correlated. The reduction of ~6 TNU and ~9 TNU with respect to the estimations which H13 and W20 respectively provide - for the almost coincident crustal area (Figure 26)- is mainly due to presence of thick sedimentary deposits composed primarily of U- and Th-poor carbonate rocks which are not taken into account in the global reference models.

**Appendix A.3 - Geoneutrino signal calculation**

The geoneutrino signals in Table 24 and Table 27 are reported as appeared in the corresponding references, or in some specific cases, are calculated using updated oscillation parameters. In this section, the approaches followed are detailed.

As in H13 the $S_{NFC}$(U+Th) are not given, we infer it from the subtraction between $S_{BC}$(U+Th) and $S_{FFC}$(U+Th), the error propagation is performed via a Monte Carlo sampling of HPEs abundances according to their PDF in order to propagate the asymmetrical uncertainties of the non-Gaussian distributions.

For KamLAND, the $S_{BC}$(U+Th) of [3] is calculated as $S_{BC}$(U+Th) = $S_{BC}$(U) + $S_{BC}$(Th) where:

$$S_{BC}(\text{U}) = \phi_{BC}(\text{U}) \cdot \frac{0.55}{0.59} \cdot \langle \sigma_U \rangle$$

$$S_{BC}(\text{Th}) = \phi_{BC}(\text{Th}) \cdot \frac{0.55}{0.59} \cdot \langle \sigma_{Th} \rangle$$

where the $\phi_{BC}(\text{U})$ and $\phi_{BC}(\text{Th})$ are the uranium and thorium geoneutrino flux given from the sum of the crustal components reported in Table 2 of [3]; 0.55 e 0.59 are the average survival probability ($\langle P_{ee} \rangle$) and the value adopted by [3] respectively; $\langle \sigma_U \rangle$ and $\langle \sigma_{Th} \rangle$ are the integrated IBD cross-section (see Section 2). The error propagation $S_{BC}$(U+Th) is performed considering $S_{BC}$(U) and $S_{BC}$(Th) fully positive correlated.

For Borexino and KamLAND, the $S_{BC}$(U+Th) of [24] and the relative uncertainties, are calculated rescaling the values reported in the reference (calculated with $\langle P_{ee} \rangle = 0.56$) for the updated $\langle P_{ee} \rangle = 0.55$.

For KamLAND and Borexino the $S_{NFC}$(U+Th), $S_{FFC}$(U+Th) and $S_{BC}$(U+Th) of [20] are obtained summing the corresponding U and Th contributions reported and considering it fully positive correlated.

For Borexino, the $S_{NFC}$(U+Th), $S_{FFC}$(U+Th) and $S_{BC}$(U+Th) as [180] and the relative uncertainties, are calculated rescaling the values reported in the reference (calculated with $\langle P_{ee} \rangle = 0.57$) for the updated $\langle P_{ee} \rangle = 0.55$.

The $S_{NFC}$(U+Th) expected at Borexino of [10] is reported as it appears in the reference while the $S_{FFC}$(U+Th) is taken from H13, since the same inputs are used. The $S_{BC}$(U+Th) is obtained summing the $S_{NFC}$(U+Th) an $S_{FFC}$(U+Th) contribution with the error propagation performed via a Monte Carlo sampling of HPEs abundances according to their PDF in order to propagate the asymmetrical uncertainties of the non-Gaussian distributions.

# Nomenclature

| Symbol | Definition | Unit/Value |
|---|---|---|
| $a(K)_X$ | Abundance of Potassium in the reservoir X | ng/g or mg/g |
| $a(Th)_X$ | Abundance of Uranium in the reservoir X | ng/g or mg/g |
| $a(U)_X$ | Abundance of Thorium in the reservoir X | ng/g or mg/g |
| $A_{CT}$ | Surface area - continents | $10^6$ km$^2$ |
| $A_{OC}$ | Surface area - oceans | $10^6$ km$^2$ |
| A10 | Arevalo, 2010 | - |
| BC | Bulk Crust | - |
| BSE | Bulk Silicate Earth | - |
| $c_0$ | Velocity of light vacuum | 299,792,458 m/s |
| ch | Chondrites | - |
| C | Secular cooling – Earth | TW |
| $C_X$ | Secular cooling of the reservoir X | TW |
| CC | Continental crust | - |
| CJPL | China Jinping Underground Laboratory | - |
| CNO | Carbon Nitrogen Oxygen | - |
| CT | Central Tile (NFC of Borexino) | - |
| CTF | Counting Test Facility | - |
| CMB | Core Mantle Boundary | - |
| DC | Delayed Coincidence | - |
| DM | Depleted mantle | - |
| $E_H$ | Energy of heat production | MeV |
| $E_{\bar{\nu}}$ | Energy of antineutrino | MeV |
| $E_{max}$ | Maximal energy of the emitted antineutrino | MeV |
| EM | Enriched Mantle | - |
| FV | Fiducial Volume | - |
| FFC | Far Field Crust | - |
| $f_C$ | Core-mantle differentiation factor | - |
| $f_D$ | Enriching factor due to volatilization | - |
| h | Specific isotopic heat production | W/kg |
| h' | Elemental specific heat production | W/kg |
| H | Radiogenic heat – Bulk Earth | TW |
| $H_X$ | Radiogenic heat in the reservoir X | TW |
| H13 | Huang et al. 2013 | - |
| HPE | Heat Producing Element | - |
| IB | Inner Balloon of KamLAND detector | - |
| IBD | Inverse Beta Decay | - |
| IC | Inner Core | - |
| ID | Inner Detector | - |
| IV | Inner Vessel | - |
| J10 | Javoy et al., 2010 | |
| JJ13 | Jackson and Jellinek, 2013 | - |
| JK14 | Javoy and Kaminski, 2014 | |
| JS | Japan Sea | |
| JUNO | Jiangmen Underground Neutrino Observatory | - |

| Symbol | Definition | Unit/Value |
|---|---|---|
| L | Distance travelled by the antineutrino from its emission point | m |
| LAB | Lithosphere–Asthenosphere Boundary | - |
| LLSVP | Large Low Velocity Province | - |
| LC | Lower Crust | - |
| LK07 | Lyubetskaya and Korenaga, 2007 | - |
| LM | Lower Mantle | - |
| LNGS | Laboratori Nazionali del Gran Sasso | - |
| LS | Lithosphere | - |
| M | Mantle | - |
| $M_x$ | Mass of the BSE | kg |
| $M_C$ | Mass of the Core | kg |
| $M(K)_X$ | Mass of Potassium in the reservoir X | kg |
| $M(Th)_X$ | Mass of Uranium in the reservoir X | kg |
| $M(U)_X$ | Mass of Thorium in the reservoir X | kg |
| $M_V$ | Mass of the volatilized material | kg |
| $m_p$ | Mass of parent nuclide | kg |
| $m_d$ | Mass of daughter nuclide | kg |
| MC | Middle Crust | - |
| MLP | Multi-Layer perceptron | - |
| MOHO | Mohorovicic discontinuity | - |
| MORB | Mid Ocean Ridge Basalts | - |
| MS95 | McDonough and Sun, 1995 | - |
| N | Number of antineutrinos emitted per decay of the parent nucleus | decay$^{-1}$ |
| $N_p$ | Number of proton targets available in the detector | - |
| $N_U$ | Number of U geoneutrino events | - |
| $N_{Th}$ | Number of Th geoneutrino events | - |
| $N_{geo}$ | Number of total geoneutrino events | - |
| NFC | Near Field Crust | - |
| OBD | Ocean Bottom Detector | - |
| OC | Outer Core | - |
| OV | Outer Vessel | - |
| OD | Outer Detector | - |
| OIB | Ocean Island Basalts | - |
| OP08 | O'Neill and Palme, 2008 | - |
| p.e. | Photoelectrons | - |
| P$_{ee}$ | Electron antineutrino survival probability | - |
| PC | Pseudocumene | - |
| PDF | Probability Density Function | |
| PO07 | Palme and O'Neill, 2007 | - |
| PO14 | Palme and O'Neill, 2014 | - |
| PMT | Photomultiplier Tube | - |
| Q | Integrated terrestrial surface heat power | TW |
| q$_{CT}$ | Mean heat flux - continents | mWm$^{-2}$ |
| q$_{OC}$ | Mean heat flux - oceans | mWm$^{-2}$ |
| Q$_{CT}$ | Heat power - continents | TW |
| Q$_{OCS}$ | Heat power - oceans | TW |

| Symbol | Definition | Unit/Value |
| --- | --- | --- |
| RLE | Refractory Lithophile Elements | - |
| RR | Rest of region (NFC of Borexino) | - |
| SED | Sedimentary layer | - |
| SNO+ | Sudbury Neutrino Observatory | - |
| SSS | Stainless Steel Sphere | - |
| $S_X(U+Th)$ | Geoneutrino signal from U and Th in the reservoir X | TNU |
| $Sp_i(i, E_{\bar{\nu}})$ | Energy spectra of the produced geoneutrino of the i-th HPE | - |
| T | Exposure time | - |
| $T_{1/2}$ | Half life | Myr |
| $T_C$ | Condensation temperature | K |
| T02 | Turcotte, 2002 | - |
| T04 | Turcotte, 2004 | - |
| TNU | Terrestrial Neutrino Unit | - |
| $U_R$ | Urey ratio | - |
| UC | Upper crust | - |
| ULVZ | Ultra Low Velocity Zone | - |
| UM | Upper Mantle | - |
| $v_p$ | Seismic velocity of primary compressional waves | km/s |
| $v_s$ | Seismic velocity of secondary shear waves | km/s |
| $V_X$ | Seismic velocity in the reservoir X | km/s |
| W18 | Wang et al., 2018 | - |
| W20-C2 | Wipperfurth et al. 2020 based on Crust 2.0 | - |
| W20-C1 | Wipperfurth et al. 2020 based on Crust 1.0 | - |
| W20-L1 | Wipperfurth et al. 2020 based on Litho 1.0 | - |
| $X_{iso}$ | Natural isotopic abundance | - |
| $\lambda$ | Decay constant | $s^{-1}$ |
| $\epsilon_{\bar{\nu}}$ | Antineutrino production rates for unit mass of the isotope | $kg^{-1} s^{-1}$ |
| $\epsilon'_{\bar{\nu}}$ | Antineutrino production rates for unit mass at natural isotopic abundance | $kg^{-1} s^{-1}$ |
| $\Phi_i$ | Unoscillated geoneutrino flux of the i-th HPE | $cm^{-2} s^{-1}$ |
| $\theta_{12}, \theta_{13}$ and $\theta_{23}$ | Mixing angles between neutrinos eigenstates | rad |
| $\delta m^2$ and $\Delta m^2$ | Square mass differences between neutrinos eigenstates | MeV |
| $\eta$ | Detector efficiency | - |
| $\sigma$ | IBD cross section | $cm^2$ |
| $\rho$ | Mass density | $g/cm^3$ or $kg/m^3$ |
| $\bar{\nu}_e$ | Electron-flavoured antineutrino | - |